\documentclass[letterpaper]{JHEP3}
\usepackage{epsfig}
\usepackage{amsmath,amssymb,graphicx}
\usepackage{pstricks}
\newcommand\fverb{\setbox\pippobox=\hbox\bgroup\verb}
\newcommand\fverbdo{\egroup\medskip\noindent%
			\fbox{\unhbox\pippobox}\ }
\newcommand\fverbit{\egroup\item[\fbox{\unhbox\pippobox}]}
\newbox\pippobox

\newcommand{\beq}{\begin{eqnarray}}
\newcommand{\eeq}{\end{eqnarray}}
\newcommand{\be}{\begin{equation}}
\newcommand{\ee}{\end{equation}}

\title{Baryon Number in Warped GUTs : \\ Model Building 
and (Dark Matter Related) Phenomenology} 

\author{ Kaustubh Agashe\\
         Department of Physics and Astronomy, John Hopkins University,
Baltimore, MD 21218-2686 \\
         E-mail: \email{kagashe@pha.jhu.edu}}
\author{G\'eraldine Servant \\
         High Energy Physics Division, Argonne National Laboratory, Argonne,
        Illinois 60439, \\
        Enrico Fermi Institute, The University of Chicago,
         Chicago, Illinois 60637,  \\
          Service de Physique Th\'eorique, CEA Saclay, F91191 Gif--sur--Yvette, France\\ 
        E-mail: \email{servant@spht.saclay.cea.fr}}

\preprint{ANL-HEP-PR-04-18\\
 EFI-04-06\\
 SPhT-T04/139\\
hep-ph/0411254}

\abstract {In the past year, a new non-supersymmetric framework for 
electroweak symmetry breaking (with or without Higgs) involving 
$SU(2)_L \times SU(2)_R \times U(1)_{B-L}$ in higher dimensional 
warped 
geometry has been suggested. In this work, we embed this gauge structure into 
a GUT such as $SO(10)$ or Pati--Salam. We showed recently (in hep-ph/0403143) 
that in a warped GUT, 
a stable Kaluza--Klein fermion can arise as a consequence of imposing 
proton stability. 
Here, we specify a complete realistic model  where this particle is a weakly interacting 
right--handed neutrino,
and present a detailed study 
of this 
new dark matter candidate, providing relic density and detection predictions. 
We discuss 
phenomenological aspects associated with the existence of 
other light ($\lesssim$ TeV) KK fermions (related
to the neutrino), whose lightness is a direct 
consequence of the top quark's heaviness.
The AdS/CFT interpretation of this construction is also presented. 
Most of our qualitative results do 
not depend on the nature of the breaking 
of the electroweak symmetry provided that it happens near the TeV brane. }
\keywords{Theories beyond the SM, Physics of the Early Universe} 

\begin{document}
\section{Introduction}
\label{sec:intro}

Five years ago, Randall and Sundrum (RS) \cite{rs1} 
proposed a solution to the gauge hierarchy problem which does not rely on supersymmetry but instead makes use of extra dimensions. Their background geometry is a slice of five-dimensional Anti-de-Sitter space with curvature scale $k$ of order the Planck scale. 
Due to the AdS warping, an exponential hierarchy between the mass scales at the two ends of the extra dimension is generated. The Higgs is localized at the end point (denoted the TeV or IR brane) where the cut-off is low, thus its mass is protected,
 whereas the high scale of gravity is generated at the other end (Planck or UV brane). 
In their original set-up, all standard model (SM) fields are localized on the TeV brane.
In this case, the effective UV cut-off for gauge
and fermion fields, in addition to the Higgs, is a few TeV. This leads to dangerous unsuppressed processes such as flavour changing neutral currents (FCNCs) and proton decay.
Of course, one can always tune the coefficients of higher-dimensional operators to be small
so that  phenomenological issues such as flavor structure, 
gauge coupling unification, proton stability, and compatibility with 
electroweak precision tests become
sensitive to the UV completion (at a scale of a few TeV) of the 
original RS effective field theory.

An alternative and more attractive solution is that only the Higgs is localized on the TeV brane (that is all that is needed to solve the hierarchy problem) and SM gauge fields and
fermions live in the bulk of AdS$_5$ \cite{hewett, pomarol1, neubert, gp}. 
An interesting aspect of promoting fermion fields to be bulk fields is that it provides 
a simple mechanism for 
generating the Yukawa structure without fundamental hierarchies in the 5-dimensional RS action
\cite{neubert, gp, huber}.
Furthermore, the same mechanism automatically protects the theory from 
excessive FCNC's \cite{gp, huber}.
There is also a strong motivation for having gauge fields in the bulk of AdS. It has been shown that in this case, gauge couplings still ``evolve'' logarithmically
\cite{pomarol2, running, continorunning, choi}. 
This leads to the intriguing possibility of constructing models 
which preserve unification at the usual (high)
scale $\sim 10^{ 16 }$ GeV
and at the same time possess Kaluza--Klein (KK) excitations at the TeV scale \cite{pomarol2, gns, rsgut}.  Indeed, while the proper distance, $r_c$, between the two branes is of order $1/M_{Planck}$, the masses of the low-lying KK excitations of 
bulk fields are of order TeV.

Despite these virtues, it has been realized that for the theory to 
pass the electroweak precision tests 
without having to push the IR scale  too high (larger than $\sim 10$ TeV
\cite{hewett, EWPM}), 
an additional ingredient was needed: a custodial isospin symmetry, like there is in the SM.
As pointed out in \cite{custodial1} and as it can be understood 
from the AdS/CFT correspondence, for the dual CFT/$4D$
Higgs sector to enjoy a {\it global} custodial symmetry, there should be a {\it gauge} custodial isospin symmetry
in the RS bulk. This means that the gauge group of the electroweak sector should be enlarged to $SU(2)_L \times SU(2)_R \times U(1)_{B-L}$. Thanks to this new symmetry, the IR scale, given by $k e^{-k\pi r_c}$, and which also corresponds to the first KK mass scale,
can be lowered to 3 TeV and still be consistent with electroweak precision constraints\footnote{Brane kinetic terms for 
gauge and fermion fields 
\cite{braneterm}
could also help in lowering the IR scale.}. This
is a major step in diminishing the little hierarchy problem in RS. 
This gauge symmetry has also been used in Higgsless models 
in warped geometry \cite{custodial2}. 

A major generic problem in RS models, as well as in many extensions of the SM,
has to do with baryon number violation. 
A source of baryon-number violation 
in any RS model is 
higher-dimensional
operators suppressed by a low cut-off near the
TeV brane. 
One solution to forbid these dangerous operators 
is to impose (gauged) baryon-number symmetry \cite{gns, rsgut}.

However, when contemplating the possibility of a grand unified theory
(GUT), there is 
additional proton decay via $X,Y$ exchange between quarks and 
leptons from the same multiplet. 
So, the question arises: how can baryon-number symmetry be consistent with a GUT?
The answer is to
break the $5D$ GUT by boundary conditions (BC)
\cite{orbifoldnew, bhn, jmr} in such a way that 
SM quarks and leptons come from different multiplets \cite{bhn, jmr}\footnote{Thus, there is no
$4D$ GUT to cause inconsistency between the GUT and the baryon symmetry.}.
Concretely, the $5D$ multiplet with quark zero-mode contains lepton-like states,
but with only KK modes: this whole multiplet can be assigned baryon-number
$1/3$. 
The $5D$ GUT partners which
do not have zero modes couple
to SM quarks via the exchange of TeV mass $X,Y$ KK modes without causing phenomenological 
problems. 
Similarly, the multiplet with lepton zero-mode has
KK quark-like states carrying zero baryon-number.

We see that the KK  GUT partners of SM fermions are exotic since they carry 
baryon-number, but no color or vice versa.
To be precise, these KK fermions (and also $X$, $Y$ gauge bosons) are
charged under a $Z_3$ symmetry which is a
combination of color and baryon-number. SM 
particles are not
charged under this $Z_3$. This implies that the
lightest $Z_3$-charged particle (LZP) is stable 
hence a possible dark matter (DM) candidate if it is neutral \cite{us}. To repeat, this is  
a consequence of requiring baryon number symmetry.
This is reminiscent of 
SUSY, where imposing $R$-parity 
(which distinguishes between SM particles and 
their {\em SUSY} partners, just like the $Z_3$ symmetry
above distinguishes SM particles from their $5D$ {\em GUT} partners)
to suppress proton decay 
results in the lightest supersymmetric particle (LSP) being stable.

Of course, to be a good DM candidate, the LZP
has to have the proper mass and interactions.
In SUSY, if the LSP is a neutralino with a weak scale mass, then it has
weak scale interactions and it is a suitable
WIMP.
As we will show in detail, the LZP 
is 
a GUT partner of the top quark and, 
as a consequence of the heaviness of the top quark, its mass
can be $O(100)$ GeV, meaning that
it can be naturally much lighter than the other KK modes (which have a mass of
a few TeV).

The interactions of the LZP depend on its gauge quantum numbers.
As mentioned above, a custodial isospin gauge symmetry
is crucial in
ameliorating the little hierarchy problem in RS by allowing
KK scale of a few TeV. We will consequently
concentrate on GUTs which contain this gauge symmetry. 
This leads us to discard warped $SU(5)$ models, the only ones 
which had been studied in detail so far and we will instead focus  on 
non supersymmetric $SO(10)$ and Pati--Salam gauge theories in warped space.
In Pati-Salam or $SO(10)$ GUT, the
LZP can have gauge quantum numbers of a RH neutrino. In this case, the 
LZP has no SM gauge interactions. It interacts by the exchange of
heavy (a few TeV), but strongly coupled non-SM gauge KK modes (with no
zero-mode). This, combined with its weak-scale mass,
implies that annihilation and detection cross-sections
are of weak-scale size, making it a good DM candidate \cite{us}.

An alternative solution for suppressing proton decay is to impose
a gauged lepton number symmetry. 
In this case, we do not obtain a stable particle (unlike the case with baryon-number
symmetry). This is similar to 
SUSY, where imposing lepton number only (instead of $R$-parity) suffices
to suppress proton decay, but then the LSP is unstable. 
In this paper, we focus on imposing baryon-number symmetry to solve
the proton decay problem since it gives a DM candidate, but we will discuss the alternative solution at the end of section \ref{subsection:gaugedlepton}. In any case, there is 
no fundamental ``stringy'' reason or naturalness argument to choose one 
possibility over the other.
 
In this article, we develop on the toy model presented in \cite{us}. We start by 
reviewing 
what the baryon number violation problem is in RS.  We then introduce the implementation of  the baryon number symmetry in warped GUT and show how this leads to a stable KK particle. Next we explain why this particle can be much lighter than 3 TeV. From section 5 to 8 we discuss model building associated with  Pati--Salam and SO(10) gauge groups in higher dimensional warped geometry.
Sections 9 and 10 detail the interactions of the KK right-handed neutrino, section 11 the values of GUT gauge couplings. We present predictions for the dark matter relic density in section 12, for direct detection  in section 13, and  indirect detection in section 14. 
Collider phenomenology of other light KK partners of the top quark
is treated in section 15.
 Section 16 discusses issues related to baryogenesis before we finally present our conclusions. 
This construction has a nice AdS/CFT interpretation which is reviewed separately in an appendix.
Other technical details can be found in the appendices as well.
 
 \section{Baryon number violation in Randall--Sundrum geometries}
 
 Let us start by reviewing
what the baryon number violation problem is in higher dimensional warped geometry.
 We work in the context of RS1 \cite{rs1} where the extra dimension is an orbifolded circle of radius $r_c$
with the 
Planck brane at $\theta=0$ and the TeV brane at 
$\theta = \pi$. The geometry is a compact slice of AdS$_5$, with curvature scale
 $k$ of order $M_{ Pl }$, the $4D$ Planck scale, with metric \cite{rs1}:
\begin{equation} 
 ds ^2 =  e^{-2k r_c |\theta|} \eta_{\mu \nu} dx^{\mu} 
dx^{\nu} + r_c^2 d \theta ^2 =  \frac{1}{ ( kz )^2 } \big[ \eta_{ \mu \nu }
d x ^{ \mu } d x^{ \nu } + ( d z ) ^2 \big], 
\label{metric}
\end{equation}
where in the last step it has been written in terms of the coordinate $z \equiv  \; e^{k r_c |\theta|}/k$ and 
\begin{equation}
\left( z_h \equiv \frac{1}{k} \right) 
\leq  z \leq \left( z_v \equiv \frac{ e^{ k \pi r_c } }{k} \right).
\end{equation}
The Planck brane is located at $z_h$ and the TeV brane at $z_v$.
We take $z_v \sim \mbox{TeV}^{-1}$, i.e. $ k \pi r_c  \sim \log \left( M_{Pl} / \hbox{TeV} \right)  \sim 30 $ 
to solve the hierarchy problem. As already said in the introduction, all SM gauge fields and fermions
are taken to be bulk fields. Only the Higgs (or alternative dynamics for EW symmetry breaking) needs to be localized on or near the TeV brane to solve the hierarchy problem.

\subsection{Bulk fermion: $c$ parameter and Yukawa couplings}
\label{fermionmodel}

The general 5-dimensional bulk lagrangian for a given fermion $\Psi$  is:
\begin{equation}
{\cal L}_{fermion}=\sqrt{g}\left( i\bar{\Psi}\Gamma^M D_M \Psi-\epsilon( \theta ) k \; c_{\Psi}
\bar{\Psi} \Psi + \epsilon ( \theta ) 
\frac{ a^{ \prime } }{ \sqrt{ \Lambda } } \Sigma \bar{\Psi} \Psi \right)
\label{fermions}
\end{equation}
where $\epsilon( \theta )$ is the sign function and appears if we compactify on a $Z_2$ orbifold rather than just an interval. Even though it will seem
that we are adding a mass term, $c_{\Psi}$ is compatible with a massless zero
mode of the 4D effective theory \cite{neubert,gp}. Zero modes are identified with the SM fermions.
The $c$ parameters control the localization of the zero modes and
offer a simple and attractive  mechanism for obtaining
hierarchical $4D$ 
Yukawa couplings without hierarchies
in $5D$ Yukawa couplings  \cite{neubert, gp}. 
4D Yukawa couplings depend very sensitively (exponentially for $c > 1/2$) 
on the value of $c$.
In short (see wavefunction in
Eq. \ref{zero}), light fermions have $c > 1/2$ (typically between 0.6 and 0.8) and 
are localized near the Planck brane. Their 
$4D$ Yukawa couplings are suppressed because of the small overlap of their wave functions 
with the Higgs on the TeV brane. 
Left-handed top and bottom quarks are close to $c = 1/2$ 
(but $< 1/2$) -- as 
shown in reference \cite{custodial1}, $c_{\tiny{t_L,b_L}} \sim 0.3-0.4$ is necessary to
be consistent with $Z \rightarrow \bar{b}_L b_L$
for KK masses $\sim 3-4$ TeV, whereas for KK
mass $\sim 6$ TeV, $c_{\tiny{t_L,b_L}}$ can be as small as $0$. 
Thus, in order to obtain $O(1)$ top
Yukawa,
the right-handed top quark must be localized 
near the TeV brane:
\begin{eqnarray}
c_{ t_R } & \lesssim & 0
\end{eqnarray}
%
As we will see later, this fact is very crucial for our DM scenario to work.
The right-handed bottom quark is localized near the Planck brane
($c > 1/2$)
to obtain the $m_t/m_b$ hierarchy.
With this set-up, FCNC's 
from exchange of 
both gauge KK modes and  ``string states'' (parametrized
by higher-dimensional flavor-violating local operators in our effective field theory) are also 
suppressed. See references \cite{gp, huber, bphysics} 
for details. The last term in (\ref{fermions}) will generate an additional bulk mass term if the bulk scalar field $\Sigma$ gets a vev. This effect will be discussed later in section \ref{bulkbreaking}.

\subsection{Effective four-fermion operators}

 The dangerous baryon number violating interactions come from effective 4-fermion operators, which, after dimensional reduction lead to \cite{gp}: 
\begin{eqnarray}
\int dy \ d^4x \ \sqrt{-g} \ \frac{\overline{\Psi}_i\Psi_j\overline{\Psi}_k\Psi_l}{M_5^3}\sim 
\int d^4x \ e^{\pi k r_c(4-c_i-c_j-c_k-c_l)} \ \frac{ {\overline{\psi}}_i^{(0)} \psi_j^{(0)}
 {\overline{\psi}}_k^{(0)} \psi_l^{(0)}}{m_{Pl}^2} 
 \label{dangerous}
\end{eqnarray}
where $i,j,k,l$ are flavor indices and the $\psi^{(0)} $ are the 4D zero mode fermions identified with the SM fermions.
To obtain a Planck or GUT scale suppression of this operator
(as required by the limit on proton lifetime),  the $c$'s have to be larger than 1, meaning that zero mode fermions should be very close to the UV brane.
Unfortunately, this is incompatible with the Yukawa structure,
 which requires that all $c$'s be smaller than 1 according to the previous subsection. 
 
\subsection{Additional violations due to KK GUT gauge boson exchange}

When working in a GUT, there is an additional potential problem coming from the
exchange of grand unified gauge bosons, such as X/Y gauge bosons. 
In a warped GUT, these gauge bosons 
have TeV and not GUT scale mass and mediate fast proton decay.
It turns out that all TeV KK modes and therefore X/Y TeV KK gauge bosons are 
localized near the TeV brane. Their 
 interactions with zero mode fermions will be suppressed only if fermions are localized very close to the Planck brane, again requiring that $c$'s be larger than 1. 
This problem arises in any GUT 
theory where the X/Y gauge bosons propagate in extra dimensions with size 
larger than $M_{GUT}^{-1} \sim 1 / \left( 10^{ 16 } \hbox{GeV} \right)$.
A simple solution to this problem suggested by  \cite{bhn, jmr} 
is to 
break the higher-dimensional GUT by boundary conditions (BC) (or on branes) so that
there is no $4D$ GUT and 
SM quarks and leptons 
can come from different GUT multiplets.
Concretely, this means that BC  are not the same for all components of a 
given (gauge or fermion) GUT multiplet so that only part of the fields in a multiplet acquire zero modes, which are identified with SM particles. 
While this circumvents the problem of baryon number violation due to KK X/Y exchange
(since $X$, $Y$ gauge bosons
do not couple to two 
SM fermion zero-modes), one still has to cure the baryon number violation due to the effective operator (\ref{dangerous}). This is done by imposing an additional symmetry. 
In the $SU(5)$ models of  
\cite{gns, rsgut}, an additional  $\tilde{U}(1)$ symmetry is imposed and usual baryon number corresponds to a linear combination of hypercharge and this additional $\tilde{U}(1)$.
Our approach in the following is slightly different. 
The additional $U(1)$ we impose really corresponds to baryon number.

\section{Imposing baryon number symmetry $U(1)_B$}
\subsection{Replication of fundamental representations and boundary breaking of the GUT}
It is clear that
for baryon number symmetry to commute with the grand unified gauge group, we need to 
replicate the number of fundamental representations so that we can 
obtain quarks and leptons
from different multiplets. 
In any case, 
we saw previously that SM quarks and leptons have to come from different fundamental representations and that at least a doubling of representations was needed to avoid the 
existence of a vertex involving a SM quark, a SM lepton and a TeV X/Y type of gauge boson
leading to fast proton decay. 
So, we choose to break GUT by BC.
Thus, BC breaking of $5D$ GUT not only
gets rid of proton decay by $X$, $Y$ exchange, but
also allows us to implement baryon number symmetry
by assigning each multiplet a baryonic charge of the SM fermion contained in it.
We need at least three fundamental representations to be able to reproduce the SM baryonic charges -1/3, +1/3 and 0.  
One may dislike the fact that in these models, SM quarks and leptons are no more unified. However, there are still motivations for considering a unified gauge symmetry. First, this provides an explanation for quantization of charges \cite{gns, rsgut}. 
Second, it allows unification of gauge couplings at high scale \cite{gns, rsgut}. In addition, one may see this splitting as a virtue
since 
$SU(5)$ quark-lepton mass relations which are inconsistent with 
data are no more present.
Let us discuss GUT breaking by boundary conditions more explicitly.

The unified gauge symmetry is broken by boundary conditions  reflecting
the dynamics taking place on the Planck and TeV branes.
As a simplification, this is commonly modelled by either Neumann ($+$)
or Dirichlet ($-$) BC\footnote{for a comprehensive description
 of boundary conditions of  fermions on an interval, see \cite{Csaki:2003sh}.} in orbifold compactifications. 
 5D fermions lead to two chiral fermions in 4D, one 
of which only gets a zero mode to reproduce the chiral SM fermion. 
SM fermions are associated with ($++$) BC (first sign is for Planck brane, second
 for TeV brane).
The other chirality is ($--$) and does not have zero mode. 
 In the language of orbifold boundary conditions, this involves 
  replacing the usual $Z_2$ orbifold projection by a 
$Z_2 \times Z^{ \prime}_2$ 
orbifold projection, where $Z_2$ corresponds to reflection about the Planck brane
and $Z^{ \prime }_2$ corresponds to reflection about the TeV brane.
The breaking of the unified gauge group to the SM 
is achieved by assigning on the Planck brane Neumann boundary conditions
for $\mu$-components of SM gauge bosons  and
Dirichlet boundary conditions for GUT gauge bosons  which are not SM gauge bosons.
On the TeV brane, all gauge bosons have Neumann boundary conditions. 
Non standard gauge bosons therefore have ($-+$) boundary conditions. Similarly, 
fermionic GUT partners  of subsection 2.3 which do not have zero modes have 
($-+$) boundary conditions. 

\subsection{$Z_3$ 
symmetry}
As soon as baryon number is promoted to be a conserved quantum number, the following transformation becomes a 
symmetry:
\begin{equation}
\Phi \rightarrow e^{ 2 \pi i \left( B  - \frac{ n_c - \bar{ n }_c }{3} 
\right) } \Phi
\end{equation}
where $B$ is baryon-number of  a given field $\Phi$ (proton has baryon-number $+1$) and  
$n_c$ ($\bar{n}_c$) is its number of colors (anti-colors).

SM particles are clearly not charged under $Z_3$. However, exotic states such as colored grand unified gauge bosons and most KK fermions with no zero modes 
($(-+)$ BC) are charged under $Z_3$
since they have the ``wrong'' combination of color
and baryon number.
For instance, 
since all fermions within a given GUT multiplet are assigned the same $B$, 
that of the zero-mode within that multiplet, 
the multiplet with the SM quark contains lepton-like KK states with $B = 1/3$
(denoted by ``prime'', for example, $L^{ \prime }$ is the GUT partner of SM
$d_R$ etc). 
Similarly, there are quark-like states carrying $B = 0$ in the multiplet with the SM lepton,
like $d^{ \prime }_R$, the GUT partner of SM $L$.
Also, colored $X$, $Y$ have $B = 0$.
As a consequence of the $Z_3$ symmetry, the lightest $Z_3$ charged particle (LZP) cannot decay into SM particles and is stable.

\subsection{Breaking gauged baryon number symmetry}

We need to gauge $U(1)_B$ in the bulk
since quantum gravity effects do not 
respect global
symmetries. 
Note that
$4D$ black holes (BH) violate $B$ at the Planck scale.
However, 
in RS, we expect the presence of $5D$ BH of TeV mass localized near the TeV 
brane \cite{Giddings:2000ay}, leading to {\em TeV} scale violation of $B$.
Basically, the effects of $5D$ BH can be parameterized by 
higher-dimensional operators suppressed by the local
$5D$ gravity scale.
We require the gauging of $B$ to protect against such effects.
Since
$5D$ fermions are vector-like, the 
$5D$ $U(1)_B$ 
gauge theory is not anomalous. However, once the orbifold projection is implemented, 
we have to worry about anomalies
from SM (zero-mode) fermions. Spectators are added 
on the Planck brane to cancel these anomalies. They are
vector-like under the SM (no pure SM anomalies) and chiral under $U(1)_B$
to cancel pure $U(1)_B$ and SM $\times U(1)_B$ anomalies (see \cite{rsgut} for 
a similar procedure in the case of warped $SU(5)$)\footnote{
Reference \cite{Lee:2002mn} also makes use of a gauged $U(1)$ symmetry to supress baryon-number violation
in a supersymmetric model with a flat extra dimension and a low fundamental scale.}.

This gauge symmetry has to be broken otherwise it would lead to the existence of a new
massless gauge boson.
We break B spontaneously
on the Planck brane  so that the $U(1)_B$ gauge boson and the spectators  
get heavy.
As a result, any 
baryon number violating operators will 
have to be localized on the Planck brane. 
Naively, we are safe since we get Planck-scale suppression for 
the operators giving proton decay, for example, $Q_L^3 L_L$.
However, there is a subtlety, namely, a restriction on how $B$ is broken as follows.

If $B$ is broken by a scalar field with arbitrary baryonic charge,
then the mass term $\bar{ L } \hat{L}^{ \prime }$
is allowed on the Planck brane,
where $\hat{L}^{  \prime }$ refers
to  the $5D$ Dirac partner of $L^{ \prime }$ from the multiplet with $d_R$
zero-mode. 
Even though the lowest $L^{ \prime }$ KK modes are localized
near the TeV brane and the zero-mode of $L$ is localized near the Planck brane,
this
mixing between the zero-mode of $L$ and KK mode of $L^{ \prime }$
results in a
sizable coupling (roughly proportional to the Yukawa) of $X,Y$ to the SM lepton (which has now an admixture of
$L^{ \prime }$) and $d_R$. Similarly, the mass term
$\bar{Q} \hat{Q}^{ \prime }$, where $Q^{ \prime }$ is from the multiplet
with $u_R$ zero-mode, is allowed. This leads to a coupling of $X,Y$ to
SM $Q$ and $u_R$. Then, $X,Y$ exchange leads to fast proton decay.

In order to forbid
such proton decay, we require that B is not broken
by $1/3$ or $2/3$ unit. It turns out that $\Delta B \neq 1/3, 2/3$ is 
enough to guarantee the stability of the LZP.
To see this,
suppose that the LZP is a color singlet with $B = 1/3$ 
(it will be the case in our model but this argument can be generalized). 
Since a color singlet SM final state
can only have integer $B$, 
$\Delta B \neq 1/3, 2/3$ implies that the LZP cannot decay into SM states.
Of course, some symmetry has to enforce $\Delta B \neq 1/3, 2/3$. 
For example, we can simply impose the $Z_3$ symmetry 
which clearly implies that $\Delta B \neq 1/3, 2/3$ and that the LZP is
stable.
$Z_3$ is imposed for proton stability and the existence of  
a stable particle is a spin-off (just like in
the MSSM).

Note that if $\Delta B = 1$, then, while the LZP is 
absolutely stable, proton decay
is still allowed via, for example, the  $Q_L Q_L Q_L L_L $ operator. However,
as mentioned above, these operators are allowed only
on the Planck brane hence are suppressed by the Planck scale.
The point is that $5D$ BH near the TeV brane (which
were the cause of the problem) cannot violate $B$.
Indeed, from the $5D$
point of view,  $B$ is an unbroken gauge symmetry near the TeV brane: there 
are KK modes of $B$ gauge boson,
even though there is no zero-mode. 
The only location where $B$ is not a gauge 
symmetry and where BH can violate $B$ is the Planck brane. The scale
suppressing these operators is the $5D$ gravity scale at the Planck 
brane which is $\sim 10^{18}$ GeV.
Below the lightest KK mass, 
the $4D$ effective theory has an accidental $B$ conservation 
like in the SM (whereas $Z_3$ is an exact symmetry).
$B$ can be understood as a {\em global}
 symmetry at low energy and we expect that anomalous 
sphaleron processes are present so that baryogenesis can be achieved despite the 
existence of an underlying $5D$ {\em gauged} $B$ symmetry.

\section{Who is the lightest $Z_3$ charged particle?}
\label{sec:who}
We have gained confidence that consistent (as far as baryon number violation is concerned) non supersymmetric\footnote{If the model is supersymmetric, the Higgs can be localized on the Planck brane  as well as the fermions so that the Planck scale
suppression of 
baryon number violation can be achieved and it may not be necessary to impose baryon number symmetry. However, in these models, one loses the geometrical explanation 
for the Yukawa structure.  
See subsection \ref{subsec:warpedSUSY} for more comments.} warped GUT theories can exhibit a stable KK particle. We are interested in identifying this state since it has crucial consequences for cosmology and collider phenomenology. 
The literature so far has dealt with a single
 KK scale $\gtrsim$ 3 TeV, making it difficult to  
observe KK states in RS at high-energy colliders. 
This is because most of the work on the 
phenomenology of Randall-Sundrum geometries have
focused on a certain type of boundary conditions for fermionic 
fields.  In this work, we emphasize the interesting
consequences of boundary conditions which do not lead to zero 
modes but on the other hand may lead to very light observable Kaluza-Klein states.

Recall that $Z_3$-charged particles are $X, Y$ type gauge bosons
(with $(-+)$ BC) and most $(-+)$ fermions.
We now compare their spectrum.

\subsection{$(-+)$ KK fermions can be 
very light}

\begin{figure}[h]
\begin{center}
\includegraphics[height=11cm]{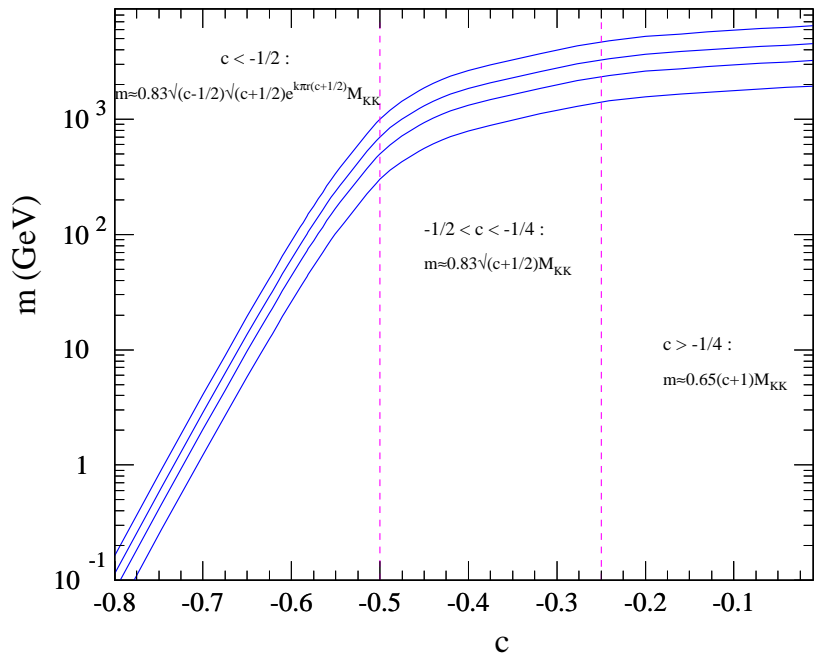}
\end{center}
\caption[]{Mass of the lightest ($-+$) KK fermion as a function of $c$ for different values of the KK gauge boson mass. From bottom to top, $M_{KK}=$3, 5, 7, 10 TeV.}
\label{fig:mLZP}
\end{figure}

When computing the KK spectrum of fermions 
one finds that for $c < 1/2$ the lightest KK fermion with $(-+)$ BC
is lighter than the lightest KK gauge boson:
\begin{equation}
m z_v  \approx\left\{ \begin{array}{ll}
  \frac{ \pi}{2} (1 + c )
\;  \  \  \hbox{for} \; \  c \gtrsim - 1/2\\
   2 \sqrt{ \alpha ( \alpha + 1 ) }  
\left( z_h/z_v \right)^{ \alpha } \ll 1 
\;  \  \  \hbox{for} \; \ c \lesssim - 1/2.\\
	\end{array} \right.
\label{KKfermion-+}
\end{equation}
where $\alpha = | c + 1/2 |$ and  $z_v = e^{ k \pi r_c } /k $.
See section \ref{fermionkk} for details. 
Here, $c$ refers to that of the $(++)$ zero-mode with identical
Lorentz helicity from the same multiplet (see below for a more
precise convention for $c$). 
For comparison, the mass of the lightest KK gauge boson (which we denote 
as KK scale of the model, $M_{ KK }$) is given by 
\begin{equation}
M_{KK} \approx z_v^{-1} 3 \pi /4
\label{MKK}
\end{equation}
for both
$(++)$ and $(-+)$ BC. Note the particular 
case  $c < -1/2$, for which the mass of this KK fermion is exponentially
smaller than that of the gauge KK mode.
We plot in Fig. \ref{fig:mLZP} the mass of the lightest ($-+$) KK fermion as a function of $c$ and for different values of $M_{KK}$.
There is an intuitive argument for the lightness of
the KK fermion (see also section \ref{cftlight} for its CFT interpretation):
for $c \ll 1/2$, the zero-mode of the  fermion with $(++)$ boundary condition
is localized near the TeV brane. 
Changing the boundary condition to $(-+)$
makes this ``would-be'' zero-mode massive, but since it is localized near the  TeV
brane, the effect of changing the boundary condition on the Planck
brane is suppressed, resulting in a small mass for the would-be zero-mode.

Let us take a detour on the chiralities of a KK fermion.
We realize SM fermions (zero-modes) as left-handed (LH) under the Lorentz group:
for example, the ${\bf 16}$
of $SO(10)$ contains the conjugate of $u_R$ etc.
For the $5D$ mass or the value of $c$ of a given multiplet, we will 
henceforth
use the
convention such that  if $c > (<) 1/2$, the LH zero-mode with $(++)$ BC
is localized near the Planck (TeV) brane.

As showed above, the $(-+)$ LH KK state is lighter than the gauge KK states
for $c < 1/2$ (and exponentially light for $c < -1/2$).
The KK mode being a Dirac fermion, its Dirac partner with  $(+-)$ BC and RH chirality (denoted by ``hat'', for example,
$\hat{L}^{ \prime }$) is also light (since the two helicities
obviously have the same spectrum). We
can show that the ``effective'' $c$
(i.e., the $c$ appearing
in equations of motion) for the RH helicity is {\em opposite} to that
of the LH helicity.
This implies that the $(+-)$ {\em left}-handed  KK states (and also their $(-+)$ RH
partners) are lighter than the gauge KK states for  $c > -1/2$ (exponentially light for $c > 1/2$).
For instance, we will consider later on a model where $SO(10)$ is broken
on the TeV brane in which case {\em left}-handed GUT partners of SM fermions (i.e., with
same chirality
as SM fermions) will have $(+-)$ BC\footnote{whereas in the
model with $SO(10)$ broken on the Planck brane,
they had $(-+)$ BC.} so that LH $(+-)$ KK partners of {\em light} SM fermions
(which have $c > 1/2$) will be exponentially light.

For simplicity, sometimes (as we did in the plot above)
we will refer to the LH chirality only (i.e. the same
Lorentz helicity as the zero-mode), but it is understood
that we mean the Dirac fermion.
The consideration of the other chirality of the $(-+)$ fermion gives another
intuitive understanding of its lightness as follows.
Changing BC on the Planck brane (where
$SO(10)$ is broken) from $(++)$ to $(-+)$ adds an extra (RH) $(+-)$ 
chirality which is localized 
near the Planck brane for $c \ll 1/2$ since the
change of BC is a small perturbation\footnote{For $c \gtrsim 1/2$
 this change of BC is not a small perturbation so that the
added helicity is not localized near the Planck brane.}.
Then, the
small overlap of the two chiralities (the
$(-+)$ LH chirality, i.e., would-be-zero-mode
is localized near the TeV brane) explains the small mass of the  $(-+)$ fermion.

\subsection{The LZP is likely to belong to the multiplet containing the SM right-handed top}
%
%
%
We have seen that 
$(-+)$ KK fermions are lighter than gauge KK states
for $c < 1/2$ so that the LZP is a $(-+)$ fermion from the
bulk multiplet having the smallest $c$ provided  $c < 1/2$
(see Fig. \ref{fig:mLZP}).
Recall that the smallest c is that of $t_R$
(see subsection \ref{fermionmodel}).
Hence, the LZP comes from the multiplet wich contains the $t_R$ zero mode. 
Moreover, its $c$ can be close to $-1/2$ so that  $m_{LZP} \ll$ TeV is possible.

At tree level and before any GUT breaking, all fields within a 
GUT multiplet have the same $c$. Loop corrections and bulk breaking of the GUT
 will lift the degeneracy between these KK masses. In the absence of a detailed loop calculation, we are unable to predict the mass spectrum and we will be guided by phenomenological requirements: the LZP should be colorless and electrically neutral if it is to account for dark matter.
In  Pati-Salam,  where the gauge group is $SU(4)_c \times SU(2)_L \times SU(2)_R$, bulk fermions are
${\bf (4, 2)}$ of $SU(4)_c \times
SU(2)_R$ and ${\bf (4, 2)}$ of $SU(4)_c \times
SU(2)_L$. 
So the LZP has gauge quantum numbers of
a right-handed (RH) lepton doublet since $\tilde{b}_R$ is neutral under $Z_3$ 
(tilded fermions 
denote $SU(2)_R$ partners of SM fermions 
and do not have zero modes). $e^{ \prime}_R$ can be 
heavier than $\nu^{ \prime }_R$ due to electroweak loop corrections
to KK masses (primed fermions denote $SU(4)_c$ partners and do not have zero modes). 

In  $SO(10)$, there are
additional  $Z_3$-charged quark-like states in the {\bf 16} GUT multiplet
containing $t_R$. 
These are probably 
heavier than $\nu^{ \prime }_R$ due to QCD loop corrections. Additional
$Z_3$-charged lepton-like states can again be heavier
than $\nu^{ \prime }_R$ due to electroweak loop corrections.
$\nu^{ \prime }_R$ is actually the 
only viable dark matter candidate. Indeed, it is well known that TeV {\em left}-handed neutrinos are excluded by direct detection experiments because of their large coupling to the $Z$ gauge boson \cite{Servant:2002hb}.
To ensure that $\nu^{ \prime }_R$ is the LZP (if  electroweak corrections are not enough)
we can make use of bulk 
breaking of the unified gauge group. This easily allows for splitting in $c$'s of the different component of the GUT multiplet (see section \ref{bulkbreaking}).

We are now ready to discuss in more details model-building issues. 
We start with the unified gauge symmetry in the bulk of AdS$_5$.
The gauge group can then be broken on the branes by boundary conditions or in the bulk 
by giving a vev to a scalar field.
As seen previously,  we are forced to break the GUT by boundary conditions to prevent proton decay.
In addition, we will find it useful 
to break it also in the bulk by a small amount.
For simplicity, we will start with the Pati--Salam model. We will then extend it to $SO(10)$ which can accomodate gauge coupling unification, just like $SU(5)$ as shown in reference \cite{rsgut}. 

\section{Pati-Salam model}

In the background of Eq. (\ref{metric}),
the lagrangian for our model reads:
\begin{equation}
S=\int d^4x \ dz \sqrt{g} \ 
\left( {\cal L}_{gauge}+
{\cal L}_{fermion} + 
{\cal L}_{ UV } \ \delta(z-z_h)+{ \cal L }_{ IR } \
\delta(z-z_v)\right)
\end{equation}
\noindent
${\cal L}_{gauge} + {\cal L}_{fermion}$ is the bulk lagrangian.
${\cal L}_{fermion}$ is given in Eq.~\ref{fermions}. We now
 focus on ${\cal L}_{gauge}$:
\begin{eqnarray}
{\cal L}_{gauge}&=&\sqrt{g}\big(-\frac{1}{4} Tr {W_L}_{   MN }
W_L^{  MN }
-\frac{1}{4} Tr {W_R}_{ MN } W_R^{MN} \nonumber\\
&& -\frac{1}{4} Tr F_{ MN}
F^{ MN} 
+ | D_M \Sigma |^2 - V (\Sigma) + \frac{ a_i }{ \Lambda^{ 3/2 } } 
\Sigma {F_i}_{  MN } F_i^{MN }\big)
\label{bulk}
\end{eqnarray}
where the indices are contracted with the bulk metric $g_{MN}$. ${W_L}_{ MN }$, ${W_R}_{  MN }$ and 
$F_{MN}$ are the field strengths for, respectively,  $SU(2)_L$, $SU(2)_R$
and $SU(4)_c$.  
$\Sigma$ is a scalar transforming under the Pati-Salam gauge symmetry. Its 
sole purpose is to spontaneously
break Pati-Salam to the SM gauge group at a mass scale below $k$. Specifically,
$\langle \Sigma \rangle \equiv v_{\Sigma}^{ 3/2}$ so that non standard gauge fields acquire
a bulk mass $\sim M_{ GUT } \sim g_{ 5D } v_{\Sigma}^{3/2}$.
The higher-dimensional operator 
coupling $\Sigma$ to the gauge fields 
gives threshold-type corrections 
to the low-energy gauge couplings (see Eq. \ref{g4D})
and is suppressed by $\Lambda$, the $5D$ cut-off of the RS effective field theory. 
We will discuss the motivation for this bulk breaking of GUT in
section \ref{bulkbreaking}.

${\cal L}_{ UV }$ includes the necessary fields to spontaneously break
$U(1)_R \times
U(1)_{B-L}$ to $U(1)_Y$ on the UV brane
and ${\cal L}_{ IR }$ contains the SM Higgs field,
a {\em bi}doublet of $SU(2)_L \times SU(2)_R$ (there is no Higgs triplet):
\begin{equation}
{\cal L}_{ IR }= {\cal L}_{Higgs} + {\cal L}_{Yukawa},
\end{equation}
${\cal L}_{Yukawa}$  generates Yukawa couplings for
fermions, it will be given in Eq.~\ref{Yukawa} and
\begin{equation}
{\cal L}_{Higgs}=
\sqrt{-g_{ IR }} \left( D_\mu H \big[ D^\mu H \big] ^{ \dagger } -V( H )\right).
\label{LHiggs}
\end{equation}
$g_{IR}$ is the induced flat space metric in the IR brane.  
After the usual field redefinition of $H$ \cite{rs1},
Eq.~(\ref{LHiggs}) takes its canonical form:
\begin{equation}
{\cal L}_{Higgs}= D_\mu H \big[ D^\mu H \big] ^{ \dagger } -V( H )
\label{LHiggscan}
\end{equation}
%
with $\langle H \rangle = 
\left( 
\begin{array}{c}
0 \\
v / \sqrt{2}
\end{array}
\right)$, $v \approx 250$ GeV.

We assume that brane-localized kinetic terms for bulk
fields are of order loop processes involving bulk couplings 
and are therefore neglected in our analysis. 

\subsection{Breaking of Pati--Salam on the UV brane}

$SU(4)_c \times SU(2)_L \times SU(2)_R$ is first broken
to $SU(3)_c \times SU(2)_L \times U(1)_R \times U(1)_{B-L}$\footnote{Here, we keep the usual standard appellation ``$B-L$'' denoting the extra $U(1)$ contained in Pati--Salam and $SO(10)$, however, it is clear that the  ``$B$'' in  ``$B-L$'' has nothing to do with the extra baryon number symmetry $U(1)_B$ we impose to protect proton stability.}  by assigning the following 
boundary conditions to the $\mu$-components of the gauge fields 
\cite{orbifoldnew, bhn, jmr}. 
\[
\begin{array}[t]{ccc}
 & \textrm{UV} & \textrm{IR}  \\
X_s & - & + \\
W_{R \; \mu}^{1,2}  & - & + \\
\hbox{other gauge fields} & + & + 
\end{array}
\]
This can be done by either orbifold BC or more general BC which approximately 
correspond to ($-+$) BC. 
On the other hand, the 
breaking of $U(1)_R\times U(1)_{ B - L }
\rightarrow U(1)_Y$ cannot be achieved by orbifold BC. 
There are two linear combinations of $W_{ R \; \mu}^3$ and $V_{ \mu }$, where
$V_{ \mu }$ denotes the $( B - L )$ gauge boson. 
One,  $B_\mu$, has ($++$) BC and is the hypercharge 
gauge boson, whereas the orthogonal combination, denoted by $Z^{ \prime }$, is spontaneously broken due to 
its coupling to a Planckian vev on the UV brane, 
which mimics $(-+)$ BC to 
a good approximation. 
\begin{equation}
Z^{ \prime }_\mu \equiv \frac{ g_{ 5 \; R } W_{ R \; \mu }^3 - \sqrt{ 3/2 } g_{ 5 \; c } 
V_{ \mu } }{ \sqrt{ g_{ 5 \; R } ^2 + 
\left( \sqrt{ 3/2 } \; g_{ 5 \; c } \right)^2 
} }  \qquad  \qquad  
B_{ \mu } \equiv  \frac{ \sqrt{ 3/2 } \; g_{ 5 \; c } W_{ R \; \mu}^3 + g_{ 5 \; R } 
V_\mu }{ \sqrt{ g_{ 5 \; R }^2 + \left( \sqrt{ 3/2 } \; g_{ 5 \; c } \right)^2 } }
\end{equation}
The electroweak covariant derivative reads
\begin{eqnarray}
D_M = 
\partial_M-i (  g_{ 5 \; L } W^a_{ L \; M }
\tau_{a \, L} + g_{ 5 \; R } W^a_{ R \; M } \tau_{a \, R} + \sqrt{ 3/2 } \; g_{ 5 \; c } 
V_M ( B - L) / 2 )
\end{eqnarray}
where $g_{ 5 \ c }$, $g_ { 5 \; L }$
and $g_{ 5 \; R }$ are the $5D$ gauge couplings of $SU(4)_c$, $SU(2)_L$ and $SU(2)_R$, 
respectively and the $\sqrt{3/2}$ factor in the coupling of $V$ comes from $SU(4)_c$ normalization.
 In terms of $Z^{\prime}$ and $B$, the five dimensional 
electroweak covariant derivative is now 
\begin{eqnarray}
D_M =  \partial_M -i \Big[ g_{ 5 \; L } W_{ L \; M }^a \tau_{a \, L} + 
g_{ 5 \; R } W^{ 1,2 }_{ R \; M } \tau^{1,2}_{R} + 
g_{5 \; Z^\prime} Z^{\prime}_M Q_{Z^\prime} + 
  g_5^{\prime} B_M ( \tau^3_R + 1/2 ( B - L ) ) \Big]
\label{bzD}
\end{eqnarray}
The couplings of the hypercharge
($Y = \tau_R^3 + (B-L)/2)$ and $Z^\prime$ gauge bosons are
\begin{equation}
g^{ \prime }_5= \frac{ \sqrt{ 3/2 } \; g_{ 5 \; c } \; 
g_{ 5_R } }{ \sqrt{ g_{ 5 \; R }^2 + \left( \sqrt{ 3/2 } \;
g_{ 5 \; c} \right)^2 } } \  \  \ , \  \  \
g_{ 5 \; Z^{ \prime } }= \sqrt{ g_{ 5 \; R }^2 + \left( \sqrt{ 3/2 } \;
g_{ 5 \; c } \right)^2 }
\end{equation}
Also, the charge under $Z^\prime$ and the mixing angle between $V$ and $W_R^3$ read
\begin{equation}
Q_{Z^\prime}= \tau^3_R-\sin^2 \theta^\prime \; Y, \ \ \ \ \ \ 
\label{QZprime}
\sin \theta^{ \prime } = \frac{ \sqrt{ 3/2 } \; g_{ 5 \; c } }
{ g_{ 5 \; Z^{ \prime } } }
\end{equation}

\subsection{Bulk fermion content}
The usual RH fermionic fields are promoted to doublets of $SU(2)_R$. Quarks and leptons
are unified into the {\bf 4} of $SU(4)_c$. However, the SM zero modes originate from different multiplets.
Indeed,  since we are breaking 
$SU(2)_R$
symmetry through the UV orbifold, one component of $SU(2)_R$ doublet
must be even and have a zero-mode
while the other component must be odd and not have a zero-mode.
Thus, 
$u_R$ and $d_R$ as well as $e_R$ and $\nu_R$ will have to come from different 
$SU(2)_R$ doublets.
Consequently, we are forced to a first doubling of the number
of $({\bf 4,2})$'s of $SU(4)_c \times SU(2)_R$. 
Since we are also breaking $SU(4)_c$ through the UV orbifold,
a second doubling is required in such a way that 
from the ${\bf 4}$ of $SU(4)_c$, only the
quark must be even and the color singlet must be odd, or
vice versa. This is
the usual procedure of obtaining quarks and lepton zero-modes
from different $SU(5)$ bulk multiplets in orbifolded GUT scenarios 
\cite{bhn, jmr}.
Concerning $({\bf 4,2})$ of $SU(4)_c \times SU(2)_L$, they are doubled only once, 
again to split quarks from leptons, i.e., in order to guarantee that $X_s$ does not
couple SM quarks to SM leptons (just as for $({\bf 4,2})$'s of $SU(4)_c \times SU(2)_R$
above).
To summarize, we have per generation\footnote{Henceforth, only the 
chirality  with the same
transformation as the SM under the
Lorentz group
will be discussed (except in section \ref{cftqual} and \ref{fermionkk}) since the
other chirality is projected out by  $Z_2$ symmetry.}, four types of $({\bf 4,2})$ under
$SU(4)_c \times SU(2)_R$, denoted by $F_R$, and two types
of $({\bf 4,2})$ under $SU(4)_c \times SU(2)_L$, denoted by
$F_L$:
\begin{equation}
F^q_{R \; 1}  =  \begin{pmatrix} u_R \\ \tilde{d}_R \\ e_R^{ \prime } \\ \nu_R^{ \prime } \\ 
\end{pmatrix}  , \
F^q_{R \; 2} = \begin{pmatrix}  \tilde{u}_R \\ d_R  \\ e_R^{ \prime } \\ \nu_R^{ \prime } \\
\end{pmatrix}  , \
F^l_{R \; 1}  =   \begin{pmatrix}  u_R^{ \prime } \\ d_R^{ \prime } \\ e_R \\ \tilde{\nu}_R \\
\end{pmatrix}  , \
F^l_{R \; 2}  =   \begin{pmatrix}  u_R^{ \prime } \\ d_R^{ \prime } \\ \tilde{e}_R \\ \nu_R \\
\end{pmatrix}  , \
F^q_L =   \begin{pmatrix}  u_L \\ d_L \\  e^{ \prime }_L\\ \nu^{ \prime }_L  \\
\end{pmatrix}  , \
F^l_L  =   \begin{pmatrix}  u^{ \prime }_L \\ d^{ \prime }_L \\   e_L \\  \nu_L \\
\end{pmatrix} 
\label{table}
\end{equation}
The untilded and unprimed particles are the ones to have zero modes, i.e. 
they are $(+,+)$. The extra fields (again, tildes denote $SU(2)_R$ partners 
and primes denote
$SU(4)_c$ partners) 
needed to complete all representations are 
$(-+)$ 
since breaking of $SU(2)_R \times SU(4)_c$ is on the Planck brane. 
Strictly speaking,
on an orbifold, $u_R$ and $e_R$ from the same multiplet 
(and similarly $d_R$ and $\nu_R$) are 
forced to have same BC.
So, for example, $e^{ \prime }_R$ in $F_{ R \; 1 }^q$ is $(++)$ to begin 
with, but we assume
that it has a Planckian (Dirac) mass with a Planck brane localized 
fermion which mimics $(-+)$ BC
to a good approximation (a similar assumption holds for $\nu^{ \prime 
}_R$ in $F_{ R \; 2 }^q$,
$u^{ \prime }_R$ in  $F_{ R \; 1 }^l$ and $d^{ \prime }_R$ in $F_{ R \; 
2 }^l$).

To each $({\bf 4,2})$, 
we assign the baryon-number corresponding to that of its zero-mode.
$U(1)_B$ commutes with Pati-Salam and we repeat that it should not  be confused with the ``$B - L$'' subgroup of Pati-Salam. Note that tilded particles are not 
``exotic'' (no $Z_3$ charge). 
Only primed particles carry an exotic baryon number and hence have $Z_3$ charge.

As for the Yukawa
couplings to the Higgs, 
they are necessarily localized on the IR brane:
\begin{eqnarray}
{\cal L}_{Yukawa}  =  \sqrt{ - g_{IR} } 
H \left( \lambda_{ u \; 5} F^q_L F^q_{R \; 1} + 
\lambda_{ d \; 5} F^q_L F^q_{R \; 2}
+ \lambda_{ e \; 5} F^l_L F^l_{R \; 1} 
+ \lambda_{ \nu \; 5} F^l_L F^l_{R \; 2}
\right) 
\label{Yukawa}
\end{eqnarray}
Note that because $u_R$ and $d_R$ zero-modes arise from
different $SU(2)_R$ doublets, we are able to give them 
separate Yukawa couplings
without violating $SU(2)_R$ on the IR brane.\\

\subsection{On an interval (instead of an orbifold)}

If we were to break Pati-Salam to the SM by more general boundary conditions
\cite{Csaki:2003dt},
the splitting of the
$SU(2)_R$ doublet and ${\bf 4}$ of $SU(4)_c$ would a priori
not be forced by
consistency of BC. But, we could not impose baryon-number consistently
in a GUT if we do not split $4$ of $SU(4)_c$. So, at least, quark/lepton splitting 
in the ${\bf 4}$ of $SU(4)_c$  would be necessary (by assigning Neumann/Dirichlet 
BC on the Planck brane). 
The  up-down quark isospin splitting could still be achieved 
(without doubling of representations) for light fermions
localized near the Planck brane thanks to different kinetic terms on the Planck brane
where $SU(2)_R$ is broken. This cannot work for top-bottom since $t_R$ has to be localized
near the TeV brane where
$SU(2)_R$ is unbroken  and $b_R$ is localized
near the Planck brane: thus, the splitting of the top/bottom $SU(2)_R$ doublet would also 
be necessary. Whether the splitting of $e_R$ and $\nu_R$ zero-modes (to obtain
different Dirac masses for charged leptons and neutrinos in case
Planck brane kinetic terms are not enough to do the splitting)
is required by phenomenology
depends on the mechanism for generating neutrino masses.

\section{Going to $SO(10)$}

\subsection{Extra gauge bosons, relations between gauge couplings 
and 
larger fermion multiplets}
\label{SO(10)}

When extending the gauge group to $SO(10)$, there are additional
gauge bosons, $X$, $Y$, $X^{ \prime }$ and $Y^{ \prime}$, which are given $(-+)$ 
BC\footnote{ On an orbifold, just as in the case of the breaking of
$U(1)_{ B - L } \times U(1)_R 
\rightarrow U(1)_Y$ in Pati-Salam, 
some of the $(-)$ BC on the Planck brane for gauge and fermion fields
are 
(effectively) achieved by a coupling
to a Planckian vev on the Planck brane.}.
The SM Higgs is now contained in $10_H$ of $SO(10)$, assigned $B = 0$. 
The breaking of $SO(10)$ to $SO(9)$ by the vev $\langle 10_H \rangle$ leads 
to the existence on the TeV brane of a color triplet pseudo Nambu-Goldstone boson,
which will be discussed in the section 7.2.

The previous three gauge couplings are now unified $g_{ 5 c } = g_{ 5  R } = 
g_{ 5  L } \equiv g_{ 5 }$ with the following relations:
$ \sin^2 \theta^{ \prime } = 3/5$, $g_{ 5  Z^{ \prime } } = \sqrt{5/2} \; g_5$
and $g^{ \prime }_5 = g_5 \sqrt{3/5}$ so that $  \sin^2 \theta _W = 3/8$ 
 at tree level at the GUT scale.
Log-enhanced, non-universal loop corrections will modify the 
relation between the low-energy 4D 
$g^{ \prime }$ and $g$ couplings  (just as in the $4D$ SM). 
The main reason is that the zero-modes can span the entire
extra dimension up to the Planck brane where $SO(10)$ is  broken and so loops are sensitive to Planckian cut-off's leading to loop-corrected
$\sin ^2 \theta_W \approx 0.23$. On the other hand, $\sin^2 \theta ^{ \prime }$ appears only in the couplings of KK modes. Those 
receive very small non-universal loop corrections (universal
loop corrections do not modify mixing angles) since 
KK modes are localized near the TeV brane, where 
$SO(10)$ is unbroken.
Therefore, $\sin^2 \theta ^{ \prime }$ is not modified by loop corrections.
We will extend this discussion in section \ref{g10}.

For fermions, let us start with the {\em orbifold} compactification. 
In this case, we are forced by the consistency of BC to split not only
quarks from leptons and but also
 $SU(2)_L$ and $SU(2)_R$ doublets. 
In addition, we have to split the components of the $SU(2)_R$ doublet. 
Thus, each of the previous $({\bf 4,2})$ of $SU(4)_c \times
SU(2)_R$ and of $SU(4)_c \times
SU(2)_L$ are promoted into a full
${\bf 16}$ of $SO(10)$ with the extra states again assigned $(-+)$ BC.
This leads to six ${\bf 16}$'s per generation.
Explicitly, one ${\bf 16}$ of $SO(10)$ for each SM representation: $Q_L =
( u_L, d_L )$, 
$u_R$, $d_R$, $L_L = ( e_L,
\nu_L )$,
$e_R$ and $\nu_R$:

$$
{\bf 16_{u_R} }  =   \begin{pmatrix} u_R \\ \tilde{d}_R \\ e_R^{ \prime }\\ \nu_R^{ \prime } \\
L^{ \prime }_L  \\
Q^{ \prime }_L \\
\end{pmatrix}  , 
{\bf 16_{d_R} }  = \begin{pmatrix} \tilde{u}_R \\ d_R \\ e_R^{ \prime } \\ \nu_R^{ \prime } \\
L^{ \prime }_L  \\
Q^{ \prime }_L \\
 \end{pmatrix}  ,
{\bf  16_{e_R} }  =  \begin{pmatrix} u_R^{ \prime }\\ d_R^{ \prime } \\ e_R \\ \tilde{\nu}_R \\
L^{ \prime }_L  \\
Q^{ \prime }_L \\
\end{pmatrix}  , 
{\bf  16_{\nu_R} } =  \begin{pmatrix} u_R^{ \prime } \\ d_R^{ \prime } \\ \tilde{e}_R \\ \nu_R
\\ L^{ \prime }_L\\
Q_L^{ \prime } \\
\end{pmatrix}  , 
{\bf 16_{Q_L}}  =  \begin{pmatrix} 
Q_L \\ L^{\prime}_L \\
u^{ \prime }_R \\ d^{ \prime }_R \\ e^{ \prime }_R \\ \nu^{ \prime }_R\\
\end{pmatrix}  , 
{\bf  16_{L_L} }  =  \begin{pmatrix}  
Q^{\prime}_L \\ L_L \\
u^{ \prime }_R \\ d^{ \prime }_R \\ e^{ \prime }_R \\ \nu^{ \prime }_R
\end{pmatrix}  
$$
The  last two lines of the first four multiplets (and the last four lines of the last two multiplets) are the extra states in going from $(4,2)$ of Pati-Salam to ${\bf 16}$ of
$SO(10)$.

Like in Pati-Salam, breaking $SO(10)$ {\em on an interval}
(by assigning Dirichlet/Neumann BC for gauge bosons)
does not necessarily force us to split fermion multiplets
(either quark-lepton splitting, 
$SU(2)_L$--$SU(2)_R$ doublet splitting or splitting
within a $SU(2)_R$ multiplet). 
But, phenomenologically, like in Pati-Salam, 
we have to obtain SM quarks and 
leptons 
from different ${\bf 16}$'s to suppress proton decay and split $SU(2)_L$ and $SU(2)_R$ 
quark doublets to assign baryon number. 
And again,
we also need to split $t_R$ and $b_R$ in a realistic model. This would lead to 
three 16's per generation:  one 16 for
$SU(2)_L$ quark doublet, one for $SU(2)_R$ quark doublet and one 16
for leptons, which is what we presented in \cite{us},
plus an extra 16 to split $t_R$ and $b_R$. 
Imposing lepton number symmetry, as discussed below,
further requires to split  $SU(2)_L$ and $SU(2)_R$ lepton doublets.
This would amount in thirteen 16's in total. 
We will discuss the impact of this large number of representations on the 
loop corrections to
gauge couplings in subsection \ref{subsec:strong}.

\subsection{Lepton Number Symmetry}
\label{leptonnumbersymmetry}

Left and right-handed
leptons could be obtained from the same ${\bf 16}$ (as we did in our toy example \cite{us}). However, in
a realistic model, we are forced to split them for the following reason. 
If $SO(10)$ is unbroken in the bulk, Majorana masses
for SM $\nu_L$ cannot be written on the TeV brane since
the $L_L L_L H H $ operator
is forbidden by the $B - L$ gauge symmetry (and similarly, 
bulk Majorana masses, i.e., $\nu_R \nu_R$ operator for RH neutrinos are not allowed).
However, we will break $SO(10)$ in the bulk for reasons presented  in section
\ref{bulkbreaking}. In this case, $B-L$ is also broken in the bulk (in
general) and the operator $L_L L_L H H$ is
allowed. This gives Majorana masses for SM $\nu_L$ of roughly the same
size as charged lepton masses since the effective UV cut-off
suppressing this operator is of order  TeV,
with some, but not much suppression from GUT breaking.
In addition, bulk Majorana masses for right-handed neutrinos are also allowed and spoil
the seesaw mechanism of 
reference \cite{seesaw}. In short,  lepton-number is violated at low scale.
To remedy this problem, we have to impose a bulk gauged 
lepton-number symmetry  {\em in addition to} the baryon-number symmetry. 
We can break it spontaneously on the Planck
brane, just like we do with baryon-number which would restrict Majorana masses for $\nu_R$ to be 
written on the Planck brane only, as required for see-saw mechanism 
for neutrino masses of reference
\cite{seesaw}\footnote{However, notice that there is no 
analog of the $Z_3$ symmetry associated with 
baryon-number since there is no analog of unbroken color invariance for leptons}.

SM left and right-handed leptons come
from different ${\bf 16}$'s with lepton numbers $+1$ and $-1$ and other
${\bf 16}$'s and Higgs are assigned zero lepton-number.
For simplicity, the toy example we 
presented in \cite{us}
did not invoke splitting of $SU(2)_R$ multiplet nor splitting of left and 
right-handed leptons.

\section{Bulk breaking of unified gauge symmetry}
\label{bulkbreaking}

\subsection{In Pati-Salam and $SO(10)$}
\label{subsection:bulkbreakingPS}

We are willing to invoke  
the bulk breaking of GUT
 via the scalar $\Sigma$ (see Eq. \ref{bulk})
in both Pati-Salam and $SO(10)$ for the following reasons: 

\begin{itemize}
\item The Yukawa coupling $\lambda _{t \; 5} H b_L \tilde{b}_R$
(see Eq. \ref{Yukawa}) leads to a mass term of the type $ m_t b_L^{ (0) } \tilde{b}_R
^{ (1) } f \left( c_R \right)$, where $f (c ) \approx \sqrt{ 2 / ( 1 - 2 c ) }$ 
(for $c > -1/2$)
where we used Eq. \ref{lambda4Dtop} and the wavefunction of $\tilde{b}_R^{ (1) }$ given in
appendix \ref{fermionkk}. There is also a KK mass,
$m_{ \tilde{b}_R^{ (1) } } \tilde{b}_R^{ (1) }  \hat{ \tilde{b} }^{ (1) }_R$,
where $\hat{ \tilde{b} }^{ (1) }_R$ is the $5D$ KK partner of $\tilde{b}_R^{ (1) }$.
The  mixing between $b_L^{ (0) }$ and $\hat{ \tilde{b} }^{ (1) }_R$ results in a shift in
the coupling of 
$b_L$
to $Z$ of order  $\sim m_t^2 
f \left( c_R \right)^2 / m^2_{ \tilde{b}_R^{ (1) } }$, using
$f \left( c_R \right) \sim 1$ (the analysis is
similar to
the $\nu^{ \prime }_R - \nu^{ \prime }_L$ mixing in section \ref{nulrmixing}). For this shift 
to be $\lesssim 1 \%$,
$\tilde{b}_R^{ (1) }$ 
needs to be heavier than $\sim 1.5$ TeV, meaning that the
$c$ for $b^{ \prime }_R $ should be $\gtrsim -1/4$ if the
gauge KK mass $M_{KK}\approx$ 3 TeV (see spectrum in section \ref{fermionkk}).
In the absence of bulk breaking, the $c$'s for all components of 
$t_R$ multiplet are the same and  $\nu^{ \prime }_R$ will have to be
heavier than $\sim 1.5$ TeV also
which restricts the viable parameter space for the LZP to account for dark matter.

\item As mentioned above, we want 
to ensure that $e^{ \prime}_R$ in Pati-Salam (and other lepton-like states
in $SO(10)$) is heavier than $\nu^{ \prime }_R$ in case electroweak 
corrections were not large enough to achieve the required splitting.
A small amount of bulk breaking of $SU(2)_R$ and Pati-Salam 
allows us to split the $c$'s of the $(-+)$ fermions
in $t_R$ multiplet and thus to address the above two issues.
To be precise, choose $c$ for $\nu^{ \prime }_R$ to be
smaller than that of $\tilde{b}_R$ and $e^{ \prime }_R$.
We will give details on the size of splitting in $c$'s in  section \ref{bulkbreakingsize}.

\item Bulk breaking of $SU(2)_R$ is also used to get a 
contribution to the Peskin-Takeuchi $T$ parameter of order $\sim 0.3$ as required 
to fit electroweak data \cite{custodial1}: $T_{bulk} \sim \frac{1}{2} \; M_{ GUT }^2 / k^2$, 
where $M_{ GUT }$ is the bulk mass of $W_R^{ \pm }$. If $M_{ GUT } / k \sim 1/2$, then
$T_{ bulk } \sim 0.1$. Loop effects 
can generate the remaining contribution to $T$ \cite{custodial1}. 
If $M_{ GUT } \sim k$, we get a too large  $T_{ bulk } \sim 0.5$ --
this is another reason, independent of
unification considerations in $SO(10)$
(see section \ref{bulkbreakingsize}), to assume that $M_{ GUT } < k$.

\end{itemize}

\subsection{Specificities of $SO(10)$}

In $SO(10)$ there are {\em additional} reasons to invoke bulk breaking:
\begin{itemize}
\item To achieve gauge coupling unification \cite{rsgut} (see section \ref{section:gaugecouplings}). 

\item 
To make the Higgs triplet, charged under $Z_3$,
heavier than $\nu^{ \prime }_R$. Indeed, 
we do not want it to be the LZP. 
As a colored particle, it is not a suitable dark matter candidate. 
Without bulk breaking and at tree-level,
it  is the massless (pseudo) Nambu-Goldstone boson coming from the
breaking of $SO(10)$ to $SO(9)$ by the Higgs vev
(recall that BC's on the TeV brane do not break $SO(10$)).
$SO(10)$ being broken also on the Planck brane (by BC), loop corrections will give it a mass
which may be too small, of order $\alpha_s M_{KK}^2/\pi$. 
With bulk breaking, the Higgs triplet gets
a tree-level mass via the operator $\Sigma {\bf 10}_H {\bf 10}_H$. For sufficient bulk GUT 
breaking, this mass is larger
than the one-loop induced mass so that Higgs triplet can be heavier than 
$\nu^{ \prime }_R$.

\item To make some $Z_3$-charged particles such as
$X$, $X^{ \prime }$, $Y$, $Y^{\prime}$ or $Q^{ \prime }_L, L_L^{ \prime }$ 
from the multiplet with $ t_R^{ (0)}$ or
$\nu^{ \prime }_R$, $u^{ \prime }_R$ from the multiplet
with  $Q_L^{ (0) }$, decay before Big Bang Nucleosynthesis (BBN).
In the absence of bulk breaking,  they can only decay via
 very higher-dimensional operators
so that their decay width may be too small, 
as explained in the next section. Note that (non-SM) 
Pati-Salam gauge boson ($X_s$) and Pati-Salam partners of zero-mode fermions
decay easily as mentioned below.
\end{itemize}

\subsection{Decay of KK particles (other than the LZP)}

Clearly, $Z_3$-charged particles eventually decay into the LZP.
In Pati--Salam, $Z_3$ charged particles decay into $\nu^{ \prime}_R$ easily:
$e^{ \prime }_R$ from $t_R^{ (0) }$ multiplet
decays into LZP $+ W^{ \pm }_R$, followed by $W^{ \pm }_R$ mixing with 
$W_L^{ \pm }$ zero-mode due to EW symmetry breaking:
the coupling $e^{ \prime }_R  W^{ \pm }$ LZP is similar to the coupling 
of the LZP to $Z$ induced via $Z-Z^{ \prime }$ mixing after EW symmetry breaking
(see section \ref{ZZprimemixing}), of order
$\sim g k \pi r_c 
m_W^2 / M_{ KK } ^2 \sim g/30 $ for $M_{ KK } \sim 3$ TeV.
$X_s$ decays fast into
$t_R^{ (0) }$ and $ \nu^{ \prime }_R$. 
$Z_3$ charged fermions from 
other multiplets can decay into the zero-mode from that multiplet and virtual
$X_s$: for example, $u^{ \prime }_R$ from multiplet with $e^{ (0) }_R$
decays into $\tilde{ \nu }_R + X_s$ followed by decays of $X_s$ and
$\tilde{ \nu }_R \rightarrow W^{ \pm } e^{ (0) }_R$ (the last decay occurs
via
$W^{ \pm }_R - W_L ^{ \mp }$ mixing).

Finally, {\em tilded} particles, not charged under $Z_3$,
decay into their $SU(2)_R$ partners
which have zero-modes and 
KK mode $W^{ \pm }_R$ which again mixes with zero-mode of
$W_L^{ \pm }$.
Tilded particles can also decay into $SU(2)_L$ doublet and Higgs as 
follows. As mentioned before
(section 7.1), there is a Yukawa coupling $\lambda_{t} f  ( c_{ t_R} ) H 
\tilde{b}_R (t, b)_L$
which results in the decay $\tilde{b}_R \rightarrow b_L H^0, t_L W^+_{ 
long. }$
-- this dominates over the  decay into $t_R W^+$ (which is suppressed
by $W_R^{ \pm } - W_L^{ \mp }$ mixing).


In contrast with Pati--Salam, decays in $SO(10)$ of the {\em non}-Pati-Salam $Z_3$-charged
particles (both fermions and gauge bosons) into $\nu^{ \prime }_R$ 
are problematic 
in the 
absence of bulk breaking. Indeed,  there is no short path for this decay.
Specifically,
$X$, $X^{ \prime }$, $Y$, $Y^{ \prime }$ or $Q_L^{ \prime }$ and 
$L_L^{ \prime }$ from $t_R^{ (0) }$
multiplet cannot decay
into the LZP via gauge interactions. The reason is that, while there are
$t_R^{ (0) } - X - Q^{ \prime }_L$ and $t_R^{ (0) } X^{ \prime } L_L^{ \prime }$ couplings,
there are no $t_R^{ (0) }- X_s - Q^{ \prime}_L$ (or
$L^{ \prime }_L)$ 
couplings and also no $t_R^{ (0) } - LZP - X$ or
$X^{ \prime }$ couplings. 

Thus, the
decays of these particles have to go through higher-dimensional operators, 
and, in order for these operators not to be suppressed by the Planck scale, they
have to be $B$-conserving. For example, operators such as
$\left( Q^{ \prime }_L Q_L Q_L L \right) \nu^{ \prime \; c }_R L H$ 
and $\left( L_L^{ \prime } Q d^c_R \nu^c_R \right) \nu^{ \prime }_R \bar{L} H$
from $\left( {\bf 16}^4 \right) \times \left( {\bf 16} {\bf 16} {\bf 10_H}
\right)$ will lead to decays of $Q^{ \prime }_L$ and $L^{ \prime }_L$ into
LZP.
They break the usual lepton-number,
but do not generate Majorana masses
since $L_L L_L H H$ on the TeV brane or $\nu_R \nu_R$ in the bulk are
forbidden by the unbroken
bulk $B - L$ gauge symmetry. Thus, in the absence of bulk breaking, we do not need to impose
lepton number to forbid these masses. However, these
operators result in $5$-body decays of $Q^{ \prime }_L$ and $L^{ \prime }_L$ into LZP
with amplitude suppressed by $6$ powers of the
KK mass since it is a dimension-$10$ operator and can result in lifetimes
longer than the BBN epoch. 

Let us give an
estimate for the decay width:
$\Gamma \sim v^2 ( \Delta m )^{ 11 } / \Lambda^{ 12 } / \left( 4096 \; \pi^7 \right)$,
where $\Delta m$ is the mass splitting between $Q^{ \prime }_L$
and the LZP which is small since they have the same $c$, 
$4096 \; \pi^7$ comes from the $5$-body phase-space and 
$\Lambda$ here is the {\it warped-down }
string scale of order a few TeV.
For $\Delta m \sim 0.2 \, m_{ LZP } \sim 100$ GeV and 
$\Lambda \sim 3$ TeV, we get 
$\Gamma \sim 10^{-22}$ GeV and a lifetime of $\sim 10^{ -2 }$ sec.
However, the lifetime is extremely sensitive to $\Delta m$ and $\Lambda$: 
for example, with
$\Delta m \sim 10$ GeV, we get a lifetime of $\sim 10^9$ sec.
Similarly, decays of $\nu^{ \prime }_R$ and $u^{ \prime }_R$ from the multiplet
with $Q^{ (0) }_L$ might be suppressed: their masses are $\sim$ few TeV so that
phase-space suppression is smaller (i.e., $\Delta m$ is larger), but
still the decay can occur after BBN since, for example, $\Lambda$ can be larger.

Let us now recall why there is a potential danger from late decays of TeV mass particles.
Particles decaying after BBN can ruin successful predictions of 
abundances of light elements. Decay products inject photons and electrons 
into the plasma which can dissociate light elements. This leads to a lifetime 
dependent bound on the 
quantity $m \times Y$, where  $m$ is the mass of the decaying particle and $Y=n/s$, 
where $n$ is the number density that this particle would have today if it had not 
decayed and $s$ is the entropy density today. The strongest bound is for lifetimes of the 
order of $10^8$s and reads $m\times Y< 10^{-12}$ GeV \cite{Cyburt:2002uv}.
The standard relic density calculation of cold massive particles leads to 
\begin{equation}
\label{abundanceformula}
m\times Y\sim \frac{x_F\sqrt{45}}{\sqrt{\pi g_*}M_{Pl}\langle\sigma v\rangle}\sim 3\times 10^{-19}\frac{x_F}{\sqrt{g_*}\langle\sigma v\rangle} \ \ \mbox{GeV}^{-1}
\end{equation}
For a relic behaving as a WIMP, we expect $x_F\sim 25$. 
If it accounts  for dark matter then $\langle\sigma v\rangle\sim 10^{-9}$ GeV$^{-2}$ and
$m\times Y\sim7.5\times 10^{-9}$ GeV. We see that even if the light 
KK states we are considering contributed to the final energy density of 
dark matter by only one percent or one per mil (after they decay into the LZP), they could
 be dangerous  if they decay late, i.e. after
BBN.  To suppress any potential danger coming from the late decay of these next-to-lightest 
$Z_3$ charged particles (NLZP), 
we invoke bulk breaking of $SO(10)$ which we discuss next.

\subsection{Decays of NLZP's with bulk 
breaking of $SO(10)$}  

In the presence of $SO(10)$ bulk breaking,
decays of $Q^{ \prime }_L$ and $L_L^{ \prime }$ from the $t_R$ multiplet
into the LZP easily take place thanks to
$X^{ \prime } - X_s$ and $Y - Y^{ \prime }$ mixing due to
\begin{eqnarray}
{\cal L}_{ IR } & \ni & \sqrt{ -g _{ IR } } \left(
\frac{ b }{ M_S^2 } 
\langle 16_{ \Sigma } \rangle D^{ \mu } \langle 16_{ \Sigma } \rangle D_{ \mu }
\langle 10_H \rangle + D^{ \mu } \langle
10_H \rangle D_{ \mu } \langle 10_H \rangle \right)
\label{Xmixing}
\end{eqnarray}
where 
$\langle 16_{ \Sigma } \rangle$ is in SM singlet component and
the covariant derivatives 
give gauge fields, $X$, $X^{ \prime }$ and $X_s$. The first term leads
to $X^{ \prime } - X_s$ mixing and hence to the decays 
\begin{eqnarray}
t_L^{ \prime } & \rightarrow & X^{ \prime } \nu^{ \prime }_R
  \stackrel{ \hbox{\tiny{via mixing}}  }{ \rightarrow } 
t_R^{ (0) } \overline{\nu}^{ \prime }_R \nu^{ \prime }_R \; \;  \ \ \mbox{and} \ \ 
\nu^{ \prime }_L  \rightarrow  {X^{ \prime }}^* t_R^{ (0) }
  \stackrel{ \hbox{\tiny{via mixing}}  }{ \rightarrow }  t_R^{ (0) } \bar{t}_R
^{ (0) } \nu^{ \prime }_R
\nonumber
\end{eqnarray}
whereas
their $SU(2)_L$ partners decay as ($X$, $Y$ and $Y^{ \prime }$ cannot mix with $X_s$
 due to their different electric charge)
\begin{eqnarray}
b_L^{ \prime } & \rightarrow & t_L^{ \prime } W_L^{ - } \rightarrow  t_R^{ (0) } \overline{\nu}^{ \prime }_R \nu^{ \prime }_R W^{ - }_L \nonumber \\
\tau_L^{ \prime } & \rightarrow & \nu^{ \prime }_L W_L^{ \pm }
\nonumber \\
 & \rightarrow & \nu^{ \prime }_R t^{ (0) }_R \bar{t}_R^{ (0) } W_L^{ \pm }
\end{eqnarray}
Similarly, the 2nd term in Eq. (\ref{Xmixing})
gives $Y - {Y^{ \prime }}^*$ mixing resulting in other
decay chains (using the $Y'-X'-W_L$ coupling).
We can estimate these decay widths as follows.
Naive dimensional analysis (NDA) size for $b$ is $ \sim \lambda_{ 5 } \Lambda$ 
(as expected since
it is a coupling of Higgs) resulting in a  $X_s - X^{ \prime }$ mixing term of order $\sim M_{ GUT }^2 
\left( \lambda_5 k v \right) / \Lambda$, 
where  $M_{ GUT }$ and $\Lambda$ are actually the warped-down values since this operator is on the TeV brane. We used the fact that
wavefunctions  for gauge KK modes at the TeV brane are $\sim \sqrt{k}$  (see
appendix A.1). 
Using Eq. \ref{lambda4Dtop}, we get $\lambda_5 k v \sim 500$ GeV with $c$ for
$(t, b)_L \sim 0.4 $ and $c$ for $t_R \sim -1/2$.  Then, the coefficient of the
$4$-fermion operator for the decay of, say, $\nu^{ \prime }_L$, is $\sim 
g_{ \mbox{\tiny SM}} ^2 k \pi r_c M_{ KK }^4 \times (X^{ \prime} - X_s$) mixing. We used the fact
that the couplings of the gauge KK mode to KK fermions and $t_R^{ (0) }$
are enhanced by $\sim 
\sqrt{k \pi r_c }$ compared to $g_{ \mbox{\tiny SM } }$
 (see section 9.1). Assuming 
 $m_{\nu^{ \prime }_L} > 2 m_t + m_{ LZP }$, we obtain
$\Gamma \sim$ (above coefficient)$^2 \times$ 
$\left( 
m_{ \nu^{ \prime }_L } - 2 m_t - m_{ LZP } \right)^5 / \left( 64 \pi^3 \right)$,
where $64 \pi^3$ is from the $3$-body phase-space. 
For $g_{ SM } \sim 1/2$, 
$m_{ \nu^{ \prime}_L } \sim 1$ TeV and $m_{ LZP } \sim 200$ GeV, $M_{ GUT } / k
\sim 1/2$
we get
$\Gamma \sim 10^{ -8}$ GeV and a lifetime $\sim 10^{ -17 }$ sec. 

Similarly,  $\nu^{ \prime }_R$ and $u^{ \prime }_R$ from the multiplet
with $Q^{ (0) }_L$ can decay into $Q^{ (0) }_L$ + ($X^{ \prime },Y^{ \prime }$) or ($X$,$Y$), followed
by mixing with $X_s$. $\nu^{ \prime }_R$ and $u^{ \prime }_R$
have masses of a few TeV so that their lifetimes are even shorter than above.

\subsection{Size of bulk breaking and splitting in $c$}
\label{bulkbreakingsize}

Having seen the motivation for 
bulk breaking, we now show what is its 
natural size. 
The splitting in $c$ (due to last term of Eq. \ref{fermions})
is given by $ (k\Delta c)\sim a^{ \prime } 
\langle \Sigma\rangle/\sqrt{ \Lambda }$ 
(where $a^{\prime}$ is defined in Eq.~\ref{fermions}).
The NDA sizes for coupling of $\Sigma$ to gauge fields (see Eq.~\ref{bulk}) and fermions are
$a \sim a^{ \prime }\sim g_5 \sqrt{ \Lambda } $ 
leading to $\Delta c \sim g_5 v_{\Sigma}^{3/2}/k$. We previously saw that the bulk mass 
for
$X,Y$ is $M_{GUT}\sim g_5 v_{\Sigma}^{3/2}$ so that
\begin{eqnarray}
\Delta c & \sim & {M_{GUT}}/{k}.
\end{eqnarray}
The size 
of $ {M_{GUT}}/{k}$ can be inferred from the requirement of gauge coupling unification:
 NDA size for 
the bulk threshold correction $\Delta$ in $1/g_{ 4D }^2$ (see Eq. \ref{g4D}),
from the  higher-dimensional operator in Eq. (\ref{bulk}) is
$\sim k \pi r_c / g_5^2 \times M_{ GUT } / \Lambda $. The size of this correction 
should be  $\sim 20 \%$ (and not larger) to accomodate unification
\cite{rsgut}. Using $k \pi r_c / g_5^2  \sim 1/g_{ 4D }^2 \sim O(1)$, we get
$M_{ GUT } / \Lambda \sim 1/5$.
Of course, this argument is not valid for Pati-Salam.
The splitting in $c$ is then given by
$\Delta c \sim (1/5)\times \Lambda / k$ where $\Lambda>k$ is required for calculability.
We also require that $M_{GUT}/k<1/2$ so 
that we can use the small GUT breaking 
approximation as follows. 
There are one-loop {\em non}-universal corrections to $1/g_{ 4D}^2$ (see Eq. \ref{g4D})
from GUT-scale splittings in masses.
For example, the splitting between (mass)$^2$ of $X,Y$ gauge bosons and SM  KK  gauge bosons
is $O \left(
M_{ GUT } ^2 / k^2 \right)$ so that these one-loop corrections  
have a size $\sim C \; \frac{M_{ GUT } ^2}{ k^2} \; \frac{k \pi r_c} {8 \pi^2}$, 
where $C$ is the Dynkin index of the bulk $X/Y$ 
gauge fields  \cite{rsgut}.
For $M_{ GUT }  / k \sim 1/2$, these result in
$\Delta _i$'s $\sim C / 8$ which is about what we require
for unification. Whereas, 
for $M_{ GUT } \sim k$,  $\Delta_i$'s 
$\sim C /2$ which spoils unification -- 
to repeat, we tolerate $\Delta \sim 1/5$\footnote{Note that
$\Delta$ from higher-dimensional operator 
can be small even for $M_{ GUT } \sim k$ 
as long as $M_{ GUT } < \Lambda$.}.
Combining the above two arguments, we get
\begin{eqnarray}
0.2 & \lesssim \Delta c & \lesssim 1/2
\end{eqnarray}
This size is enough to obtain the splitting in mass between
KK particles from the $t_R$ multiplet as required in section 7.1.
Explicitly, $c$ for GUT partners of $t_R$ is given by 
$c_{ t_R } \pm \Delta c$ with $c_{ t_R } \lesssim 0$ and we have seen that the
mass of the $(-+)$ fermion is very sensitive to $c$ for $c \sim -1/2$. 
Thus, 
$\nu^{ \prime }_R$ (assuming its $c$ is the smallest)
can be  significantly lighter than other
$Z_3$-charged GUT partners of $t_R$ and ensured to be the LZP.
Also, $\tilde{b}_R$ can be easily heavier than $1.5$ TeV (as constrained experimentally by
$Z \rightarrow b \bar{b}$), while at the same time 
$m_{\nu^{ \prime }_R} <$ TeV (which is the preferred mass range in order  
to obtain the correct relic density).

\section{Other models}
Before discussing 
the interactions of the LZP and showing that it is a good DM candidate,
we briefly mention other related models.

\subsection{$SO(10)$ breaking on the TeV brane}
An alternative possibility is to break $SO(10)$ to $SU(3)\times SU(2)_L
\times SU(2)_R \times U(1)_{ B - L }$ 
on the TeV brane, using 
$(+ -)$ boundary conditions for the other gauge fields of $SO(10)$:
we choose not to break $SU(2)_R$ by BC on the TeV brane in order to preserve the custodial symmetry.
Thus, $SU(2)_R \times U(1)_{B-L}$ should be broken to $U(1)_Y$ on the UV brane
(as in the previous model). Extra fermionic states 
with the same chirality as zero-modes (i.e., SM fermions) 
are also $(+-)$ while they are $(-+)$ for the other chirality.
We can still define a $Z_3$ as before. The LZP now comes from the multiplet with the 
{\em largest} $c$, namely the multiplet with one of the 
light fermions having $c > 1/2$ as explained  in section 4.1. 
As usual, due to bulk GUT breaking, we can assume that the LZP is $\nu^{ \prime }_R$.
Annihilation of the LZP via $Z^{ \prime }$ exchange (for $(-+)$ chirality), 
which will be described in the next section, is
the same in the two models. However, the one via $X_s$-exchange is negligible in
this model since the 
zero-mode which couples to the LZP via $X_s$
is now localized near the UV brane (cf. the previous model,
where this channel is important since $t_R$ is localized close
to the TeV brane). 
Concerning the coupling of the LZP to the $Z$ (playing an important role in annihilation and elastic scattering and which will be described in the next section), the one occurring 
via $Z^{ \prime } - Z$ mixing (for $(-+)$ chirality) is the same in
the two models and the one via $\nu^{ \prime }_R
- \nu^{ \prime }_L$ mixing (for $(-+)$ chirality) 
is also similar, except that the $5 D$ Yukawa entering this
coupling is that of the {\em light} fermion.

The Higgs multiplet is still a bi-doublet of $SU(2)_L \times SU(2)_R$ but there is no Higgs triplet since $SO(10)$ is broken on the TeV brane. As far
as unification of couplings goes,
if there is no bulk GUT breaking, the ``would-be'' zero-modes of $X, X_s$ etc.
get a mass
$\sim M_{ KK } / \sqrt{ k \pi r_c }$ which spoils unification. However, with
bulk breaking, these modes get a mass of $\sim M_{ GUT }$ so that unification is similar
to the previous model (see reference \cite{rsgut}).

\subsection{Warped SUSY $SO(10)$}
\label{subsec:warpedSUSY}

If the model has supersymmetry in the bulk, the Higgs can be localized near the Planck brane since
 SUSY protects its mass. Thus, SM fermions can also be localized very close to the Planck brane
($c \gg 1/2$) 
so that higher-dimensional baryon-number violating operators are suppressed by Planckian scales. There is no longer a need to impose baryon-number symmetry. There will be no stable KK state. However, there is still a possibility to account for dark matter
 if the lightest supersymmetric particle is stable via R-parity  conservation.
Of course, one loses the explanation of the hierarchy of 
fermion masses which is one of the appealing
features of non-SUSY RS. One has to introduce small Yukawa couplings by hand. 
 If one was to address the issue of Yukawa hierarchy by delocalizing the 
fermions ($c \stackrel{<}{\sim} 1/2$), 
then a baryon number symmetry would 
be required. In addition, the Higgs would also have to be in the bulk and
should be given
almost a flat profile. Otherwise, MSSM unification will be spoiled by the modification of the
contribution of the Higgs to the running.  For recent works on warped supersymmetric $SO(10)$, 
see references \cite{Nomura:2004is}.

\subsection{$SU(5)$ model}

$SU(5)$ models do not contain a custodial symmetry and are constrained by EW precision tests.
The IR scale has to be pushed to 10 TeV or more (depending on the size of brane kinetic terms). This introduces a little hierarchy problem and also make these models less appealing since there is no hope to produce KK modes at colliders. Nevertheless, we briefly discuss this model
to see whether there can be a stable KK particle.
Suppose $SU(5)$ is broken to the SM on the Planck brane:
$X,Y$ gauge bosons are $(-+)$. If $d_R$ from ${\bf \bar{5} }$ is $(++)$, i.e., has zero-mode,
then, on an orbifold, $L^{ \prime }_L$ from the same multiplet has to be $(-+)$.
Consistency of BC on an orbifold requires the same BC for $u_R$ and $e_R$, i.e., 
zero-modes for both $u_R$ and $e_R$ can come 
from the same ${\bf 10}$, but we give one of them a
Planckian mass with fermion localized on the 
Planck brane so that it is effectively $(-+)$\footnote{The same
argument applies to Pati-Salam model (as mentioned before) 
and to the $SO(10)$ model.}. So one gets
two ${\bf \bar{5}}$'s and three ${\bf 10}$'s per generation with
zero-modes for $d_R$, $L_L$, $Q_L$, $u_R$ and $e_R$, respectively.
One has to impose baryon-number. $Z_3$ again gives a stable particle. The 
only electrically and color neutral,but $Z_3$-charged particle is
$\nu^{ \prime }_L$: if it is to account for dark matter, then its mass is constrained to be at least a few tens of TeV from direct detection experiments \cite{Servant:2002hb}.

On an orbifold, it is also possible to obtain the
stability of a KK state via a discrete symmetry not related to baryon-number:
One can define $P = Z_2$-charge $\times Z_2^{ \prime }$-charge. Bulk 
interactions are
$P$-invariant even after compactification (which breaks $Z_2$ and $Z_2^{ \prime }$ 
separately but leave the product intact). Particles with zero-modes $(++)$ are $P$-even, particles
with no zero-modes $(-+)$ or $(+-)$ are $P$-odd\footnote{This parity was denoted GUT-parity
in reference \cite{gns}, but it can be present in any model with
gauge symmetry breaking on $Z_2 \times Z_2^{ \prime }$ orbifold.}.

If we assume that the 
bare lagrangian on each brane respects both $Z_2$ and $Z_2^{ \prime }$
(of course, on an orbifold, it has to respect $Z_2$ corresponding to reflection
about that brane), then all tree-interactions are 
$P$-even. Loops
cannot generate $P$-violating interactions and P-parity is exact at loop-order.
The lightest $P$-odd particle is stable
since it cannot decay into $P$-even SM particles hence can be the DM.
Again, the only candidate is $\nu^{\prime}_L $.
Note that in Pati-Salam or $SO(10)$, we cannot assume
$P$-parity since the bi-doublet Higgs couples $W_R^{ \pm }$ $(-+)$
to $W_L^{ \pm}$ $(++)$, i.e., the Higgs couplings do not preserve $P$-parity.

\subsection{$SO(10)$ model with gauged lepton number}
\label{subsection:gaugedlepton}
As we said in the introduction, 
imposing  only a (gauged) lepton number symmetry is enough to prevent proton decay,
although $\Delta B = 2$, {\it i.e } neutron-antineutron 
oscillations are still allowed but suppressed by the TeV scale. In this case, we need again to replicate representations.
On an interval, three {\bf 16}'s per generation with lepton numbers
$+1$, $-1$ and $0$ containing zero-modes for
$L_L$ and $L_R$ and all quarks, respectively, are sufficient. 
In addition, extra {\bf 16}'s for the third generation
are needed to split $b_R$ and $t_R$ as usual  and also
$(t,b)_L$ from $t_R$ and $b_R$ (due to the three different $c$'s). As in the case of
baryon-number symmetry, we add spectators on the Planck brane
and break lepton-number spontaneously on that brane.

In this alternative, we do not obtain a stable particle hence no DM
candidate. This is because there is no unbroken gauge symmetry 
under which only leptons are charged  so that there is
no analog of unbroken $Z_3$ symmetry, even if lepton number is unbroken.
The $\nu'_R$ (and other KK states) from
$t_R$ multiplet will still be light, but
$\nu^{ \prime }_R \rightarrow $ neutron $+ S$ (where
$S$ is a neutral scalar SM final state with zero lepton number) 
or proton $+ S^{\prime}$ (where $S^{\prime}$ is a charged scalar final state)
is allowed. 
Note that the above decay of $\nu^{ \prime }_R$ breaks
baryon-number by $1$ (since $\nu^{ \prime }_R$
has zero baryon-number), but this is allowed since we 
are not imposing baryon-number in this case.
The final state has to involve a proton or a neutron
which are the only SM fermionic states carrying zero lepton-number 
(recall that $\nu'_R$ has zero lepton number). For example, 
there is a coupling $\left( X^{ \prime }, Y^{ \prime } \right) 
Q^{ (0) } d^{ (0) }_R$
from a bulk interaction since zero-modes of $Q$ and $d_R$
can be obtained form same multiplet so that we get
$\nu^{ \prime }_R \rightarrow t^{ (0) }_R X_s^*$,
followed by $X_s \rightarrow d_L^{ (0) } 
d^{ (0) }_R$ (via $X^{ \prime } - X_s$ mixing). 
%
%
In this model, baryon number violating decays such as
$\left( X^{ \prime }, Y^{ \prime } \right)
\rightarrow Q^{ (0) } d_R^{ (0) }$
and $\left( X, Y \right) \rightarrow Q^{ (0) } u_R^{ (0) }$
could be observed at colliders. 

However, on an {\em orbifold},
consistency of BC will force us to split $SU(2)_L$ and $SU(2)_R$ doublet
quarks also so that we will require a larger number of {\bf 16}'s.
Recall that there is a GUT parity in the bulk in this case
(we call it P-parity in 
section 8.3)
under which all $(-+)$ states (with no zero-modes) are odd.
Hence, the lightest P-odd state (most likely $\nu^{ \prime }_R$) cannot decay via 
bulk interactions. Other light KK states can decay into it in the bulk as in our model with 
baryon-number.
P-parity can be broken by brane interactions. In fact, in $SO(10)$ or Pati-Salam
model, Higgs 
couplings are not invariant under
P-parity so that $P$-parity has to be broken 
on the TeV brane. Thus, $\nu^{ \prime }_R$ will decay via interactions
on the TeV brane.
To be concrete, the operator
$\bar{16}_Q D \! \! \! \! \! \! \! \not \; \;
16_d$, leading to
$\left( X^{ \prime }, Y^{ \prime } \right) 
\rightarrow Q^{ (0) } d_R^{ (0) }$ as before, 
is allowed {\em only} on the TeV brane\footnote{In the bulk,
such a decay is not allowed due to the
$P$-parity or equivalently, as mentioned above, since
$Q^{ (0) }$ and $d_R^{ (0) }$ are obtained from different multiplets.}. Then,
$\nu^{ \prime }_R$ can decay as before.
Or, in the absence of $X^{ \prime } - X_s$ mixing, 
$\nu^{ \prime }_R$ can decay via higher-dimensional operators on the TeV brane. 

\section{Interactions of the KK right-handed neutrino}
We are interested in computing the energy density stored in the LZP. The LZP, once it stops interacting with the rest of the thermal bath, is left as a relic.  We define $x_F = m / T_F$ where $T_F$ is the freeze-out temperature. The general formula for the contribution 
of a massive cold relic
to the energy density of the universe
is:
\begin{equation}
\Omega_{relic} h^2= \frac{s_0 \ h^2}{\rho_c M_{Pl}}\sqrt{\frac{45}{\pi g_*}}\frac{1}{\int_{x_F}^{\infty}dx \ \frac{\langle \sigma v \rangle}{x^2} }
\label{RELIC1}
\end{equation}
Here, 
$s_0$ is the entropy density today, $\rho_c$ is the critical energy density of the universe,  
$h$ is the reduced expansion rate ($H_0=h \times 100$ km s$^{-1}$ Mpc$^{-1}$) and
$g_*$, the number of relativistic degrees of freedom,
 is evaluated at the freeze-out temperature.
In the non relativistic limit, the thermally averaged annihilation cross section reads $\langle \sigma v \rangle\approx a+ b v^2$, where $v$ is the relative velocity between the two annihilating particles and Eq. (\ref{RELIC1}) becomes
\begin{equation}
\Omega_{relic} h^2= \frac{1.04 \times 10^9}{M_{Pl}}\frac{x_F}{\sqrt{g_*} }\frac{\mbox{GeV}^{-1}}{(a+ 3b/x_F)}
\label{RELIC2}
\end{equation}
where $a$ and $b$ are in GeV$^{-2}$. In the industriously studied case of neutralino dark matter, $a$ is 
smaller than $b$ because of the Majorana nature of the dark matter particle, leading to a
$p$-wave suppression of the annihilation cross section. 
In contrast, the LZP is a Dirac fermion and its cross section is not helicity suppressed.
To evaluate $\Omega_{LZP}$, we need to compute the annihilation cross section of the LZP. By definition, a WIMP 
has an annihilation cross section of the right order, $10^{-9} $ GeV$^{-2}$, leading  to the appropriate relic density to account for dark matter. 
We will now detail how our KK right-handed neutrino annihilates and explain why we expect 
it to behave as a typical WIMP.

\subsection{Estimates of cross-sections}
\label{subsec:estimates}
 We start with estimates of the couplings of the LZP and of its annihilation and elastic scattering 
cross-sections. We will then present the details in the following sections
and appendices.

\begin{figure}[h]
\begin{center}
\includegraphics[height=3.3cm]{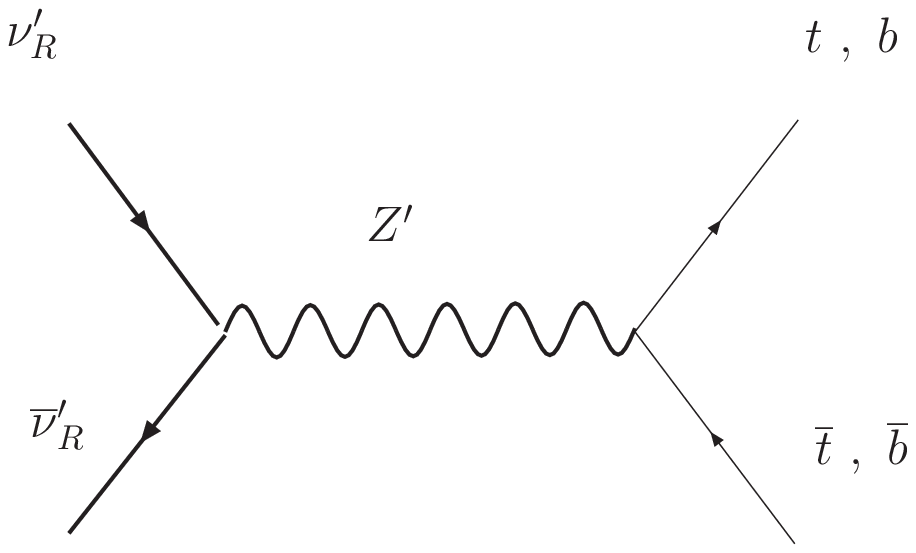}
\hspace{2cm}
\includegraphics[height=3.3cm]{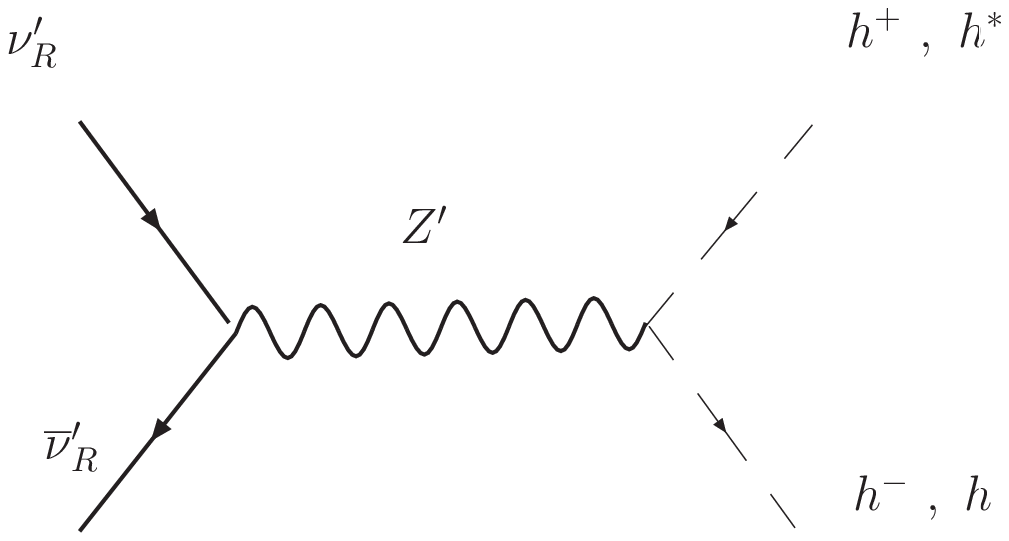}
\includegraphics[height=3.3cm]{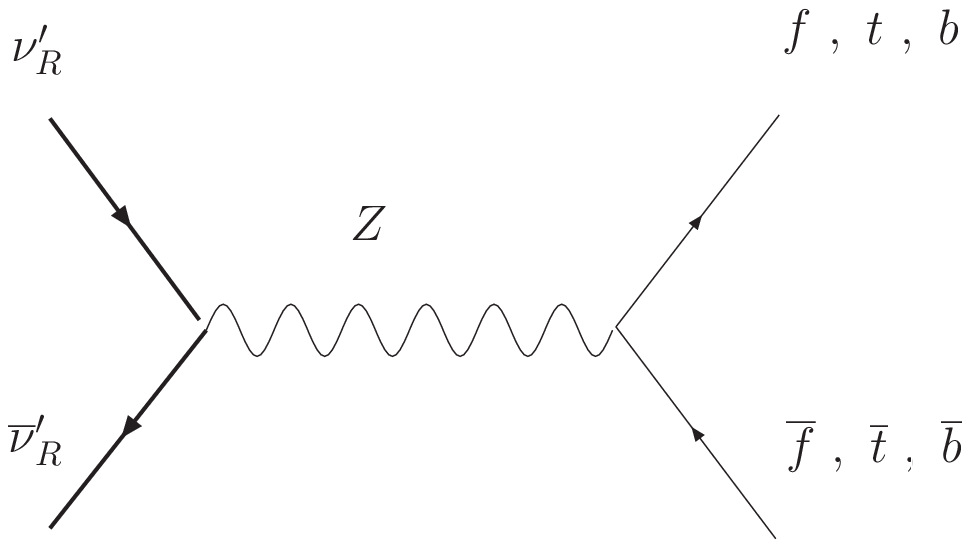}
\hspace{2cm}
\includegraphics[height=3.3cm]{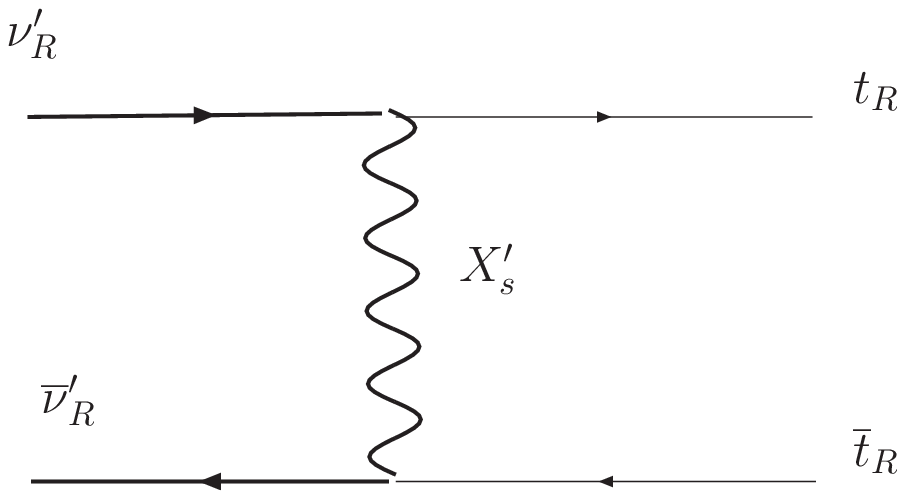}
\includegraphics[height=3.3cm]{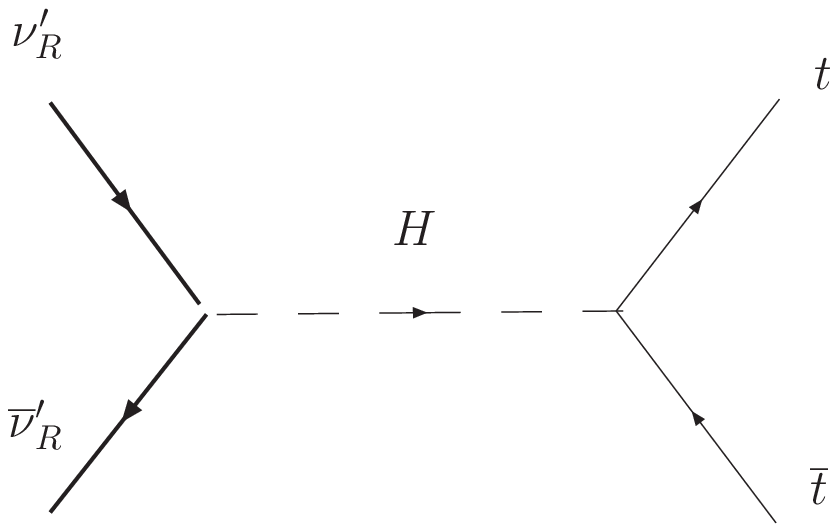}
\hspace{2cm}
\includegraphics[height=3.3cm]{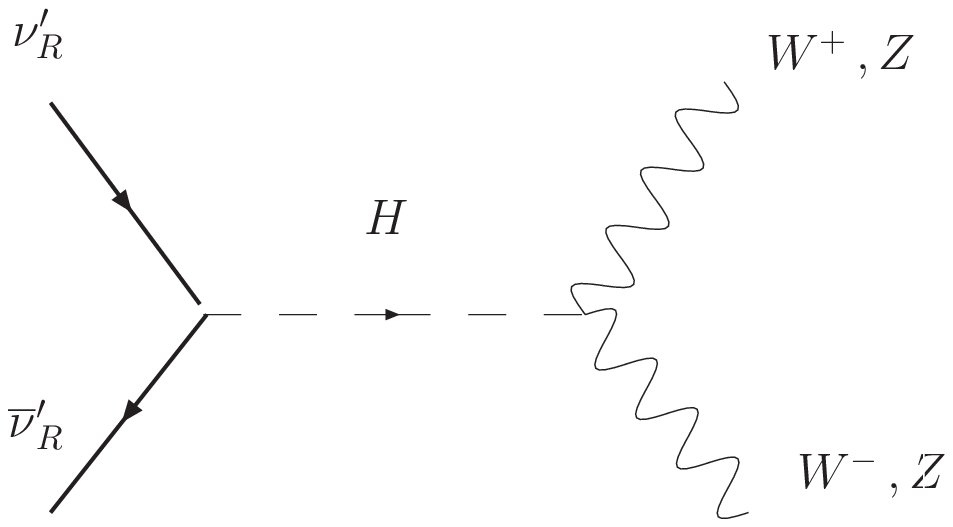}
\caption[]{LZP annihilation channels. $f$ denotes all SM fermions other than top and bottom.}
\label{fig:anni}
\end{center}
\end{figure}

All gauge and fermion KK modes, including the LZP, as well as the Higgs, the top and 
 possibly the left-handed bottom quarks, are localized near the TeV brane. Consequently, any
coupling between these particles is large.
The LZP can annihilate significantly through an $s$-channel exchange of $Z^{ \prime }$ 
gauge boson 
(into top quarks and Higgs) 
as well as a $t$-channel exchange of KK $X_s$ gauge boson into a zero mode $t_R$ as shown in 
Fig. \ref{fig:anni} (recall that the LZP is from the $t_R$ multiplet).
As explained below, those couplings are typically 5 or 6 times larger than SM couplings.
However, the particle which is exchanged has a mass of at least 3 TeV.  Effectively, the annihilation cross section has the same size as the one involving SM couplings and particles of mass of order 500 GeV. We are indeed dealing with ``weak scale" annihilation cross sections. 

In addition, we will 
show that the LZP has a significant coupling to the $Z$.  
Since the LZP can be naturally much lighter than gauge KK modes, 
$s$-channel annihilation through $Z$-exchange can also have the right size. 
 This coupling also results in a cross-section for 
direct detection via $t$-channel $Z$ exchange which is of weak-scale size.

We explain in appendix \ref{section:Higgsexchange} why we can neglect the annihilation
 through Higgs exchange in our analysis.
 
 Note that at the lowest order, the LZP cannot annihilate with itself into SM particles but only with its antiparticle, due to $Z_3$ conservation.

Let us begin by estimating the couplings of the LZP.
The $\nu^{ \prime }_R X_s t_R^{ (0) }$ coupling, appearing in the $t$-channel 
annihilation, is given
by the overlap of the three wavefunctions (see Eq. \ref{zero2kk}).
The coupling of $\nu^{ \prime }_R$ to $Z^{ \prime }$ KK modes, used in the s-channel $Z'$ annihilation, is given by Eq. \ref{3kk}.
Using the wavefunctions in Eqs. \ref{zero}, \ref{gaugekk} and
\ref{fermion-+}, we can show that $t_R^0$, $Z^{ \prime }$ and
$X_s$ KK modes
and $(-+)$ helicity of the LZP are all localized near the TeV brane. So, we expect 
the above couplings of the LZP (for $(-+)$ helicity) to have the
same size as the coupling of, say, gauge KK modes to the Higgs  on
the TeV brane. Evaluating the wavefunction of the gauge KK mode (see
Eq. \ref{gaugekk}) at the TeV brane, 
we can show that the coupling of the gauge KK
mode to the Higgs is enhanced compared to that of zero-mode gauge bosons by 
$\approx \sqrt{ 2 k \pi r_c }$ so that
we expect the above two couplings to be also
$\sim \sqrt{k \pi r_c} \times$ $4D$/zero-mode gauge coupling. A
numerical evaluation of the overlaps 
Eqs. (\ref{zero2kk}) and (\ref{3kk}) indeed confirms this expectation.
This is also expected from the CFT interpretation as explained in section \ref{cftquant}.
For $c\lesssim-1/2$, the
coupling of the other $(+-)$ helicity of the LZP to $X_s$ and $Z^{ \prime }$
is suppressed since it is localized
near the Planck brane (see appendix \ref{fermionkk}).

As mentioned above, the coupling of
$Z^{ \prime }$ to the Higgs is enhanced compared to SM couplings
(see also Eq. \ref{gaugeKKHiggs}).
Similarly,
the coupling of $t_R$ to $Z^{ \prime }$
is also enhanced by $\sqrt{ k \pi r_c }$ compared to $4D$ (``would-be'' zero-mode) gauge
coupling since both
$t_R$ zero-mode and $Z^{ \prime }$ are
localized
near the TeV brane\footnote{Again, a numerical
evaluation of the overlap of wavefunctions (Eq. \ref{zero1kk}) confirms this 
expectation and 
this is also expected from the CFT interpretation (section \ref{cft}).}. On the other
hand,
the coupling of light fermions to $Z^{ \prime }$ is
negligible since they
are localized near the Planck brane where $Z^{ \prime }$ wavefunction vanishes.
Thus, annihilation of the LZP via $Z^{ \prime }$ exchange is dominantly into
$t_R$ and Higgs (or longitudinal $W$ and $Z$).

The crucial point is that while
the gauge KK modes have a mass of a few ($3-4$) TeV, their coupling
is larger than that of 
gauge SM couplings by a factor $\sqrt{ k \pi r_c } \sim 5-6$: 
effectively the size of the interaction is like the exchange of $\sim 500-600$ GeV particles
with SM couplings.
Also, as mentioned above, $\nu^{ \prime }_R$
 can be naturally much lighter than gauge KK modes, 
 with a mass of a few hundreds of GeV. Thus, the LZP 
can naturally have ``weak-scale'' annihilation
cross-sections. 
%
%

We now explain what is the origin of the coupling of the LZP to the $Z$.
 
\subsection{Coupling to $Z$ induced by $Z-Z^{ \prime }$ mixing}
\label{ZZprimemixing}

To identify the SM electroweak gauge bosons $W$ and $Z$, we work 
in the insertion approximation for the Higgs vev as follows.
We first set the Higgs vev to zero and 
decompose the 5D $W$ and $Z$ into their zero and KK modes (i.e. the mass 
eigenstates from the effective $4D$ point of view). 
Then, we treat the Higgs vev as a perturbation:
the Higgs vev not only gives mass to zero-modes of
$W$ and $Z$, but also mixes
the zero-mode of $Z$ with KK mode of $Z$ and $Z^{ \prime }$. This
mixing is allowed due 
to the fact that the Higgs is localized on the TeV brane.
It means that the  physical $Z$ (and $W$) is dominantly the zero-mode
of $Z$, but has an admixture of KK modes of $Z$ and $Z^{ \prime }$.
We will consider the effect of this mixing at the lowest order, i.e., 
only up to ${\cal O}(v^2)$. The higher
order effects are suppressed by
$\sim v^2 g^2 k \pi r_c / M_{ KK } ^2$ in this case since the coupling 
of the Higgs to KK modes
of $W, Z$ is enhanced. Even with this enhancement, the error in our
approximation is at most $\sim {\cal O} \left(0.1 \right)$ for the KK masses
we will consider ($\gtrsim 3 $ TeV).
 On the other hand, the physical photons and gluons
are just identified with the zero modes.

The LZP being $\nu^{ \prime }_R$ does not have any direct coupling to
zero nor KK modes of $Z$. However, a coupling of $\nu^{ \prime }_R$
to the physical
$Z$ is induced via its coupling to the
$\big[ \hbox{KK mode of}
\; Z^{ \prime } \big]$-component of the physical $Z$: 
\begin{eqnarray}
g_{ Z \; I }^{ \nu^{ \prime }_R } 
& = & - \sum_n 
\frac{ g_Z Q^H_Z v^2 }{ m_n^2 }  
g_{ Z^{ \prime \; (n) } }^{ \nu^{ \prime } _R }
g_{ Z^{ \prime \; (n) } }^H
=- \sum_n
\frac{ m_Z^2 }{ m_n^2 } 
\frac{ g_{ Z^{ \prime \; (n) } }^{ \nu^{ \prime }_R } 
g_{ Z^{ \prime \; (n) } }^H }{ g_Z Q^H_Z },
\label{gLZPZviaZZprime}
\end{eqnarray}
where $m_n$ is the mass of the $n^{th}$ KK mode of $Z^{ \prime }$ and
$g_{ Z^{ \prime \; (n) } }^{ \nu^{ \prime } _R }$ and $g_{ Z^{ \prime \; (n) } }^H$
are the couplings of 
the $n^{ \hbox{th} }$ KK mode of $Z^{ \prime }$ to 
the lightest $\nu^{ \prime }_R$ KK mode and the Higgs, respectively
(see Eqs. \ref{3kk} and \ref{gaugeKKHiggs}).
Also, the
charge under $Z$ is $Q_{Z} = \tau^3_L - Q \sin^2 \theta _W$ so that
$Q^H_Z = \pm 1/2$ and 
in the second line, we have used
$m_Z^2 = g_Z^2 v^2 \left( Q^H_Z \right)^2$.
As mentioned above 
$g^{ \nu^{ \prime }_R, \; H }_{ Z^{ \prime \; (1) } } / g_{ Z^{ \prime } }
\sim \sqrt{ k \pi r_c }$, where 
$g_{ Z^{ \prime } } \equiv g_{ 5D \; Z^{ \prime } } / \sqrt{ \pi r_c }$
is the coupling of the 
``would-be'' zero-mode of $Z^{ \prime }$ just as $g_Z = 
e / \left( \sin \theta_W \cos \theta_W \right)$ 
is the coupling of the zero-mode of $Z$. 
This
results in a coupling of  $\nu^{ \prime }_R$ to $Z$ $\sim g_Z \;
k \pi r_c \; \frac{g^2_{ Z^{ \prime } } }{ g^2_Z } \times 
\frac{m_Z^2 }{ M_{KK}^2}$.
Equation (\ref{gaugeKKHiggs}) for $g_{ Z^{ \prime \; (n) } }^H$ assumes that the Higgs is 
localized on the TeV brane and will be modified in models where 
the Higgs has a profile in the bulk (see appendix \ref{sec:profileHiggs}).

As mentioned above, the coupling of the $(+-)$ helicity of the LZP to $Z^{ \prime }$ is suppressed
so that, in turn, its coupling to $Z$ induced by $Z- Z^{ \prime }$ mixing is very small.

\subsection{Coupling to $Z$ induced via $\nu^{ \prime }_R - \nu^{ \prime }_L$
mixing}
\label{nulrmixing}

There is another source of coupling of the LZP to $Z$ as follows.
We denote $5D$ Dirac KK partners of $\nu^{ \prime }_R$ (from $t_R$ multiplet)
and $\nu^{ \prime }_L$ (from $\left( t,b \right)_L$ multiplet) by $\hat{ \nu }^{ \prime }_R$
and $\hat{ \nu }^{ \prime }_L$. 
These have LH and RH Lorentz chiralities,  
respectively -- the subscript R and L denotes the fact that these are doublets
of $SU(2)_R$ and $SU(2)_L$. 
%
%
There is a Yukawa coupling of $\nu^{ \prime }_R$
and $\nu^{ \prime }_L$ to the Higgs which is the GUT counterpart of the top Yukawa:
$\lambda _{t \; 5} H \nu^{ \prime} _L \nu^{ \prime }_R$ (see Eq. \ref{Yukawa}). 
Note that only $\nu^{ \prime }_R$ and $\nu^{ \prime }_L$, i.e. $(-+)$ chiralities,
couple to the Higgs since $\hat{ \nu^{ \prime }_R }$
and $\hat{ \nu^{ \prime }_L }$ ($(+-)$ helicities) vanish on
the TeV brane.
This results in a
$\nu^{ \prime }_R - \nu^{ \prime} _L$ mass term, denoted by
$m_{ \nu^{ \prime }_L \nu^{ \prime }_R }$. Using wavefunctions 
of  KK fermions at the TeV brane (see Eqs. \ref{fermion-+TeV1} and \ref{fermion-+TeV2}),  it
is given by
\begin{equation}
m_{ \nu^{ \prime }_L \nu^{ \prime }_R } \approx\left\{ \begin{array}{ll}
 2 \lambda_{ t \; 5} k \frac{v}{ \sqrt{2} } \;
\hbox{for} \; c_{ \nu^{ \prime }_R } > - 1/2 + \epsilon, \; \hbox{where} \; 
\epsilon \sim 0.1\\
  \frac{ 2 \lambda_{ t \; 5 } k }{ f \left( c_{ \nu^{ \prime }_R } \right) }
\frac{v}{ \sqrt{2} } \;
\hbox{for} \; c_{ \nu^{ \prime }_R } < - 1/2 - \epsilon\\
	\end{array} \right.
\end{equation}
%
%
%
The $5D$ Yukawa coupling, 
$\lambda _{t \; 5}$, is related to $m_t$ as follows. Using the wavefunction of the fermionic 
zero-mode (Eq. \ref{zero}), we get,
with $c_{ L, R }$ for top quark $< 1/2 - \epsilon$, 
%
\begin{eqnarray}
\lambda_{t} & \approx & \frac{ 2 \lambda_{ t \; 5 } k }{ f \left( c_{ t_L } 
\right) 
f \left( c_{ t_R } \right) },  
\label{lambda4Dtop}
\end{eqnarray}
where
\begin{eqnarray}
f ( c ) & \approx & \sqrt{ \frac{2}{ 1 - 2 c } }
\end{eqnarray}
Therefore
\begin{equation}
m_{ \nu^{ \prime }_L \nu^{ \prime }_R } \approx\left\{ \begin{array}{ll}
m_t f \left( c_{ t_L } \right) f \left( c_{ t_R } \right)
\;
\hbox{for} \; c_{ \nu^{ \prime }_R } > - 1/2 + \epsilon\\
m_t 
\frac{ f \left( c_{ t_L } \right) f \left( c_{ t_R } \right) }
{ f \left( c_{ \nu^{ \prime }_R } \right) } \;
\hbox{for} \; c_{ \nu^{ \prime }_R } < - 1/2 - \epsilon\\
	\end{array} \right.
\end{equation}
In the following numerical estimates,
we will use $c_{ t_L } \sim 0.4$, $c_{ t_R } \sim -1/2$ leading to
 $2 \lambda_{ 5D } k \sim 3$ and also $c_{ \nu^{ \prime }_R }
\gtrsim-1/2$ 
(in the Pati-Salam symmetric limit,
$c_{ \nu^{ \prime }_R } = c_{ t_R }$) so that $m_{ \nu^{ \prime }_L \nu^{ \prime }_R }
\sim 500$ GeV.
We get the following mass matrix:
\begin{eqnarray}
\left( \bar{ \nu^{ \prime } }_R  \ \ \hat{ \nu }^{ \prime }_L
\right) M
\left( 
\begin{array}{c}
\tilde{ \nu }^{ \prime }_R  \\
\nu^{ \prime }_L
\end{array}
\right) \ \ \mbox{with} \ \ 
M & = & \left(
\begin{array}{cc}
m_{ \nu^{ \prime }_R } & m_{ \nu^{ \prime }_R \nu^{ \prime }_L } \\ 
0 & m_{ \nu^{ \prime }_L } 
\end{array}
\right)
\end{eqnarray}
The mixing angles for 
$\hat{ \nu }^{ \prime }_R - \nu^{ \prime }_L$ and
$\nu^{ \prime }_R - \hat{ \nu }^{ \prime }_L$,  obtained by 
diagonalizing $M^{ \dagger } M$ and $M M^{ \dagger }$ are denoted $\theta _L$ 
and $\theta_R$, respectively. In the limit $m_{ \nu^{ \prime }_L } \gg m_{ \nu^{ \prime }_R }$,
$m_{ \nu^{ \prime }_R \nu^{ \prime }_L }$, 
we get
\begin{equation}
\theta_L  \approx  \frac{m_{ \nu^{ \prime }_R }
m_{ \nu^{ \prime }_R \nu^{ \prime }_L } }{ m_{ \nu^{ \prime }_L }^2 } 
\ \ \mbox{and} \ \ 
\theta_R  \approx  \frac{ m_{ \nu^{ \prime }_R 
\nu^{ \prime }_L } }{ m_{ \nu^{ \prime }_L } }
\end{equation}
Explicitly,
\begin{eqnarray}
\left( \nu^{ \prime }_1 \right)_L & = & \cos \theta_L \tilde{ \nu }^{ \prime }_R
+ \sin \theta_L \nu^{ \prime }_L
\nonumber \\
\left( \nu^{ \prime }_1 \right)_R & = & \cos \theta_R \nu^{ \prime }_R + 
\sin \theta_R \tilde{ \nu }^{ \prime }_L
\end{eqnarray}
where $\nu^{ \prime }_1$ is the lightest mass eigenstate (i.e. the LZP).
Since $\nu^{ \prime }_R$ and $\tilde{ \nu }^{ \prime }_R$
do not couple to the $Z$, it is clear that the coupling to $Z$ 
induced by the above mixing is given by
\begin{eqnarray}
g_{ Z \; II }^{ 
\nu^{ \prime }_{ 1 \; L, R } 
} & \approx &  \frac{g_Z}{2}  \sin ^2 \theta_{ L, R }
\label{ZnuLnuRcoupling}
\end{eqnarray}
where $\frac{g_Z}{2}$ 
is the coupling of $\nu^{ \prime }_L$ and 
$\tilde{ \nu }^{ \prime }_L$ to $Z$.
Since $\theta_R \gg \theta_L$ (for $m_{ \nu^{ \prime }_R } \ll 
m_{ \nu^{ \prime }_L }$ which is valid for the ranges of $c$'s we consider), we will 
consider only the induced coupling of
$\left( \nu^{ \prime }_1 \right)_R$ to $Z$ and neglect the coupling of
$\left( \nu^{ \prime }_1 \right)_L$.
In the Pati-Salam symmetric limit,
$c_{ \nu^{ \prime }_L } = c_{ t_L } \sim 0.4$ 
so that 
$m_{ \nu^{ \prime \; (1) }_L }
\sim 
3/4 \pi z_v^{-1}$  (same as mass of gauge KK mode:
see Eqs. (\ref{KKfermion-+}) and (\ref{MKK})). 
So, this coupling is roughly comparable in size
to the coupling of $\nu ^{ \prime }_R$ to $Z$ induced by $Z-Z^{ \prime }$ mixing.
 
Due to bulk GUT breaking,
$c$ for $\nu^{ \prime }_L$ can be $>$ or $< 1/2$ even though
it is in the same multiplet as $(t, b)_L$. 
Hence, $\nu^{ \prime }_L$ can 
be heavier or lighter than $3/4 \pi z_v^{-1}$, resulting in a variation in
the LZP coupling to the $Z$.

We see that both induced $Z$ couplings to the $(+-)$ helicity of the 
LZP  are small. 
We will  consider
only the resultant $Z$ coupling to the
$(-+)$ helicity of the LZP. We denote this coupling by 
$g_Z^{ \nu^{ \prime }_R }$:
\begin{eqnarray}
g_Z^{ \nu^{ \prime }_R } & \equiv & g_{ Z \; I }^{ \nu^{ \prime }_R } +
g_{ Z \; II }^{ \nu^{ \prime }_{ 1 R } }
\label{LZPZ} 
\end{eqnarray}

Given this LZP -$Z$ coupling, we can estimate
the cross-section for  LZP annihilation  via $Z$ exchange
into a given pair of SM fermions
as $\sigma \sim \left( k \pi r_c \; g^2_{ Z ^{ \prime } } m_Z^2 / M_{ KK }^2 \right)^2
\times 
m_{ LZP }^2 / \left( m_{ LZP }^2 - m_Z^2 \right)^2$,
where momentum in the $Z$ propagator
is $\sim m_{LZP}$.
Clearly, 
for $m_{ LZP } \gg m_Z$, 
this cross-section is suppressed by
$\sim m_Z^4 / m_{ LZP }^4$ 
compared to $X_s$ or $Z^{ \prime }$ exchange,
but for $m_{ LZP } \ll m_Z$, 
it is the dominant annihilation channel, especially once we sum over
all the SM fermions in the final state.

We can also estimate 
its cross-section for scattering off quarks in nuclei by 
$t$-channel exchange of $Z$: $\sigma _Z \sim
\left( k \pi r_c \; g^2_{ Z ^{ \prime } } / M_{ KK }^2 \right)^2 \times 
m_{ LZP }^2$ (here $Z$ propagator gives
$1/m_Z^2$ since the exchanged momentum is $\ll m_{ LZP }$). 
Since $m_{ LZP } \sim$ few $100$ GeV, we see that 
direct detection cross-sections for 
the LZP are of weak-scale size\footnote{$Z^{ \prime }$ exchange 
is small here since
light quarks couple very weakly to $Z^{ \prime }$}.

There is also a coupling of the two chiralities of the LZP to the Higgs which will 
be used in appendix   \ref{section:Higgsexchange} to estimate annihilation via
Higgs exchange:
\begin{eqnarray}
g_H & = & 2 \lambda_{ 5D } k \; \sin \theta_L \cos \theta_R \;
\hbox{for} \; c_{ \nu^{ \prime }_R } > -1/2 + \epsilon \nonumber \\
 & \approx & \frac{ 
2 \lambda_{ 5D } k m_{ \nu^{ \prime }_R \nu^{ \prime }_L } 
m_{ \nu^{ \prime }_R } 
}
{ 
m_{ \nu^{ \prime }_L }^2 
} \; \hbox{in the limit} \; m_{ \nu^{ \prime }_L } \gg m_{ \nu^{ \prime }_R }, \;
m_{ \nu^{ \prime }_R \nu^{ \prime }_L }
\nonumber \\ 
 & \sim & \frac{ 
1.5 \; \hbox{TeV} m_{ \nu^{ \prime }_R }
}
{ 
m_{ \nu^{ \prime }_L }^2 
}, 
\end{eqnarray}
whereas for $c_{ \nu^{ \prime }_R } < -1/2 - \epsilon$, we get
$g_H = 2 \lambda_{ 5D } k / f \left( c_{ \nu^{ \prime }_R }
\right) \; \sin \theta_L \cos \theta_R$.
  
 Clearly, both  $g_H$ and $g^{ \nu^{ \prime }_R }_{ Z, \; II }$ 
depend sensitively on the Higgs profile and will be modified in 
models where the Higgs is the fifth component of a gauge boson $A_5$ 
(see appendix B) or in Higgsless models. Our numerical analysis 
will actually be done assuming that the Higgs is $A_5$.
  
\section{Effect of NLZP's and coannihilation} 
In SUSY dark matter, the effect of NLSPs can be dramatic. For instance, 
the annihilation cross section of the neutralino being helicity suppressed, 
if the NLSP is a scalar, the coannihilation cross section can control the relic density of the LSP. The situation is different for the 
lightest KK particle (LKP) in universal extra dimensions
\cite{Servant:2002aq} and will be similarly different for the LZP since we are not dealing with a Majorana particle.
However, even if coannihilation does not play a major role, the effect of NLZPs on the relic density should still be considered.  Indeed, the quantity $x_F=m/T_F\sim 25$,
where $T_F$ is the freeze-out temperature, 
of a weakly interacting particle is almost a constant
since it depends only logarithmically on the mass and annihilation cross section. Therefore, the freeze-out temperature of a particle grows linearly with its mass. The NLZPs will freeze-out earlier but the question is whether they will decay before or after the LZP freezes. If they decay before,  we do not have to consider their effect 
since their decay products will thermalize and the final relic density of the LZP will only depend on the annihilation cross section of the LZP, $ \sigma_{\mbox{\tiny LZP}}$. On the other hand, if they decay after, they will contribute to the final relic density of 
the LZP by a factor given by $\sigma_{\mbox{\tiny LZP}}/\sigma_{\mbox{\tiny NLZP}}$
(since $\Omega_{ relic } \propto 1 / \sigma_{ relic }$). 
In SUSY, 
the annihilation cross sections of squarks and sleptons are enhanced 
relative to those of the neutralino and,
unless they are degenerate with the neutralino, they decay fast into
it. Consequently, if they are heavier by 
say 20 percent (so that coannihilation does not play any role), 
their effect can be omitted. Let us check  now what happens with NLZPs.

\subsection{Relic density of other $Z_3$-charged fermions}

The other light KK GUT partners of $t_R$ 
have SM gauge interactions unlike the LZP.
We
estimate the cross-sections due to zero-mode $Z$ or gluon exchange
as follows (up to factors of $2 \pi$ from phase space):
\begin{eqnarray}
\sigma_{ Z \rightarrow f \bar{f} } \sim \frac{ g_{ SM }^4 N }{ m_{ NLZP }^2 }
\end{eqnarray}
These cross-sections are
enhanced by a factor $N \sim 20$ for $Z$ exchange and
gluon exchange due to multiplicity of final states.
In addition, NLZP's also annihilate
via $s$-channel $Z^{ \prime }$ and KK $Z$ or gluon exchange similarly to LZP:
\begin{eqnarray}
\sigma_{ KK Z, gluon, Z^{ \prime } } &
\sim & \frac{ g_{ SM }^4 \left( k \pi r_c \right)^2
m_{ NLZP }^2 }{ M_{ KK }^4 }
\end{eqnarray}
where it is assumed that $m_{NLZP} < M_{ KK }$.
Since the {\em total} LZP annihilation
cross-section for LZP is of this size,
it is clear that
the total annihilation
cross-section of the NLZP
is larger that that for LZP.
For $m_{ NLZP } \lesssim N^{ 1/4 } M_{ KK } / \sqrt{ k \pi r_c }$,
the cross-section from exchange of zero-mode $Z$ or gluon dominates.
The smallest ratio of annihilation cross-sections
of NLZP and LZP occurs for this ``critical''
mass and is
$\sim \sqrt{N}
\left( M_{ KK } / \sqrt{ k \pi r_c } / m_{ LZP } \right)^2$ which is 
$\gtrsim \sqrt{N}$ since typically
$m_{ LZP }\lesssim M_{ KK } / \sqrt{ k \pi r_c }$
-- the latter
also implies that this critical mass $\gtrsim N^{ 1/4 } m_{ LZP }$.

Depending on the mass and couplings of the NLZP, its  decay into the LZP occurs
before or after the LZP freezes out (but the decay can easily occur
before BBN in the latter case: see section 7.4). Let us
consider the important case when the NLZP decays after the LZP freezes out.
It is clear that for a wide range of NLZP masses,  the NLZP 
annihilation cross-section is $\gtrsim 10$
times that of LZP so our relic density
predictions will receive 
corrections
 $\lesssim 10 \%$. The exception is
when $m_{NLZP}$ is close to the critical mass {\em and}
$m_{ LZP } \sim M_{ KK } / \sqrt{ k \pi r_c }$
in which case the relic density can be as large
as $\sim 1/ \sqrt{N} \sim 1/4$ of the LZP and a more careful
study is required.

$Z_3$-charged fermions from other multiplets are heavier ($\sim M_{ KK }$)
so that KK $Z^{ \prime }$, $Z$, gluon exchange dominates
the annihilation with cross-sections
much larger than LZP. This results in a 
very small relic density (before their decay into the LZP).
Also, we do
not have to consider
$n= 2$ level KK states since they decay into $n=1$ very fast.

\subsection{Coannihilation}
\label{subsec:coanni}

The only important coannihilation channel is with $\tau^{ \prime }_R$, the 
$SU(2)_R$ partner of the LZP (from $t_R$
multiplet) via $s$-channel exchange of $W_R^{ \pm }$
followed by mixing of $W_R^{ \pm }$ with $W_L^{ \pm }$. 
Indeed,
the only direct LZP coupling to zero-mode fermion 
is LZP-$X_s-t_R^{ (0) }$. Thus, 
coannihilation with, say, KK $Q^{ \prime }$ from $t_R$ multiplet
into $t_R$ pairs has to go through
$X - X_s$ mixing hence is suppressed
(since the only coupling of KK $Q^{ \prime }$
to zero-mode fermion is $Q^{ \prime } - t_R^{ (0) } - X$).
Whereas, coannihilation with, say, KK $L^{ \prime }_L$ from $(t,b)_L$ multiplet
can proceed via $X_s$ exchange since there is
a KK $L^{\prime}_L - (t,b)_L^{ (0) } -X_s$ coupling. However, this co-annihilation
is 
small because $(t,b)_L^{ (0) }$ has an almost flat profile thus has small
overlap and coupling with KK $L^{ \prime }_L$ and $X_s$. In any case, KK $L^{ \prime }_L$
has mass $\sim M_{ KK }$ so that its relic density (before it decays
into the LZP) is much smaller than that of the LZP.
Recall that $\tilde{b}_R$ is heavy  ($\gtrsim 1.5$ TeV ) as well as the KK mode of $t_R$
 ($\gtrsim 3$ TeV). Moreover, they are not charged under $Z_3$ hence
decay
fast into SM states so that coannihilation with those states can also be ignored.

The coupling above
results in a prompt $2$-body decay of $\tau^{ \prime }_R$ into LZP and $W^{ \pm }$.
Therefore, unless $\tau^{ \prime }_R$ and LZP are degenerate,
$\tau^{ \prime} _R$ decays into the LZP before the LZP freezes out so that
we do not need to consider coannihilation.
If
$\tau^{ \prime} _R$ is nearly degenerate with the LZP, coannihilation could occur.
 However, this
co-annihilation cross-section is of the same size as that for LZP self-annihilation
via $Z$ exchange, hence it is smaller than the total LZP 
self-annihilation.
Also, if $\tau^{ \prime }_R$ and LZP are degenerate, 
the 
number density of $\tau^{\prime }_R$
is much smaller than that of the LZP since its mass is less than the critical
mass mentioned above. For these two reasons, it makes sense to neglect
co-annihilation in this first study.

\begin{figure}[h]
\begin{center}
\includegraphics[height=3cm,width=7cm]{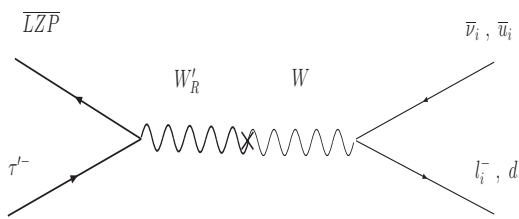}
\end{center}
\caption[]{Potentially non-negligible coannihilation channel.}
\label{fig:coanni}
\end{figure}
 

\section{Values of gauge couplings}
\label{section:gaugecouplings}

 In order to calculate the relic density and direct detection prospects of 
the LZP, we need to determine the couplings of KK modes in terms of the 
observed SM gauge couplings. This relation is somewhat
non-trivial as we will show in this section. The brief summary is that 
couplings of KK modes
(up to overlap of wavefunctions) vary from $g_s$ to $g^{ \prime }$
which are the QCD and the hypercharge gauge couplings, respectively.

\subsection{Gauge couplings in $SO(10)$}
\label{g10}

In the case of $SO(10)$ gauge symmetry in the bulk, 
the three 5D Pati-Salam gauge couplings are unified, 
$g_{ 5 \; c } 
= g_{ 5 \; L } = g_{ 5 \; R } \equiv g_5$. However, loop corrections
are crucial in relating these bulk couplings to couplings of KK and zero-modes
of gauge fields as we show in what follows.

\subsubsection{No bulk breaking of GUT}



Let us begin with the case of no bulk breaking. 
At tree-level, all zero-mode SM 
gauge couplings  are given by
$g_{ 5 } / \sqrt{ \pi r_c }$ due to 4D gauge invariance\footnote{For $W$, $Z$, 
this is true before electroweak symmetry 
breaking and for $\gamma$, $Z$, up to weak mixing angles.}.
Couplings of KK gauge modes are also
given by $g_{ 5 }$ up to factors of overlap of wavefunctions. We now study 
how loop corrections change this picture.

Loop corrections 
to couplings of gauge zero-mode and KK modes are linearly divergent. 
 Since divergences are short distance dominated, 
they can be absorbed into
renormalization of local terms, i.e.,
bulk gauge coupling and brane-localized couplings. Hence,
the 
divergences appear in couplings of gauge zero and KK modes
in the same way.
Bulk and
TeV brane-localized divergences are $SO(10)$ symmetric (since
$SO(10)$ is unbroken there), whereas Planck-brane localized divergence
is not. Recall that
brane-localized terms are neglected in our analysis.

The  finite part of 
one-loop corrections to couplings of lightest  KK modes
are mostly universal since KK modes are 
localized near the TeV brane where GUT is unbroken. 
We absorb all 
of these finite universal corrections into renormalized $g_{ 5 }$, denoted by
$g_ { 5 \; ren. }$ (which is therefore also universal) so that
one-loop corrected couplings of gauge KK modes are given (up to wavefunction
overlaps) by $g_{ 10 } 
\equiv g_ { 5 \; ren. } / \sqrt{ \pi r_c }$. 

In contrast, the
finite one-loop corrections to zero-mode gauge couplings
are log-enhanced (loops are sensitive to Planckian cut-off's since 
zero-modes span the entire extra dimension) and 
non-universal (since GUT is broken on the Planck brane).
These corrections will explain
why low-energy measured SM gauge couplings are non-universal as follows.

Given this, let us see
if we can 
extract the couplings of gauge KK modes (i.e. $g_{ 10 }$)
from measured zero-mode gauge
couplings which have the following form \cite{gns, rsgut}:
\begin{equation}
\frac{1}{ g_{ 4 \; i }^2 } = \frac{1}{ g_{ 10 }^2 } + \frac{C}{8 \pi^2 } k \pi r_c 
+ \frac{ b^{ RS }_i }{ 8 \pi^2 } \log \frac{ 
k
}
{ m_Z } + \Delta_i.
\label{g4D}
\end{equation}
The non-universal correction with $b^{ RS }_i$ is 
IR dominated and therefore calculable and is roughly the running due to
loops of SM gauge zero-modes, i.e., $b^{ RS } =$ gauge contribution to SM
$\beta$-function coefficients. This differential running is almost the same as in the SM
(up to the contribution of the Higgs in the SM which is small: running due to 
fermions in the SM is mostly
universal). 
The term with $C$ (which can be an $O(1)$ contribution in $g_{ 4 }^{-2}$), where
$C$ is given by, for example, 
$\frac{2}{3} \times$ the Dynkin index for bulk fermions, corresponds to
finite,  universal contributions (roughly from loops of KK modes)
which cannot be absorbed in
$g_{ 5 \; ren. }$ (i.e., in $g_{ 10 }$). The point is that the
finite parts of one-loop corrections to couplings
of zero-mode and KK mode are non-local (and hence are not constrained by
$5D$ gauge invariance) and so do not have to be 
identical (unlike divergent parts which have to be the same
by locality and $5D$ gauge invariance). 
The $C$-term is calculable in this case since 
bulk particle content is known (cf. next section).

In $SU(5)$ or $SO(10)$,
calculable one loop non-universal
corrections (from $b_{ RS }$) give unification of gauge couplings
to within $\sim 10 \%$, just as in the SM: 
$\Delta_i$'s denote threshold-type non-universal
corrections (tree-level or loop) which can correct this discrepancy.
In this case, $\Delta_i$ can be due to 
finite non-universal loop corrections
from localizing zero-mode fermions in the bulk as follows.
The contribution to running of zero-mode gauge coupling from loops with
zero-mode fermions  
is universal (as in SM) since even though quark and lepton zero-modes come
from
different bulk multiplets, they can be assembled into complete $SU(5)$ multiplets.
However,  KK  fermions within a multiplet are split in mass due to 
different BC on the Planck brane\footnote{For $c > 1/2$, spectrum of $(++)$ KK fermions
is given by 
$m_n z_v \approx$ zeroes of $J_{ c - 1/2 }
\approx \pi \left( n + c / 2 - 1/2 \right)$, where the last formula
is valid for  $m_n z_v \gg 1$, whereas, 
for $-1/2 < c < 1/2 - \epsilon$ (where $\epsilon 
\gtrsim 0.1$), we get $m_n z_v \approx$ zeroes of $J_{ -c + 1/2 }
\approx \pi \left( n - c / 2 \right)$, where the last formula
is valid for  $m_n z_v \gg 1$. Compare this to the spectrum for $(-+)$ fermions
in section \ref{fermionkk}.}: 
this splitting 
is negligible for $c > 1/2$ (light fermions with zero-modes
localized near Planck brane)
and $O(1)$ for $c \ll 1/2$ (for $t_R$ multiplet with zero-mode near TeV brane)
Thus, there are non-universal
threshold-type corrections to
zero-mode gauge couplings from loop
contribution of these KK modes with split masses \cite{choiflavor, gherghetta,
acs}.
These effects depend on the various $c$ parameters.
 Such non-universal effects from fermion loops do not
contribute to the couplings of KK modes since, for the loops to be non-universal,
they have to sense the Planck brane where
GUT is broken, whereas KK modes are localized near the  TeV brane.

Due to the dependency on the choice of $c$ parameters, 
these $\Delta_i$'s could result in a $\sim 10 \%$ uncertainty 
in extracting the
renormalized $5D$ gauge coupling, i.e., $g_{ 10 }$,
from the measured $4D$ gauge coupling (see Eq. \ref{g4D}).

\subsubsection{Bulk breaking of GUT}
\label{subsubsec:bulkbreaking}

The breaking of GUT by bulk scalars in  complete 
$SO(10)$ representations modifies the expression
for $4D$ gauge couplings
as follows.

First, let us assume $M_{ GUT } /k \rightarrow 0$, which means that we neglect 
for the moment the GUT-scale splitting in masses of bulk fields
(splitting of $X$, $Y$ from SM gauge fields
and also between various components of bulk scalars which break
$SO(10)$). 
We see that
the contribution from loops of KK modes of bulk scalars to the universal
$C$-term in Eq. \ref{g4D} 
depends on the  unknown representation
of bulk scalar 
which breaks $SO(10)$ while the part of the $C$-term from bulk gauge and fermion fields
is calculable.
Due to this
UV-sensitivity in
$C$, there is an $O(1)$ uncertainty 
in extracting $g_{ 10 }$ from measured couplings.
This is why we {\em allow $g_{ 10 }$ to vary 
between, say, $g^{ \prime}$ and $g_s$}.

 
Now, consider the effects of finite (small) $M_{ GUT }$.
In \cite{rsgut}, 
$\Delta_i$'s
from local  higher-dimensional operators 
(with bulk breaking of GUT), i.e., $a$-term in Eq. \ref{bulk}, 
were invoked to achieve unification.
These are UV-sensitive and uncalculable since  the representation of the 
bulk scalar 
and the coefficient of the higher-dimensional operator are
 unknown. However,
they are  local effects which can be absorbed into the renormalized bulk coupling,
i.e., in the 1st term in Eq. \ref{g4D}.
This is clearly seen from Eq. \ref{bulk}\footnote{So, 
strictly speaking, $g_{ 5 \; ren. }$ or $g_{ 10 }$
differ by $\sim 10 \%$ between the various
subgroups of $SO(10)$. We neglect this effect.}.
The point is that these effects
enter identically in couplings of zero and KK modes
so that they
do not affect the extraction of $g_{ 10 }$ from measured gauge couplings.


As mentioned in section \ref{bulkbreakingsize},
there are also finite
non-universal loop corrections of 
size $\sim \frac{C k \pi r_c }{  8 \pi^2} \frac{M_{ GUT }^2}{ k^2}$
in  zero-mode gauge coupling 
due to GUT-scale splitting in
$5D$ masses. For example, $X,Y$ have 5D mass $M_{ GUT }$ so that KK modes
are not exactly degenerate with SM KK modes. Similarly,
there are $O \left( M_{ GUT } \right)$
splittings in 5D masses for the various components of the
bulk scalar which breaks GUT. As usual, these lead to UV-sensitive corrections. 
For $M_{ GUT } < k$, these loop corrections are smaller than the $b^{ RS }$ terms
and can be incorporated in $\Delta _i$'s in Eq. \ref{g4D}.

Bulk GUT breaking being present near the TeV brane as well,
these effects are also
present in the couplings of KK modes. However, they
are non-local and so do not enter
in the same way in the corrections to 
couplings of zero-modes and KK modes.
Just like the universal $C$ term in Eq. \ref{g4D},
these loop corrections (and this contribution to 
$\Delta_i$'s)
cannot be completely absorbed into
$g_{ 5 \; ren. }$ (i.e., into $g_{ 10 }$)\footnote{unlike 
the contribution from higher-dimensional
operators.
As usual, the non-universal
corrections
to couplings of KK modes can be absorbed into $g_{ 5 \; ren. }$
which then differ by $\sim 10 \%$ between various subgroups of 
$SO(10)$.}. Thus, recalling that
these $\Delta_i$'s are 
UV-sensitive, 
they result in an additional $\sim 10 \%$ 
uncertainty in extracting
$g_{ 10 }$ from the measured $4D$ gauge coupling.

\subsection{Gauge couplings in Pati-Salam}

In Pati-Salam, the $5D$ $SU(4)_c$ (and hence the $X_s$) gauge coupling
is independent of
the $SU(2)_R$ (and hence the $Z^{ \prime }$)
bulk gauge coupling. Annihilation of the LZP depends on both
gauge couplings, whereas its direct detection (via its
induced coupling to $Z$) depends only on the latter and on the mass
of $\nu^{ \prime}_L$ from $\left( t,b \right)_L$ multiplet
as explained in section 9.3.

It is clear that at tree-level, 
the three bulk gauge couplings ($g_{ 5 \; c}$, $g_{ 5 L }$ and $g_{ 5 \; R }$)
are fixed by the three measured SM gauge 
couplings.
The analysis of loop corrections can be done in a way similar to the $SO(10)$ case.
We start with the case of no bulk breaking. As before, divergences 
in loop corrections to couplings
of KK and zero-modes of gauge fields are identical and can be absorbed into
renormalized bulk and brane couplings, but these divergences, both 
 bulk and brane-localized, are  non-universal, unlike in $SO(10)$.
The finite loop corrections to 
couplings of KK modes are also non-universal for 
the same reason. As before, we absorb divergences and finite loop corrections
to couplings of KK modes into the three $g_{ 5 \; ren. }^i$.
As far as zero-mode gauge couplings are concerned,
the $b^{ RS }$ contribution (roughly from SM gauge zero-modes) is like in $SO(10)$.
However, the
finite $O(1)$ correction in $4D$ gauge couplings 
which cannot be completely absorbed into $g_{ 5 \; ren. }$, i.e., the
$C$-term in Eq. \ref{g4D} is also non 
universal since gauge KK modes are not in complete
$SO(10)$ or $SU(5)$ multiplets, but this contribution is
calculable as before
since the bulk gauge and fermionic content is known.

With bulk breaking of Pati-Salam, 
like in the case of $SO(10)$, the $O(1)$ contribution to $C$-term 
from bulk scalars which break Pati-Salam depends on their unknown
representations,
but the 
crucial difference is that this contribution is non universal 
since the bulk scalars need not be in complete
$SO(10)$ multiplets.
As before, 
due to this $O(1)$ UV-sensitivity of $C$-term,
there is $O(1)$  
uncertainty in extracting $g_{ 5 \; ren. }$ from 
measured gauge couplings.
The difference from $SO(10)$ is that this uncertainty 
is not a uniform effect in all $g_{ 5 \; ren}^i$. 
Thus, we independently
vary each of the three analogs of $g_{ 10 }$ between $g^{ \prime }$
and $g_s$ (just as we varied $g_{ 10 }$ in the case of $SO(10)$).



\subsection{5D strong coupling scale}
\label{subsec:strong}

So far, we discussed the size of finite one-loop effects. We found that finite
universal effects, namely the $C$ term (which is roughly 
the contribution from KK modes in the loop)
can be comparable to the tree-level effect in $g_4^2$ (see Eq. \ref{g4D}).
This implies that we
need the $5D$ cut-off $\Lambda$
to be not much larger than $k$. Otherwise, the linearly divergent loop effect
which is larger than the finite effect (again, this effect is mainly from KK modes
and was absorbed into $g_{ 5 \; ren. }$) will be larger than the tree-level contribution
and 
perturbation theory breaks down completely.
The problem with
$\Lambda \sim k$ is that the
5D effective field theory (or KK) description is no longer valid.

Let us consider this issue in more detail by estimating the 5D strong
coupling scale, $\Lambda_{ strong }$, i.e., the scale at which the size of the
divergent loop contribution
becomes as large as the tree level one. Of course,
the maximal allowed cut-off scale is also $\Lambda_{strong}$.
To obtain $\Lambda_{strong}$, we equate
the tree-level $1/g^2_{5D}$ to its one-loop correction (see, for example,
\cite{Chacko:1999hg}):
$$
\frac{1}{g^2_{5D}} \sim 2 \times 2 \times 2/3 \times 10 
\frac{\Lambda_{strong}}{ 24 \pi^3}
$$
where we have considered the contribution of bulk fermions since,
due to the large number of bulk matter multiplets, we expect the fermion
effect to be large\footnote{Note that the differential
running (thus gauge coupling unification) is not modified from that in the SM
due to these large number of bulk multiplets
since the KK modes from these multiplets
are in complete SU(5) multiplets, whereas the zero-modes are exactly
the SM particles.}.
Here, we have included factors of 2/3 for fermions, 2 for 5D or Dirac fermions,
another 2 for the Dynkin index of the 16 and finally $24 \pi^3$ for the 5D loop factor.
We assumed a total number of ten bulk 16's.

Using $k \ g^2_{5D}  \sim g^2_{4D} \times \log (M_{Pl}/TeV)$, we
get $\Lambda_{strong} \sim 1.5 \ k$, which is close to, but a bit larger than $k$.
The contribution from gauge fields in the loop
is also of the same order and tends to cancel
the fermion contribution. Also, there are $O(1)$ uncertainties
in the value of $\Lambda_s$ (the above is just an estimate). Thus, 
the strong coupling scale and the cut-off scale can be
a factor of 2 or so larger (but not more) than $k$ so
that the 5D effective field theory description is valid (see \cite{acs} for
more details).
The point is that, in order to be able to neglect the effect of the
exchange of new states at the cut-off scale in our
cross section calculations (compared to KK exchange), it is clear that the
cut-off states should be heavier than KK scale, i.e. there should be a gap between
$k$
and the $5D$ cut-off.
Since the cross-sections are typically $\propto 1 / M^4$, where $M$ is the mass of the exchanged
heavy particle,  the effect of cut-off states is
suppressed by $O(10)$ even for a small gap of $\sim 2$.

\section{Dark matter relic density}

We now have all ingredients at hand to make a detailed calculation of annihilation cross sections.
We are going to present our predictions assuming the following:\\

$\bullet$ The LZP indeed comes from the multiplet with $t_R$. There is still a possibility that it comes from the multiplet with $(b,t)_L$.
The reason is that
the $c$ for $\left( t,b \right)_L$
is $\sim 0.3-0.4$ while, at the same time, it is not excluded that $c$ for $t_R$ is $\sim 0$ rather than $-1/2$ . So, if we allow for splitting $\Delta c$ up to 0.5, it may happen that the KK RH neutrino in the $(t,b)_L$ multiplet has a (negative) $c$ which is smaller than 
the $c$'s of the $(-+)$ fermions 
in the $t_R$ multiplet. This is a possibility we do not investigate here.\\

$\bullet$ We ignore coannihilation effects as well as  the NLZP's contribution to the final relic density 
 as argued in section 10. \\

$\bullet$ We assume there is no asymmetry between LZPs and anti-LZPs, at least  before freeze-out. The total dark matter energy density is given by $\Omega h^2= (n_{\mbox{\tiny LZP}}+ n_{\overline{\mbox{\tiny LZP}}}) m_{\mbox{\tiny LZP}}/\rho_c$ so that the effective annihilation cross section $\sigma $ in Eq.
\ref{RELIC1} corresponds to $\frac{1}{2}{ \sigma_{\nu'_R \overline{\nu}'_R \rightarrow SM} }$. \\

We evaluated all diagrams presented in Fig.~\ref{fig:anni}. Expressions for the cross sections are given in appendix \ref{Appendix:anni}. We fixed the Higgs mass to $m_h=$500 GeV but our results do not depend sensitively on $m_h$. We looked at the two cases $M_{KK}=3$, $6$ TeV. For each case, a range of values for the LZP--$Z$ coupling is obtained 
 by  varying 
$c$ of ${\nu^{\prime}_L}$ from $(t, b)_L$ multiplet
(see section \ref{nulrmixing}).  
We allow $g_{10}$ to be a free parameter which we vary between $g^{\prime}$ and $g_s$. The origin of the uncertainty on $g_{10}$ is explained in \ref{subsubsec:bulkbreaking}.
For $m_{\mbox{\tiny LZP}} < M_Z/2$, the LZP-$Z$ coupling is in principle constrained by the 
invisible partial width of the $Z$: 
\be
\Gamma_{Z\rightarrow \nu^{\prime}_R  \overline{\nu}^{\prime}_R}^{\mbox{inv}}=\frac{{{g_Z^{\nu'_R}}}^2\,{\sqrt{1 - \frac{4\,m_{\mbox{\tiny LZP}}^2}{{{M_Z}}^2}}}\,
    \left( -2\,m_{\mbox{\tiny LZP}}^2 + {{M_Z}}^2 \right) }{24\, \pi {M_Z} } \lesssim 1.5 \mbox { MeV}
\ee
Such a bound is almost always satisfied. It is only in the very narrow region 
with $g_{10}=g_s$ and $M_{KK}=3$ TeV that this constrains $c_{\nu^{\prime}_L}$  
to be $\gtrsim 0.3$.

\begin{figure}[h]
\begin{center}
\includegraphics[height=7cm,width=10.5cm]{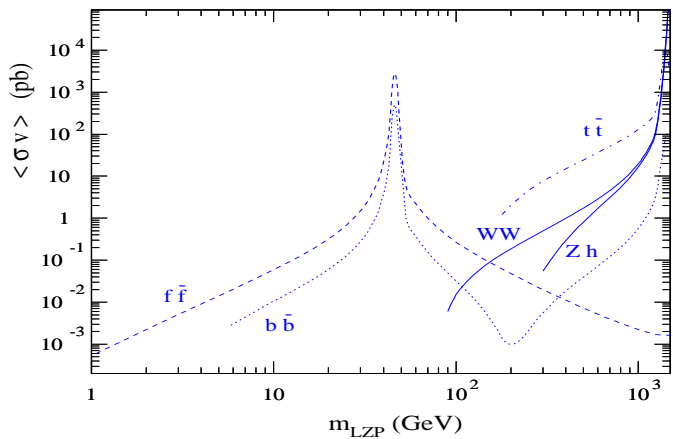}
\end{center}
\caption[]{The different annihilation channels evaluated at the freeze out temperature, for a typical choice of parameters, namely, 
$m_{KK}=3$ TeV, $c_{t_R}=-1/2$.  ``f'' denotes all SM fermions except top and bottom. }
\label{fig:compare}
\end{figure}

Figure \ref{fig:compare} shows the relative sizes of the various contributions to the total annihilation cross section for a typical choice of parameters.

To summarize,
we obtain the correct relic density for the LZP for a wide range of masses from
$10$ GeV to $1$ TeV.

\begin{figure}[h]
\begin{center}
\includegraphics[height=7.5cm,width=13.cm]{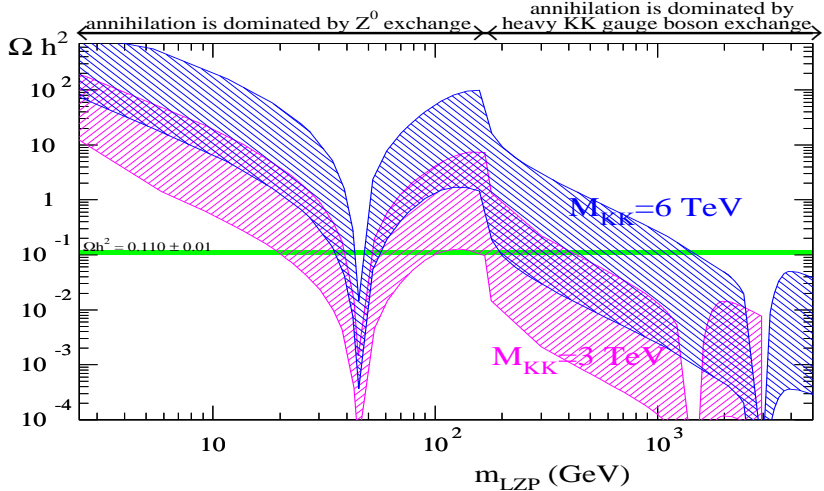}
\end{center}
\caption[]{Predictions for $\Omega_{\mbox{\tiny LZP}} \ h^2$ for two values of the gauge KK mass $M_{KK}$. Both regions are obtained by varying $g_{10}$ and $c_{\nu_L^{\prime}}$.
The kink at $m_{LZP}=m_t$ corresponds to the opening of the annihilation channel into top quarks.
 }
\label{fig:relicdensity}
\end{figure}

\section{Direct detection}

WIMP dark matter can be probed directly via its elastic scattering off nuclei in underground detectors. 
Several groups are presently carrying out direct searches for galactic halo WIMPs through their elastic scattering off target nuclei. In the absence of positive signal, these experiments set limits on the properties of WIMP dark matter (given some assumptions on halo properties and the local dark matter distribution).
Experiments such as CDMS or Edelweiss are now able to probe WIMP-nucleon cross sections of order $10^{-7}$ pb and therefore put constraints on the parameter spaces of various DM candidates.
The most stringent constraints come from spin-independent interactions. In particular, 
any WIMP with a large coupling to the $Z$ gauge boson is severely constrained. 

In a significant region of parameter space, our LZP has a large coupling to the Z as detailed in subsections  \ref{ZZprimemixing}  and \ref{nulrmixing}. Consequently, as is shown in Fig.~\ref{fig:directprediction}, its entire parameter space should be tested in near future experiments. The elastic 
scattering cross section is an important quantity as it also controls the rate at which 
particles accrete into the Earth and the Sun and so determines the signal in 
the indirect detection experiments as we will see in section \ref{sec:indirect}.

\subsection{Elastic scattering cross section}

\begin{figure}[h]
\begin{center}
\includegraphics[height=3cm,width=5.5cm]{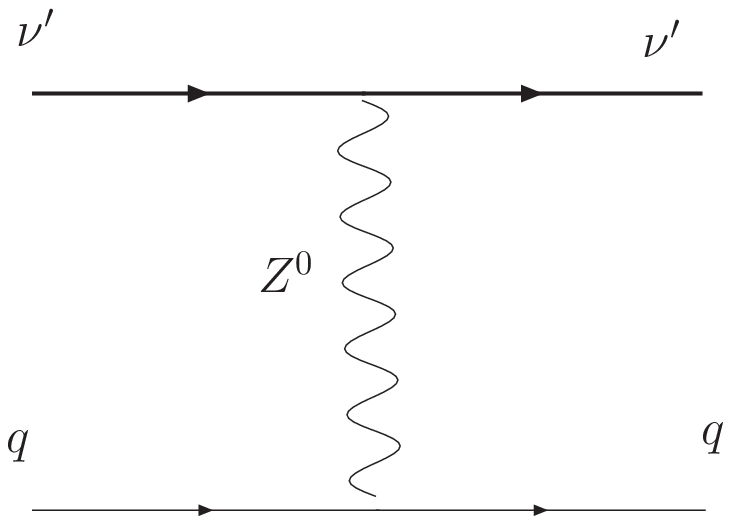}
\includegraphics[height=3cm,width=5.5cm]{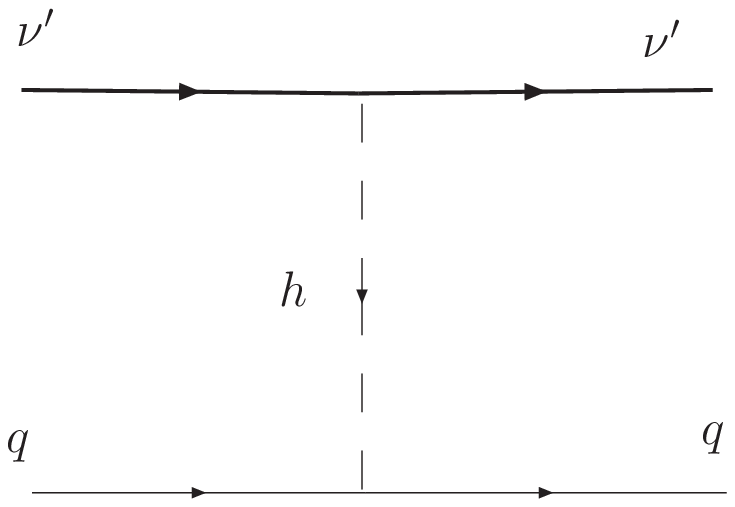}
\includegraphics[height=3cm,width=5.5cm]{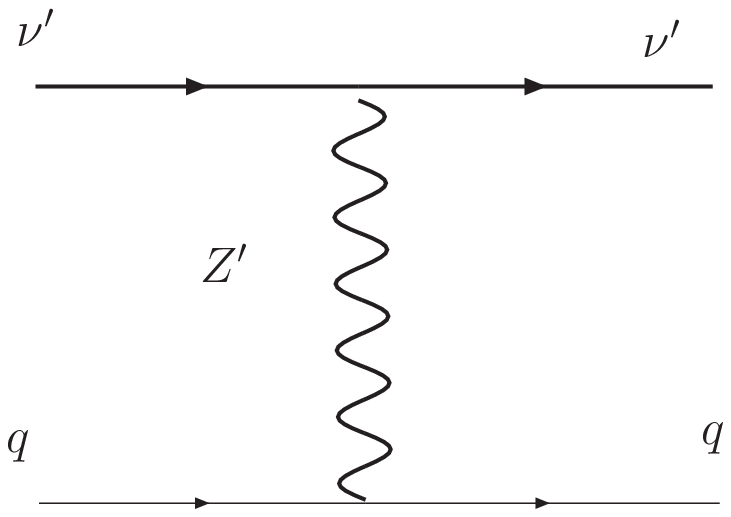}
\end{center}
\caption[]{Three diagrams potentially contributing to the elastic scattering cross section between the LZP and a quark. Effectively, only the $Z^0$ exchange contributes significantly.}
\label{fig:elastic}
\end{figure}

There are actually three potential 
diagrams contributing to elastic scattering:  t-channel $Z$, $Z^{\prime}$ or Higgs exchanges as illustrated in Fig.~\ref{fig:elastic}. 
The $Z^{ \prime }$ exchange is smaller than the $Z$ exchange since the 
coupling of light quarks to $Z^{ \prime }$ KK modes is
small and the mass of $Z^{\prime}$ is at least 3 TeV. Finally, the Higgs-exchange is suppressed by the small `Yukawa' coupling of nucleons. We checked numerically that 
the last two contributions are indeed negligible. In the following, we will focus on the Z exchange.
Note that it leads to a spin-independent (SI) interaction in contrast with 
supersymmetric dark matter where  the Majorana nature of the neutralino makes the 
$Z$-exchange contribute only in 
the much less constrained  spin-dependent interactions.
Given the ``weak'' but ``not so weak'' coupling of the LZP to the $Z$, we 
obtain
an elastic scattering
cross section for the LZP which is larger than that of the typical LSP in 
supersymmetry or of the LKP in models with universal extra dimensions \cite{Servant:2002hb}.
We detail below the calculation.

The 2-body cross section ($1+2\rightarrow 3+4$)  may be written as (${\bf p}_{1cm}$ is the momentum of particle 1 in the center of mass frame)
\begin{equation}
\frac{d\sigma}{dq^2}=\frac{1}{64 \pi s}\frac{1}{|{\bf p}_{1 cm}|^2}|\langle{\cal M}\rangle|^2 \ \ \mbox{with} \ \ { p}_{1  cm}=\frac{{ p}_{1 lab} \ m_2}{\sqrt{s}} 
\ee
$|\langle{\cal M}\rangle|^2$ is the matrix element squared in a nuclear state, summed over  final states and averaged over initial states.
Assuming particle 1 is the LZP and particle 2 is the nucleus, we get
\be
 |{\bf p}_{1 cm}|^2 s = { p}_{1 lab}^2 \ m_2^2= m_{\mbox{\tiny{LZP}}}^2 v^2 m_N^2 \ \ \mbox{and} \ \
 \frac{d\sigma}{dq^2}=\frac{|\langle{\cal M}\rangle|^2}{64 \pi v^2 m_{\mbox{\tiny{LZP}}}^2m_N^2}
\end{equation}
The elastic scattering cross section at zero momentum transfer, $\sigma_0$, is defined as 
\begin{equation}
\frac{d\sigma}{d q^2}\equiv \frac{\sigma_0}{4 \mu^2 v^2} F^2(q^2) \ \ \ \ \ \ \ \mbox{where}
 \ \ F^2(q^2=0)=1 \ \ \ \ \ \ \mbox{and} \ \ \mu=\frac{m_{\mbox{\tiny{LZP}}} m_N}{m_{\mbox{\tiny{LZP}}}+m_N} 
\end{equation}
$F^2(q^2)$ is the nuclear form factor and $\mu$ the reduced mass.
Thus,
\be
\sigma_0\equiv \frac{|{\cal \langle M \rangle }|^2_{q^2=0}}{16 \pi (m_{\mbox{\tiny{LZP}}}+m_N)^2}
\ee
In the calculation of $|{\cal \langle M \rangle }|^2$, we only keep the dominant contribution due to t-channel $Z$ exchange. In the non-relativistic limit, $q^2\ll m_Z^2$, the corresponding effective lagrangian at the quark level reads:
 \begin{equation}
\frac{ g_Z^{ \nu^{ \prime }_R } }{4 m_Z^2} \   
\left[\overline{u}_{\mbox{\tiny{LZP}}}\gamma^{\mu} u_{\mbox{\tiny{LZP}}} + \overline{u}_{\mbox{\tiny{LZP}}}\gamma^{\mu} \gamma^5u_{\mbox{\tiny{LZP}}}      \right] \times \left[ (g^q_L+g^q_R) \overline{u}_q\gamma_{\mu} u_q+ (g^q_R-g^q_L) \overline{u}_q\gamma_{\mu} \gamma^5 u_q\right]
 \end{equation}
 where
 \begin{eqnarray}
 g_L^u=\frac{(\frac{1}{2}-\frac{2}{3}\sin^2\theta_W)e}{\sin\theta_W\cos\theta_W} \ \ \ \ \   g_R^u=\frac{(-\frac{2}{3}\sin^2\theta_W)e}{\sin\theta_W\cos\theta_W} \ \ \ \
  g_L^d=\frac{(-\frac{1}{2}+\frac{1}{3}\sin^2\theta_W)e}{\sin\theta_W\cos\theta_W} \ \ \ \ \   g_R^d=\frac{(\frac{1}{3}\sin^2\theta_W)e}{\sin\theta_W\cos\theta_W}
\nonumber
 \end{eqnarray}
 The next step is to evaluate matrix elements in a nucleon state. Given the fact that $\langle \overline{u}_q \gamma^i u_q \rangle \approx 0$ and $\langle \overline{u}_q \gamma^0 \gamma^5 u_q \rangle \approx 0$, the effective four-fermion interaction reduces to :
  \begin{equation}
\frac{ g_Z^{ \nu^{ \prime }_R } }{4 m_Z^2} \ \left( \left[\overline{u}_{\mbox{\tiny{LZP}}}\gamma^{0} u_{\mbox{\tiny{LZP}}}     \right]  \left[ (g^q_L+g^q_R) \overline{u}_q\gamma_{0} u_q\right] +
 \left[\overline{u}_{\mbox{\tiny{LZP}}}\gamma^i \gamma^5 u_{\mbox{\tiny{LZP}}}     \right]  \left[ (g^q_R-g^q_L) \overline{u}_q\gamma_i \gamma^5 u_q\right] 
 \right)
 \end{equation}
 The first term will contain the operator $\langle \overline{u}_q\gamma_{0} u_q \rangle= \langle u^{\dagger}_q u_q \rangle$ which simply counts valence quarks in the nucleon. This part of the vector interaction is coherent. We then sum over nucleons in the nucleus. The second term leads to a spin-dependent (SD) interaction.
We obtain:
 \begin{equation}
|\langle{\cal M}\rangle|^2_{SI}=\frac{8 \ m_{\mbox{\tiny{LZP}}}^2 \times 4 \ m_N^2 S_{\mbox{\tiny SI}}(q) \ 4 \pi \times b_N^2 }{2 \  (2J+1)} 
 \end{equation}
  \begin{equation}
|\langle{\cal M}\rangle|^2_{SD}=\frac{4 \ m_{\mbox{\tiny{LZP}}}^2 \times 4 \ m_N^2 S_{\mbox{\tiny SD}}(q) \ 8 \pi  }{2 \  (2J+1)} 
 \end{equation}
 \begin{equation}
 b_N=Z b_p+(A-Z)b_n \ \  , \ \
b_p=\frac{g_Z^{ \nu^{ \prime }_R }\ e\ (1-4\sin\theta_W)}
{8 M_Z^2\sin \theta_W \cos \theta_W} \ \ \ \mbox{and}  \ \ \ b_n=-\frac{ g_Z^{\nu^{ \prime }_R }\ e}{8 M_Z^2\sin \theta_W \cos \theta_W}
\end{equation}
 $S_{\mbox{\tiny SI}}(q)$  and $S_{\mbox{\tiny SD}}(q)$ are the nuclear form factors defined as $S_{\mbox{\tiny SI}}(q)=(2J+1) F^2(q)/(4\pi)$  and 
$ S_{\mbox{\tiny SD}}(q)=(2J+1)\Lambda^2J(J+1) F^2(q)/\pi$ \cite{Engel:1992bf}. 
The coefficient $\Lambda$ depends on the spin $J$ of the nucleus, 
$\langle S_p\rangle$ and $\langle S_n\rangle$ the parts of the nucleus' 
spin carried by protons and neutrons.
\be
\Lambda=\frac{a_p\langle S_p\rangle +a_n\langle S_n\rangle}{J} \ \ \ , \ \ \ 
a_{p,n}=\sum_{u,d,s} b'_q\Delta^{(p,n)} q \ \ \ , \ \ \ b'_q=\frac{g_Z^{\nu'_R} (g_R^q-g_L^q)}{4 M_Z^2}
\label{LAMBDA}
\ee
As a result:
 \begin{equation}
 \sigma_0^{SI}=\frac{\mu^2 b_N^2}{\pi}=\frac{ \left( g_Z^{ \nu^{ \prime }_R } \right)^2  
\mu^2 e^2  }{64 \pi m_Z^4 \sin^2\theta_w \cos^2 \theta_w} \left[Z(1-4\sin^2\theta_w)-(A-Z)\right]^2 
 \label{sigmaSI0}
 \end{equation}
  \begin{equation}
 \sigma_0^{SD}=\frac{ 4 \mu^2 \Lambda^2 J(J+1)  }{ \pi } 
 \label{sigmaSD0}
 \end{equation}
 From now, given the fact that experiments are carried on heavy nuclei such as germanium for instance, we will ignore the SD contribution which is smaller than the SI one by a factor $1/(A-Z)^2$.

 \subsection{WIMP-nucleon cross section}

\begin{figure}[h]
\begin{center}
\includegraphics[width=7.2cm,height=8cm]{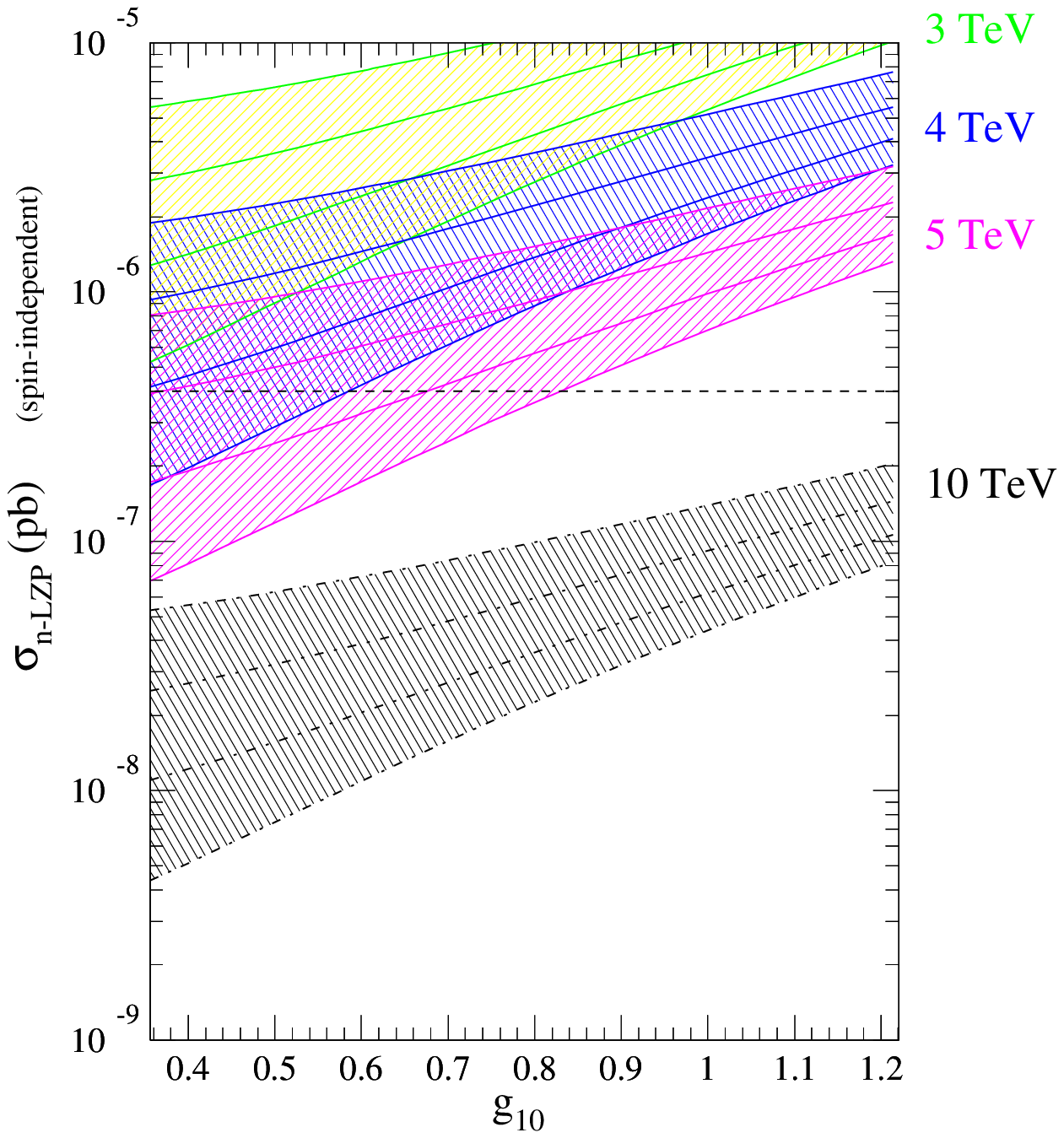}
\hspace{1.2cm}
\includegraphics[width=7.2cm,height=8cm]{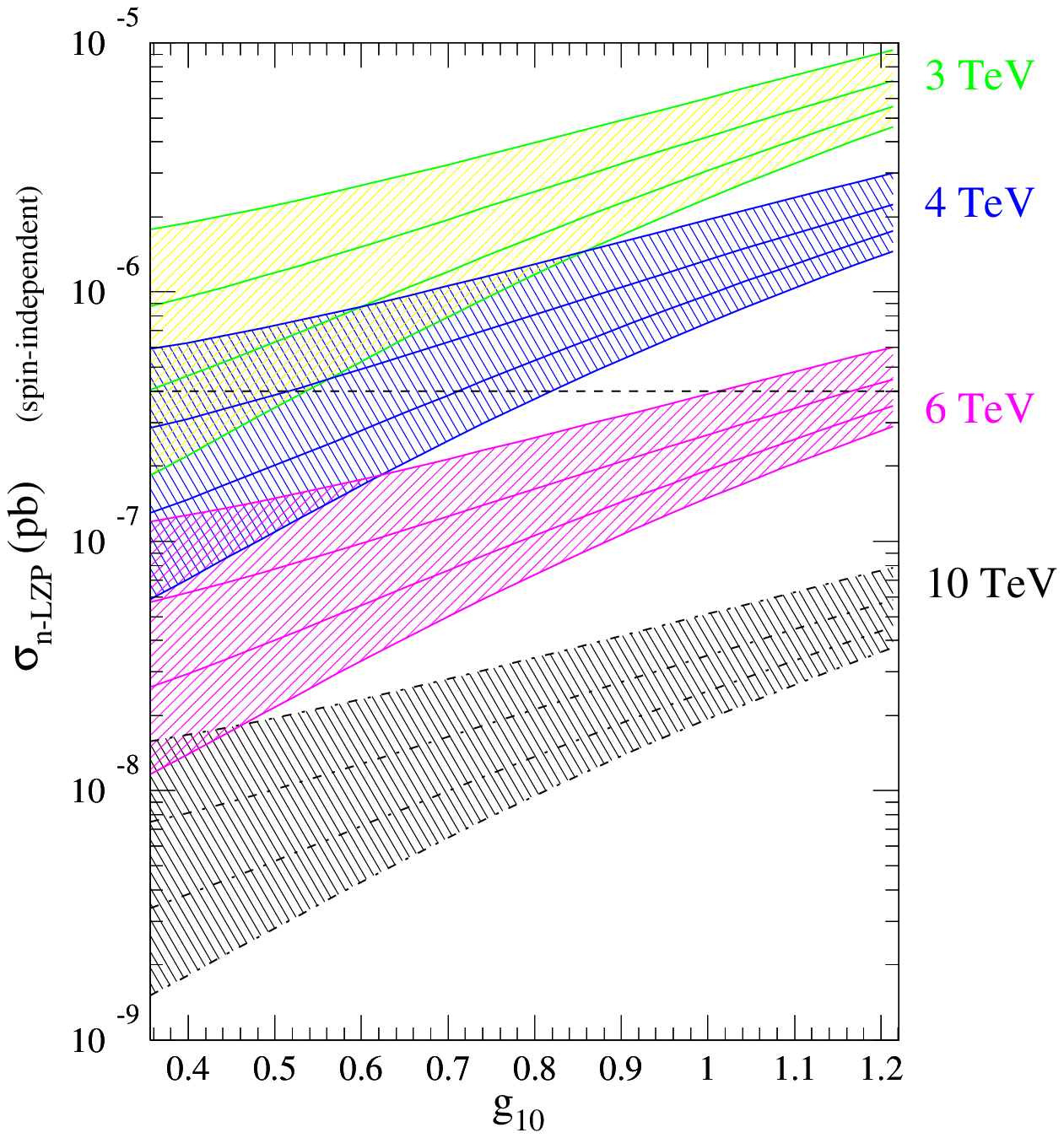}
\caption[]{ Predictions for $\sigma_{n,LZP}$. The left plot has been derived assuming that 
the Higgs is localized exactly on the TeV brane and  
for $m_{KK}=$3, 4, 5, 10 TeV. The right plot corresponds 
to the case where the Higgs is the fifth component of a gauge boson
(i.e., a PGB) with a profile in 
the bulk given by Eq. (\ref{HiggsasA5}). Those two cases lead to significantly different LZP--Z couplings. Indeed, both the Yukawa coupling of the Higgs to $\nu'_L$ and $\nu'_R$ and the $Z-Z^{\prime}$ mixings are modified by the profile (see sections \ref{ZZprimemixing} and \ref{nulrmixing}). We see that the precise value of the LZP--Z coupling (which will vary from one model of EW symmetry breaking to another) is crucial for event rates at direct detection experiments. At least, models with the Higgs localized on the TeV brane are quite constrained here. The horizontal line indicates the experimental limit \cite{Akerib:2004fq} which only applies to a range of WIMP masses.}
\label{fig:directprediction}
\end{center}
\end{figure}

The SI cross-section data is commonly normalised to a single nucleon
to compare results from different experiments (which use different target nuclei).
For sufficiently low momentum transfer, and assuming that the WIMP has the same interaction for protons and neutrons, the $A$ scattering amplitudes add up to give a coherent cross 
section $\propto A^2$. Therefore, experimentalists express their bound in terms of the WIMP-nucleon cross sections, using this $A^2$ scaling.  We follow this convention in our 
Fig. \ref{fig:directprediction}.
However, the WIMP has typically diffferent interactions between protons and neutrons
and the experimental bound has to be interpreted with a bit of care.
This is the case of Dirac neutrinos (like our LZP), where
the interactions to protons is suppressed by $(1-4\sin^2 \theta_W)^2$ and 
the interaction is essentially due to neutrons.  So, in order to obtain the WIMP-nucleon cross section, one should rather use the scaling $(A-Z)^2$. To interpret data,
one should also keep in mind that experimental limits 
use the prevailing convention of assuming that the local halo density of dark matter is 0.3 GeV/cm$^3$, and that the characteristic halo velocity $v_0$, is 220 km s$^{-1}$ and the mean earth velocity $v_E$ is 232 km s$^{-1}$. Some uncertainties are associated with these numbers.

The nucleon-WIMP cross section is defined as
\be
\sigma_{p,n}=\sigma_0\frac{\mu_{p,n}^2}{\mu^2_A}\frac{C_{p,n}}{C_A} \ \ , \ \ 
\mu_A=\frac{m_{\chi}m_A}{m_{\chi}+m_A} \ \ , \ \  \sigma_0 \propto \mu^2_A C_A
\ee
\be
\mu_{p,n}=\frac{m_{\chi}m_{p,n}}{m_{\chi}+m_{p,n}}\approx m_{p,n} 
\ee
For a scalar interaction involving a $Z$ exchange,
$C_A=(Z(1-4\sin^2\theta_W)-(A-Z))^2$. Then, using $\sin^2\theta_W=0.23120$ 
\be
C_p= (1-4\sin^2\theta_W)=0.0752 \ll 1  \ \mbox{and} \  C_n=1
\ee
which leads to
\be
 \sigma_n=\sigma_0\frac{m_p^2}{\mu^2}\frac{(1-4\sin^2 \theta_W)^2}{(Z(1-4\sin^2\theta_W)-(A-Z))^2}
\ee

\be
\sigma_p=\sigma_0\frac{m_p^2}{\mu^2}\frac{(1-4\sin^2 \theta_W)^2}{(Z(1-4\sin^2\theta_W)-(A-Z))^2}\approx 5.65\times 10^{-3} \sigma_n 
\ee

As can be seen in Fig.~\ref{fig:directprediction}, the LZP-nucleon cross section is large ($\gtrsim 10^{-10}$ pb)
so that our models will be tested in the near future.

\section{Indirect detection}
\label{sec:indirect}
Indirect dark matter searches consist in looking for products of dark matter annihilation including gamma-rays, positrons, anti-protons and neutrinos. Such signal would come from regions where the dark matter density is large, like in the center of the galaxy. One difficulty is that the expected flux depends very sensitively on the dark matter profile at the galactic center, something which is still poorly known. We also expect that WIMPs annihilate in the Sun and the Earth, in which case, uncertainties in their distribution are much more under control. We focus on this signal in the following.

\begin{figure}[h]
\begin{center}
\includegraphics[width=7.5cm,height=4.5cm]{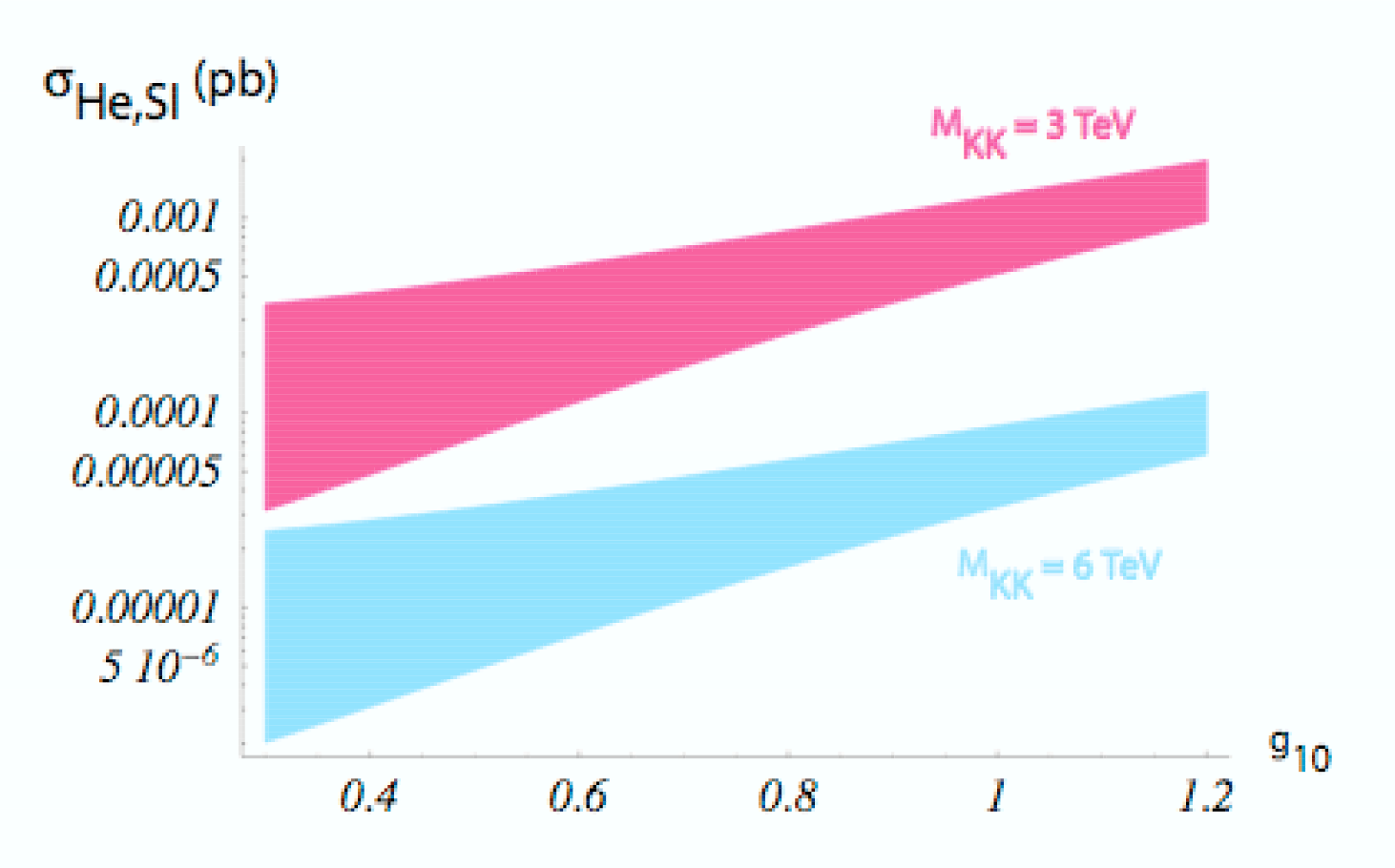}
\includegraphics[width=7.5cm,height=4.5cm]{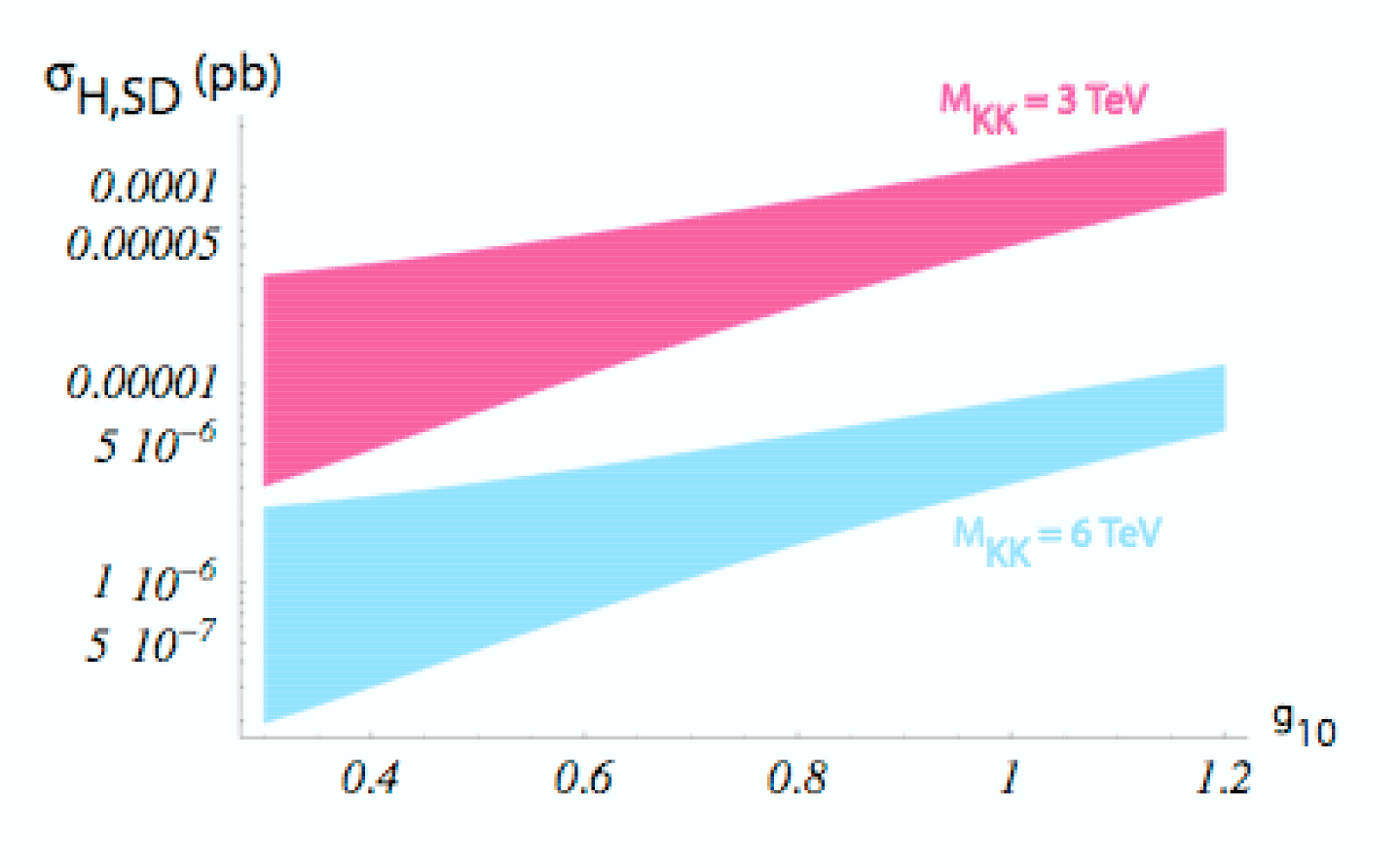}
\caption[]{ Values of $\sigma_{He,SI}$ $\sigma_{H,SD}$ for $m_{KK}=3$ TeV (red) and $m_{KK}=6$ TeV (blue). Each region is obtained by varying $c_{\nu'_L}$ in the range $[c_{t_R}-0.5,c_{t_R}+0.5]$. }
\label{fig:indirectsigmas}
\end{center}
\end{figure}

\begin{figure}[h]
\begin{center}
\includegraphics[width=9cm,height=4.5cm]{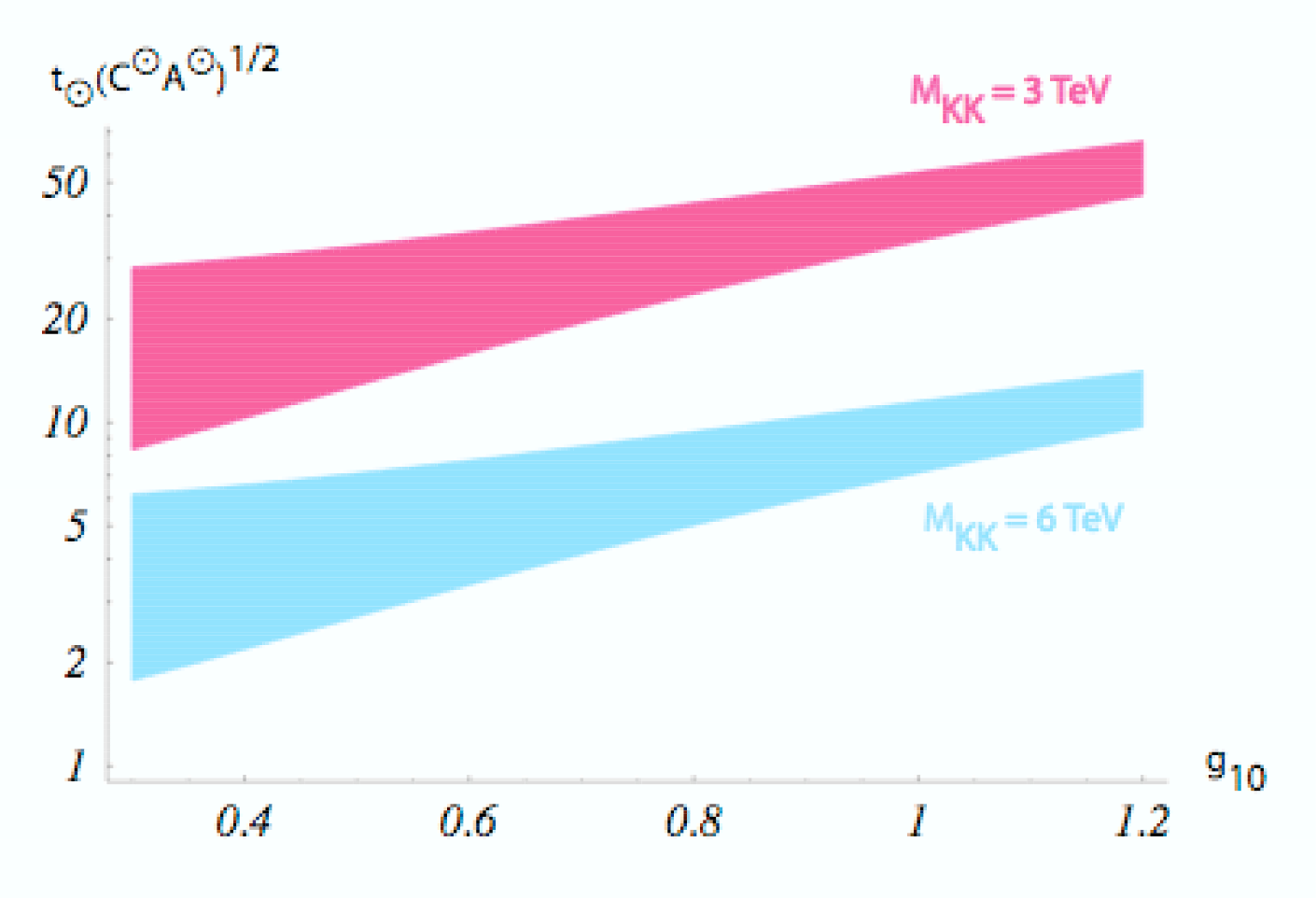}
\caption[]{ $\sqrt{C^{\odot} A^{\odot}} t_{\odot}$ evaluated at $\langle \sigma_{anni}   \ v \rangle =1 $ pb. We see that it is always larger than 1, meaning that equilibrium between capture and annihilation is reached in the Sun. }
\label{fig:equil}
\end{center}
\end{figure}

When equilibrium is reached between the rate $C^{\odot}$ at which WIMPs are captured in the Sun (determined by the WIMP-nucleus elastic scattering cross section) and the annihilation rate $A^{\odot}$, the annihilation rate in the Sun is maximized and given by \cite{Hooper:2003ui}
\begin{equation} 
\Gamma =  \frac{1}{2} \, C^{\odot} \, 
\tanh^2 \left( \sqrt{C^{\odot} A^{\odot}} \, t_{\odot} \right) \; 
\end{equation}
where $t_{\odot} \simeq 4.5$ billion years.
The equilibrium condition is
$\sqrt{C^{\odot} A^{\odot}} t_{\odot} \gg 1$. $A^{\odot}$ is given by
\be
A^{\odot} = \frac{\langle \sigma v \rangle }{V_{eff}} \ \ \ \ \mbox{where}  \ \ V_{eff}= 5.7 \times 10^{27} \mbox{cm}^3 \left(\frac{100 \mbox{GeV}}{m_{WIMP}}\right)^{3/2}
 \ee
Since we have a large elastic scattering cross section (much larger than the 
LSP in SUSY and LKP in UED) we anticipate  interesting signals. However, remember that the
Sun is mainly made of protons and scattering of the LZP with protons is suppressed. Fortunately, interactions with helium should provide observable signal.
The capture rates for spin-dependent and spin-independent interactions are given by \cite{Hooper:2003ui}
\begin{equation} 
C_{\mathrm{SD}}^{\odot} \simeq 3.35 \times 10^{20} \, \mathrm{s}^{-1} 
\left( \frac{\rho_{\mathrm{local}}}{0.3\, \mathrm{GeV}/\mathrm{cm}^3} \right) 
\left( \frac{270\, \mathrm{km/s}}{\bar{v}_{\mathrm{local}}} \right)^3  
\left( \frac{\sigma_{\mathrm{H, SD}}} {10^{-6}\, \mathrm{pb}} \right)
\left( \frac{100 \, \mathrm{GeV}}{m_{\rm{LZP}}} \right)^2 
\label{c-eq}
\end{equation} 
\begin{equation}
C_{\mathrm{SI}}^{\odot} \simeq 1.24 \times 10^{20} \, \mathrm{s}^{-1} 
\left( \frac{\rho_{\mathrm{local}}}{0.3\, \mathrm{GeV}/\mathrm{cm}^3} \right) 
\left( \frac{270\, \mathrm{km/s}}{\bar{v}_{\mathrm{local}}} \right)^3 
\left( \frac{2.6 \, \sigma_{\mathrm{H, SI}}
+ 0.175 \, \sigma_{\mathrm{He, SI}}}{10^{-6} \, \mathrm{pb}} \right) 
\left( \frac{100 \, \mathrm{GeV}}{m_{\rm{LZP}}} \right)^2 
\end{equation}
$\rho_{\mathrm{local}}$ is the local DM density. $\bar{v}_{\mathrm{local}}$ 
is the local rms velocity of halo DM particles.  
$\sigma_{\mathrm{H,SD}}$ and $\sigma_{\mathrm{H, SI}}$ are the spin-dependent 
and spin-independent, LZP-on-proton (hydrogen)
elastic scattering cross sections,  $\sigma_{\mathrm{He, SI}}$ is the 
spin-independent, LZP-on-helium elastic scattering cross section. 
 The SI cross sections are trivially obtained from Eq.~\ref{sigmaSI0}:
 \be
 \sigma_{H,SI}=\frac{ \left( g_Z^{ \nu^{\prime}_R } \right)^2 
e^2 \mu^2_H (1-4\sin^2 \theta)^2}{64 \pi M_Z^4 \sin^2 \theta \cos^2 \theta} \ \ \ , \ \ \ 
  \sigma_{He,SI}=\frac{ \left( g_Z^{ \nu^{\prime}_R } \right)^2 
e^2 \mu^2_{He} \left(2(1-4\sin^2 \theta)-2\right)^2}{64 \pi M_Z^4 \sin^2 \theta \cos^2 \theta}
 \ee
where $ \mu_{H,He}=m_{\mbox{\tiny LZP}} m_{H,He}/(m_{\mbox{\tiny LZP}}+m_{H,He})$.
It is clear that  $ \sigma_{He,SI}$ dominates by a factor $10^4$.
The SD interaction (see Eq.~\ref{sigmaSD0}) is given by
\be
 \sigma_{H,SD}= \frac{3 \ \mu^2_{H} \Lambda^2}{\pi} \ \ \ , \ \ \  \Lambda=a_p
  \ \ \ , \ \ \ 
 a_{p}=\frac{e \ g_Z^{\nu'_R} } {8 M_Z^2 \cos \theta \sin \theta} \left[ -\Delta u + \Delta d+\Delta s\right]
\ee
where
$\Delta u=0.78\pm 0.02$, $\Delta d=-0.48\pm 0.02$ and $\Delta s=-0.15\pm 0.02$ \cite{Ellis:2000ds}
 are the spins carried by the quarks u, d, s respectively in the proton.
In fig.~\ref{fig:indirectsigmas}, we have plotted $\sigma_{He,SI}$, $\sigma_{H,SD}$. In fig.~\ref{fig:equil}, we plotted $\sqrt{C^{\odot} A^{\odot}} t_{\odot}$ evaluated at $\langle \sigma_{anni}   \ v \rangle \approx 1$ pb, the value of the annihilation cross section leading to the correct relic density. This
shows that throughout the parameter region leading to the ideal relic density, the Sun always reaches equilibrium between the LZP capture and annihilation. The event rate and prospects for indirect detection will be provided elsewhere \cite{withDan}.

\section{Collider phenomenology}

A very exciting aspect of these models is the potential for discovery of KK modes at colliders. This is to be contrasted with previous studies carried out in Randall-Sundrum background, where the emphasis has been on $(++)$ type of boundary conditions, in
 which case it might be difficult to produce KK modes at the LHC 
since the KK masses have to be larger than 3 TeV.
In our $SO(10)$  model, as we already emphasized in section \ref{sec:who}, 
all KK modes  in the multiplet with $t_R^{(0)}$ (except 
KK modes for $t_R$ itself) are expected to be light because 
they have ($-+$) BC as well as
 $c$ close to $-1/2$. Since the splittings in $c$ are unknown, 
we will take these masses to be free parameters. Another interesting aspect is 
that most of the $Z_3$-charged fermions from the
$t_R$ multiplet cannot decay very easily. The reason is that
they have to eventually
decay into the LZP (and SM particles, i.e., zero-modes) and, in this multiplet, 
only $t_R$ has zero mode. As a result, these decays have to go 
through a certain number of virtual states. 
This leads to very distinctive signatures which we will present in this section.
%
%

We repeat that KK modes of ($-+$) gauge bosons might be too heavy
to be significantly produced at the LHC ($ M_{KK} \gtrsim 3$ TeV). 
Similarly, ($-+$) KK fermions coming from other multiplets (with $c\gtrsim 0$)  and 
($++$) KK fermions are considered too heavy.  
For instance, $t_R^{(1)}$ in the multiplet we are interested in is heavy.
Actually, a detailed calculation would be required 
to determine whether the strong coupling of heavy
KK modes
can compensate for the large mass suppression in the production cross-section. In contrast,
colored light KK modes from $t_R$
multiplet (with a mass $\lesssim 1$ TeV) will be copiously
produced at the LHC.
 

Before discussing collider signatures, let us mention that we do not expect any experimental constraint coming from the additional $U(1)_B$ gauge boson.
As said earlier, we couple it to a Planckian vev 
on the UV brane. This mimics $(-+)$ BC to a good approximation.
The coupling of light fermions to KK modes of $U(1)_B$ is 
negligible compared to the $4D$ would-be' zero-mode $U(1)_B$ gauge
coupling since light fermions  are localized near the Planck brane where $U(1)_B$ KK modes effectively vanish\footnote{A numerical evaluation of overlap of wave functions
confirms that this coupling
has the same size as the Yukawa couplings.}. 
Whereas, the coupling of $t_R$ to $U(1)_B$ KK modes is enhanced by $\sqrt{ k \pi r_c }$ compared to $4D$ gauge
coupling since both $t_R$ zero-mode and $U(1)_B$ KK modes are localized near the TeV brane. 
Also, the $U(1)_B$ KK modes do not mix with zero-mode of $Z$ 
(unlike $Z^{ \prime }$ KK modes) since the
Higgs does not carry $B$.
Thus, we see that KK masses of $\sim$ few TeV 
for $U(1)_B$ gauge boson are not constrained by current data.

\begin{figure}[h]
\begin{center}
\includegraphics[width=4.5cm,height=2.5cm]{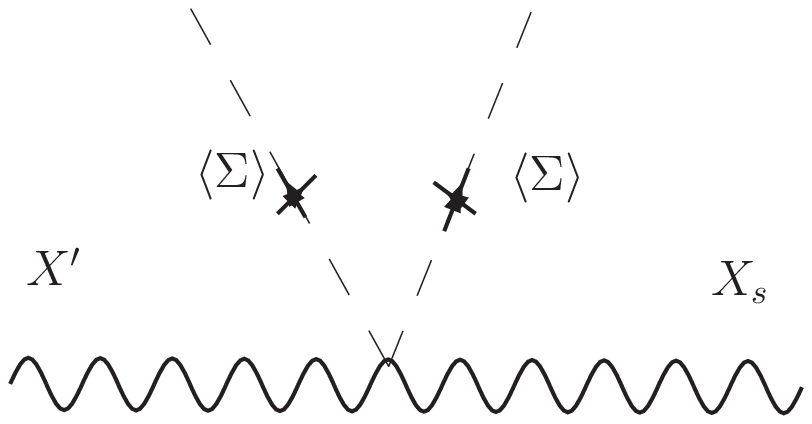}
\includegraphics[width=4.5cm,height=2.5cm]{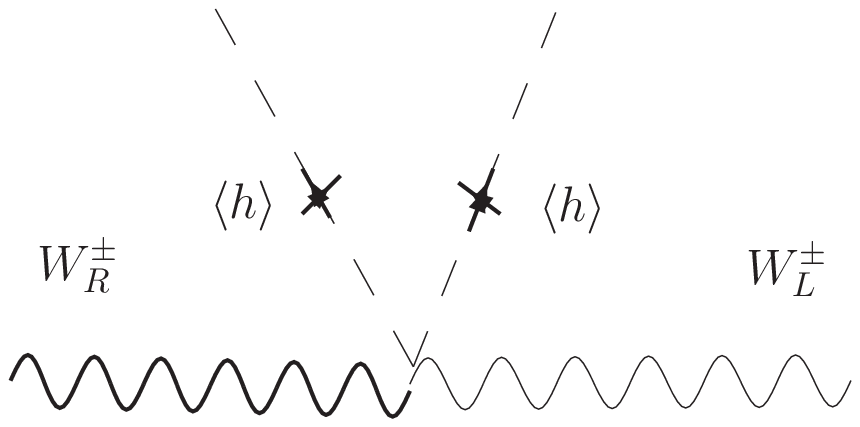}
\caption[]{Diagrams leading to $X^{\prime}-Xs$ and $W_R^{\pm}-W_L^{\pm}$ mixings.}
\label{fig:mixings}
\end{center}
\end{figure}

It is clear that in the absence of GUT bulk breaking, $SU(2)_L$-charged KK modes 
in the $t_R$ multiplet such as $b_L^{\prime}$, $t_L^{\prime}$, $\nu_{\tau,L}^{\prime}$ and $\tau_L^{\prime}$ decay  very slowly (see section 7.3). 
Thus, they will cross the detector and those which 
carry an electric charge will easily be detected due to their CHAMP 
(stable charged massive particles)-like signatures.
Colored particles hadronize and what is detected is some charged KK meson made of a light quark and a KK $b_L^{\prime}$ or $t_L^{\prime}$.
In the presence of GUT bulk breaking of the unified gauge symmetry, 
$SU(2)_L$-charged KK modes can decay due to $X^{\prime}-X_S$ mixing (see Fig.~\ref{fig:mixings}). 
The size of this mixing will depend on the profile of the GUT breaking 
scalar $\Sigma$ in the bulk. If this mixing 
is too small then decay will take place outside the detector. 
On the other hand, if $\Sigma$ has a flat profile, 
the mixing is large and the decay easily takes place in the detector, although
the lifetime is quite sensitive to the mass splitting
(as shown in section 7.4). This situation leads to very interesting signatures. 
\begin{figure}[h]
\begin{center}
\includegraphics[height=4.5cm,width=6.5cm]{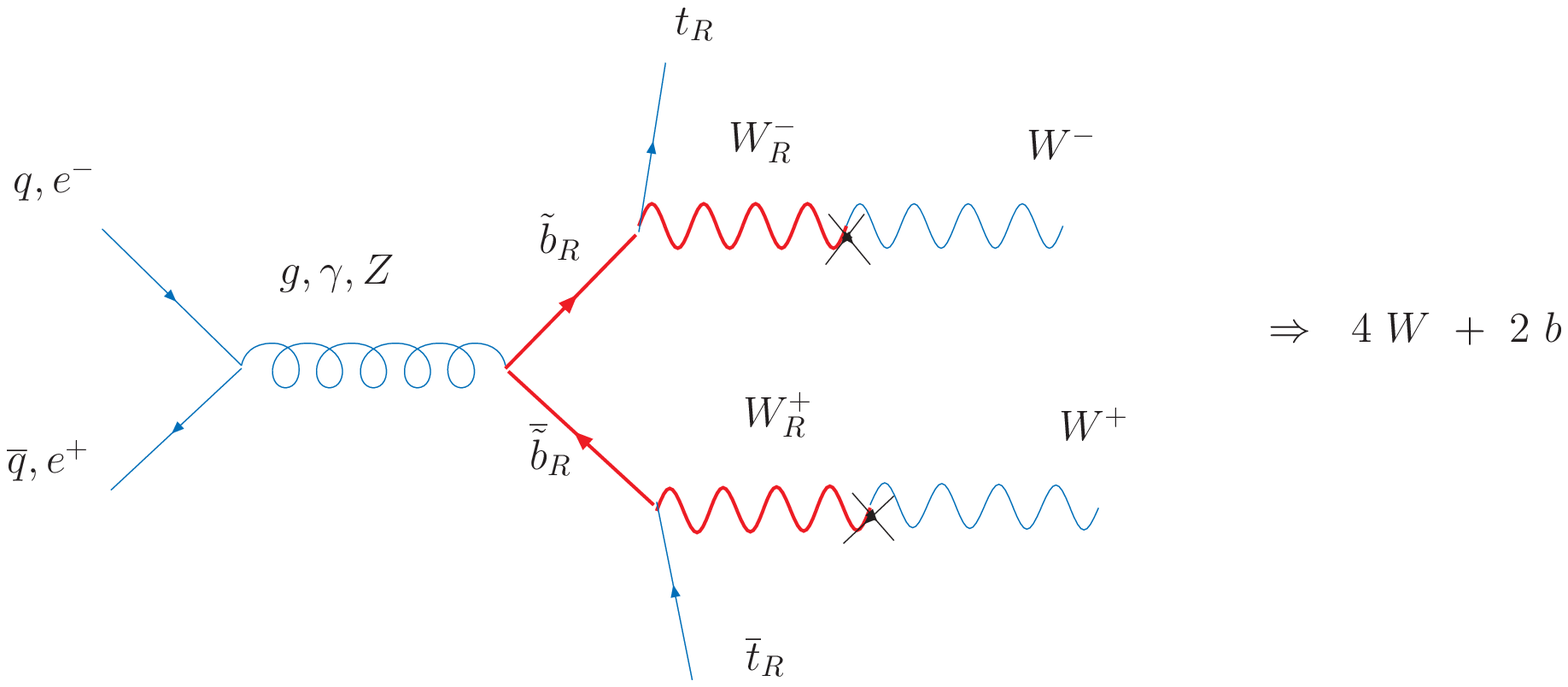}
\includegraphics[height=4.5cm,width=6.5cm]{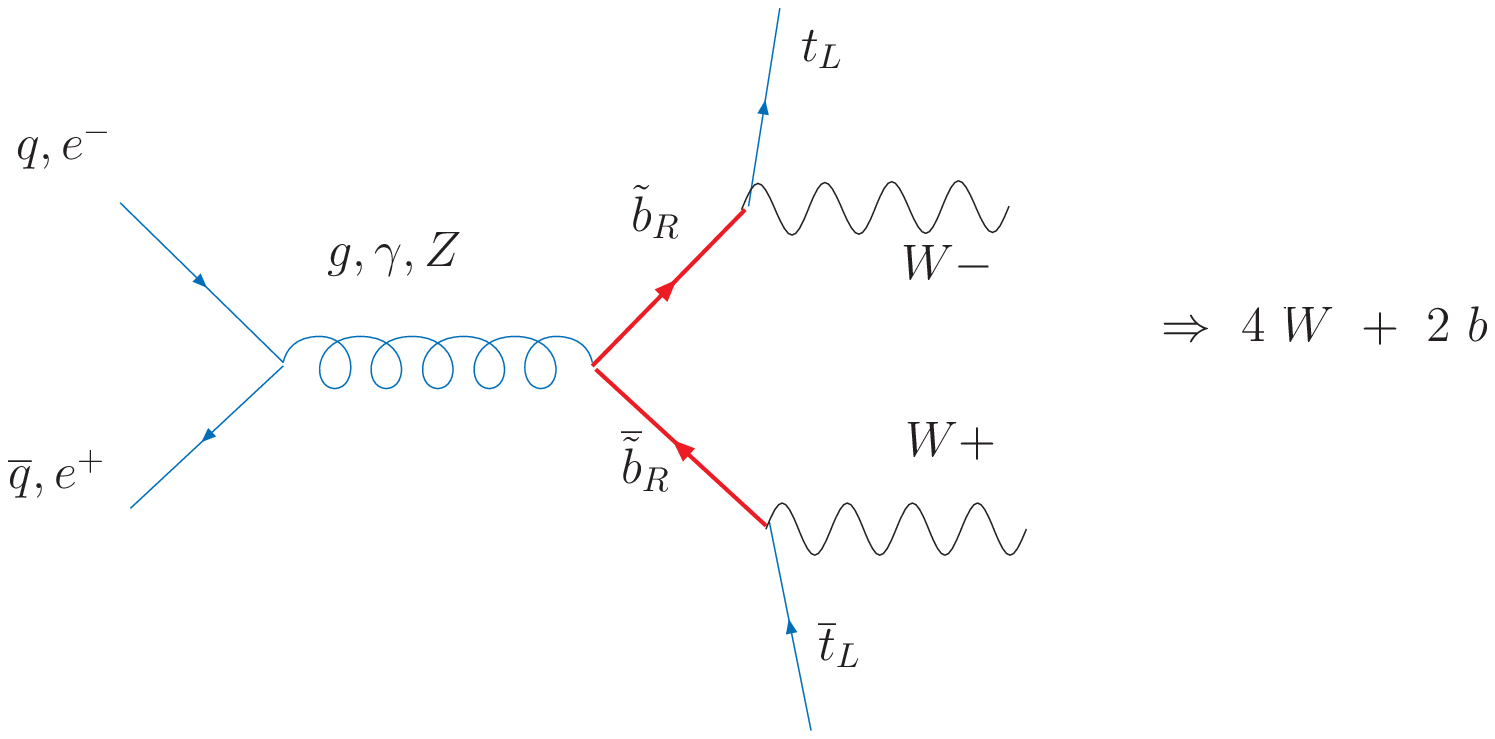}
\includegraphics[height=4.5cm,width=3.5cm]{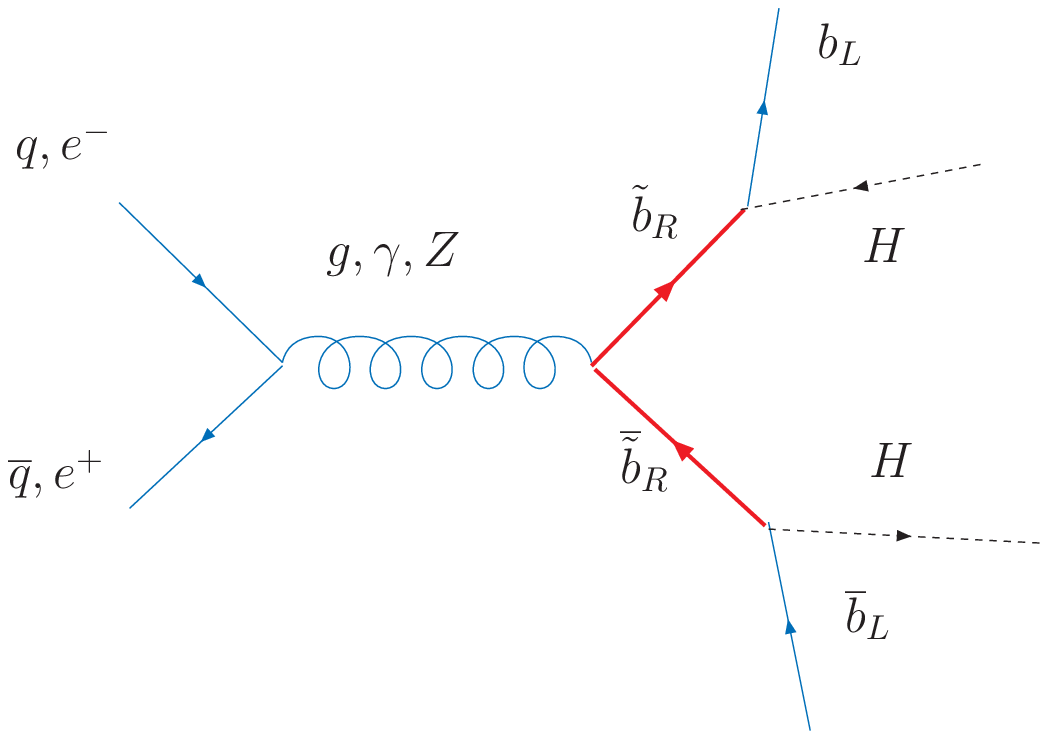}
\includegraphics[height=4.5cm]{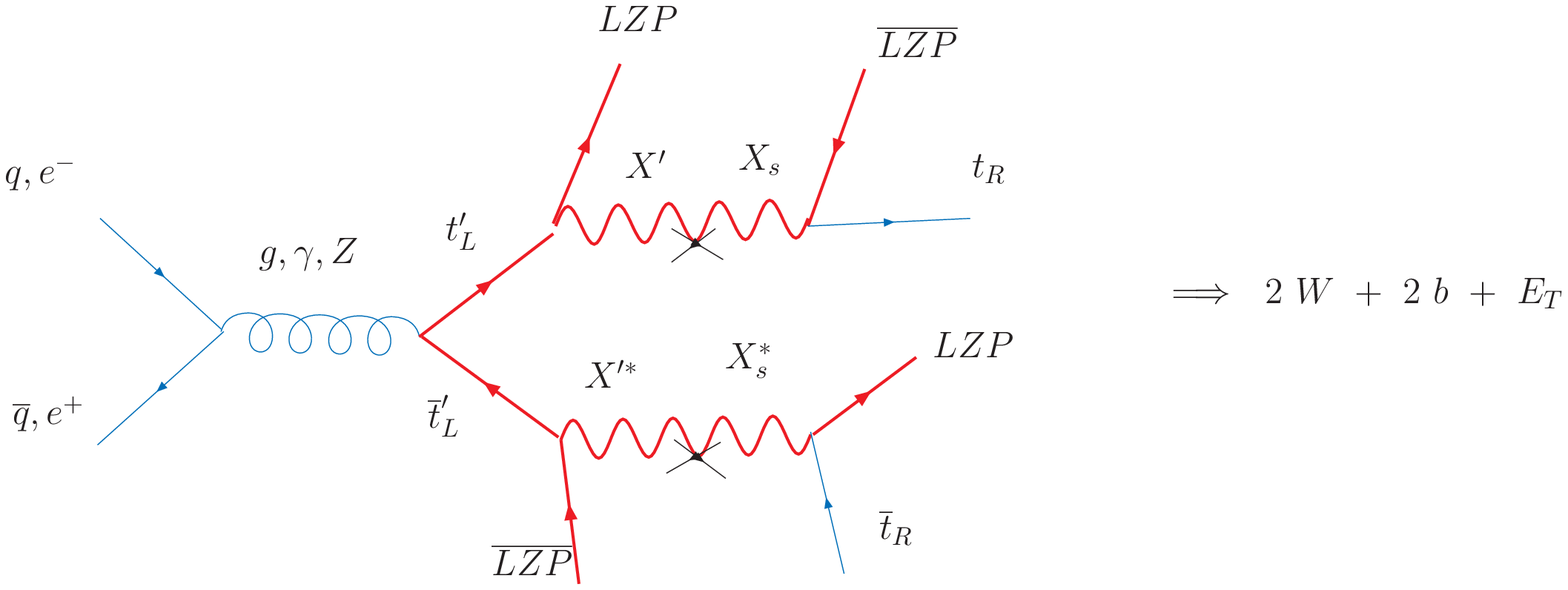}
\includegraphics[height=4.5cm]{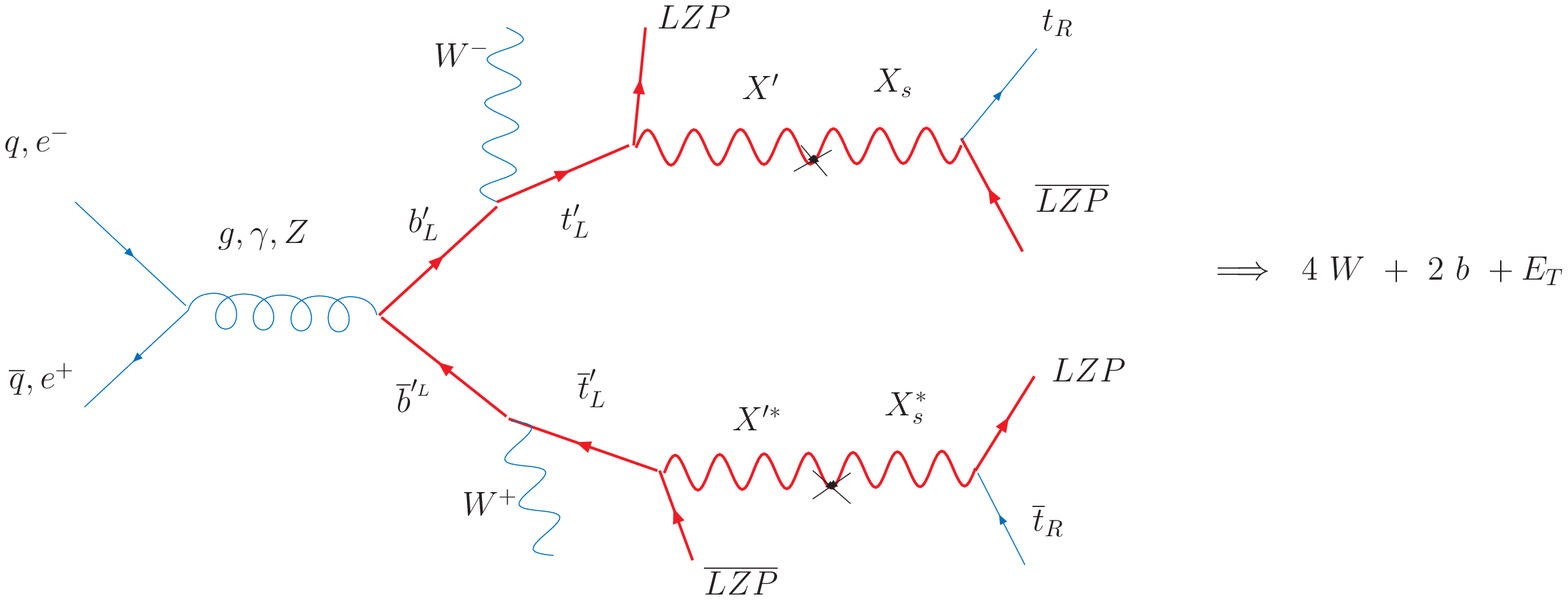}
\includegraphics[height=4.5cm]{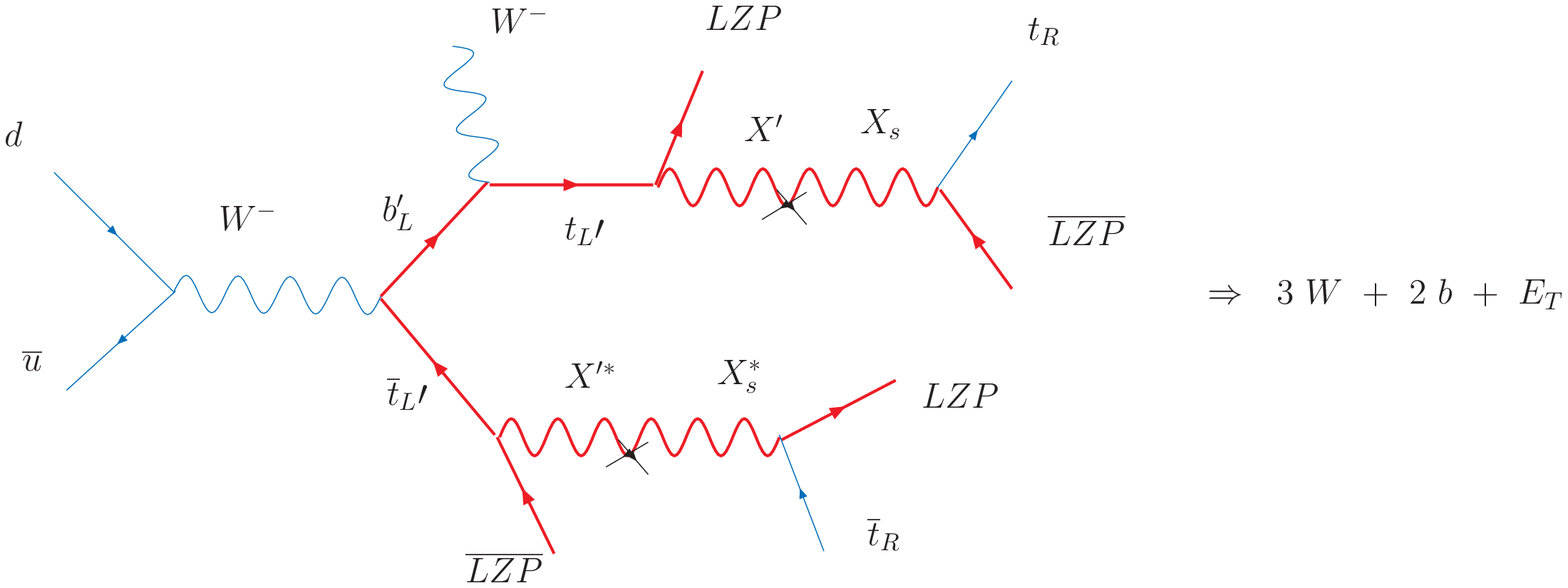}
\caption[]{Production of KK quarks}
\label{fig:colliderquarrks}
\end{center}
\end{figure}
\begin{figure}[h]
\begin{center}
\includegraphics[height=4.3cm]{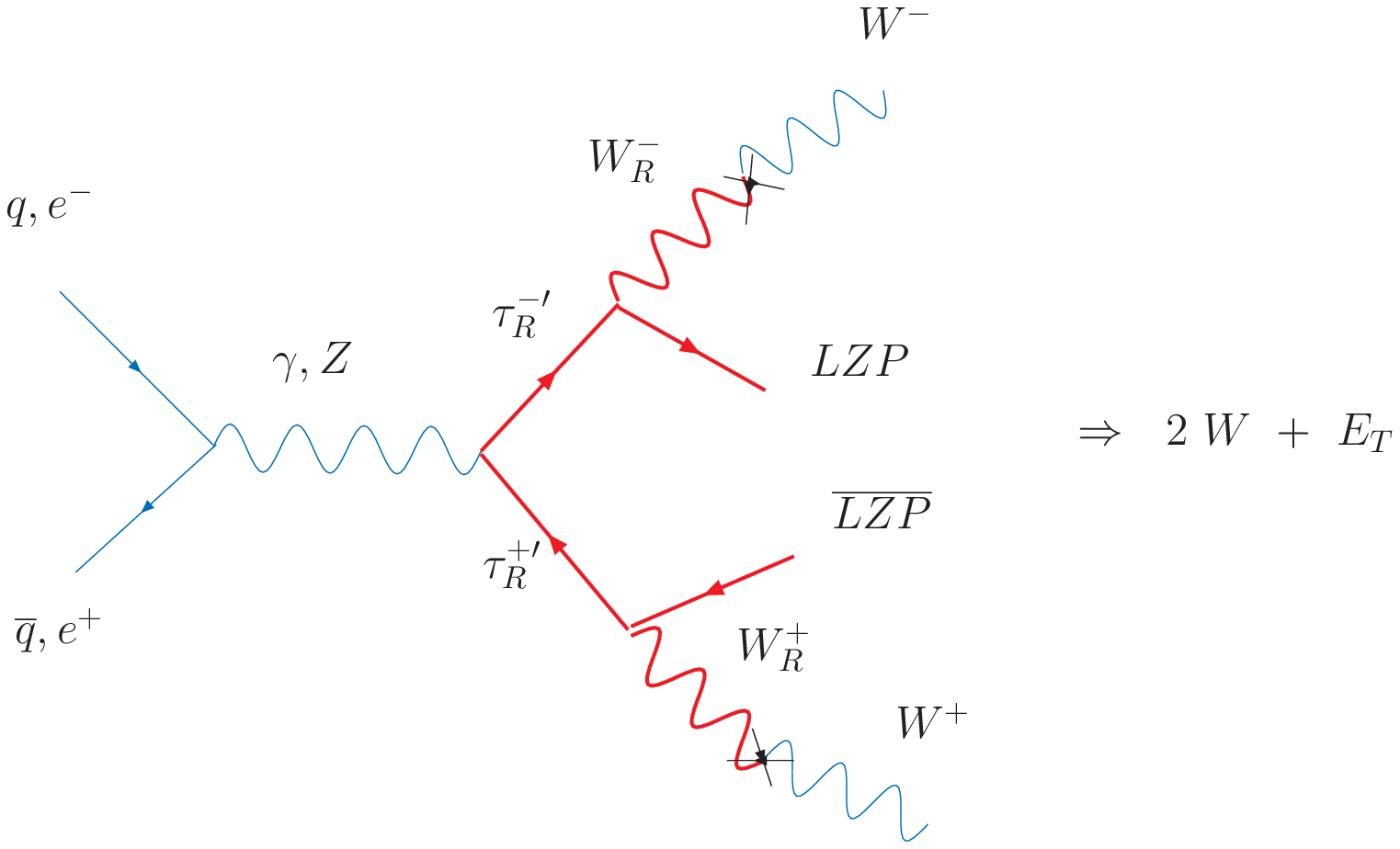}
\includegraphics[height=4.3cm]{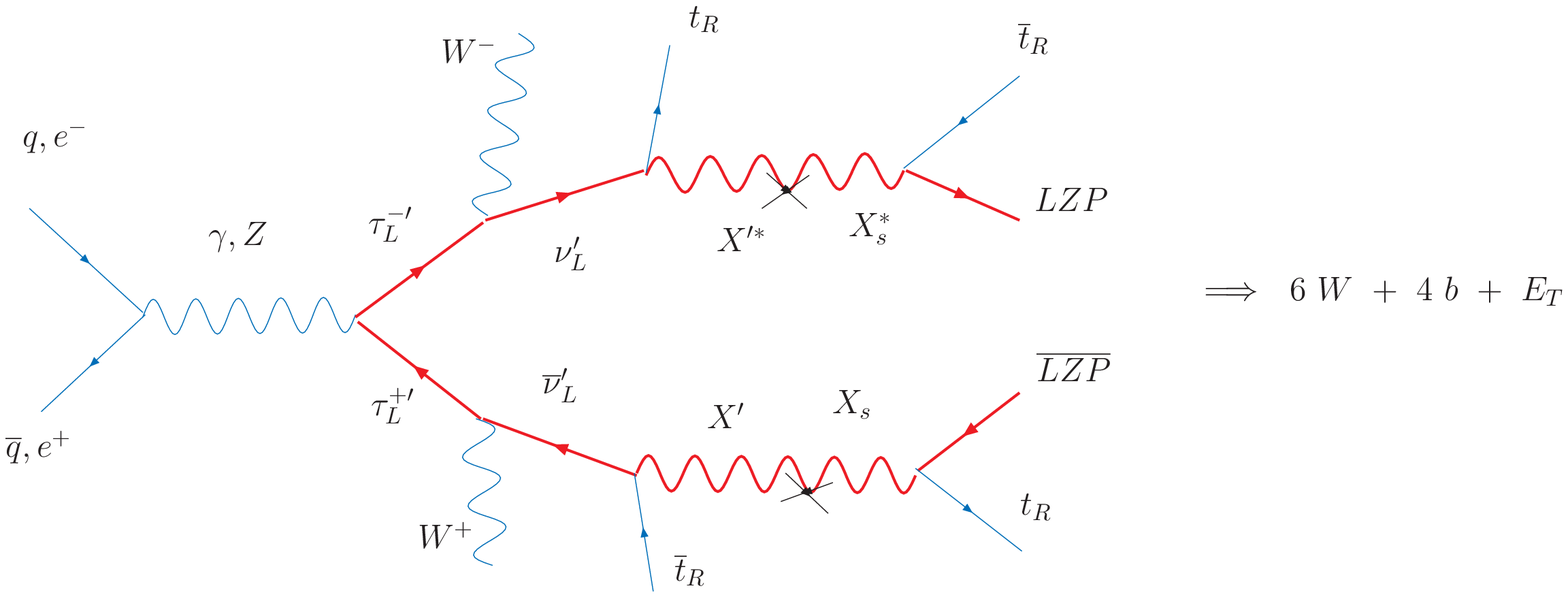}
\includegraphics[height=4.3cm]{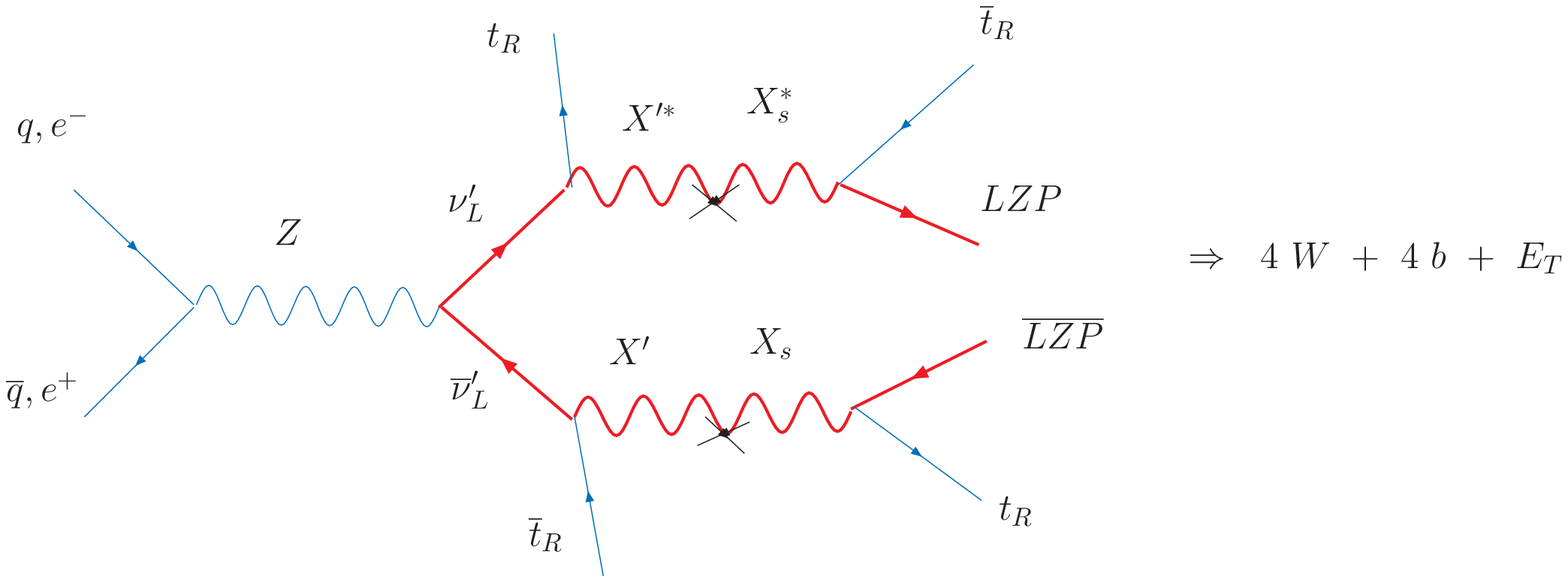}
\includegraphics[height=4.cm]{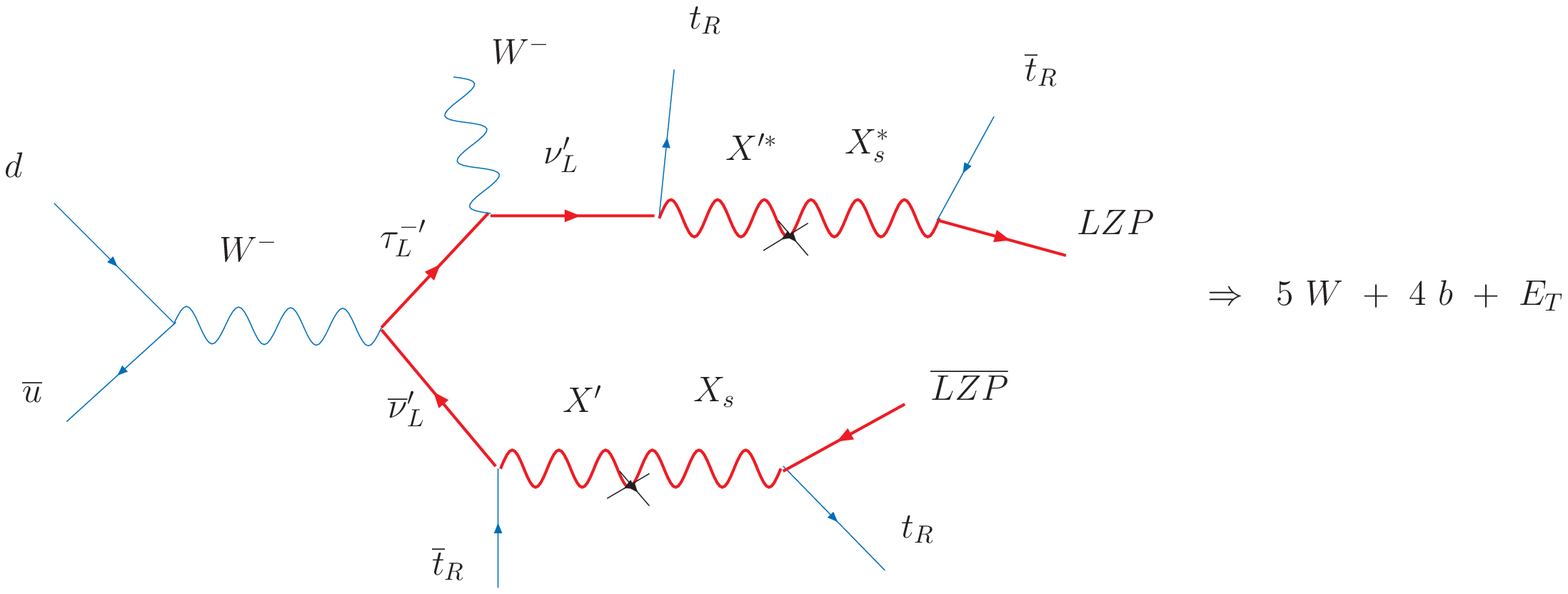}
\caption[]{Production of KK leptons. $\nu^{\prime}_L$ refers to $\nu^{\prime}_{\tau,L}$.}
\label{fig:colliderleptons}
\end{center}
\end{figure}
We now list all 
the decays which are illustrated in Fig.\ref{fig:colliderquarrks} and \ref{fig:colliderleptons}:
\begin{equation}
\tilde{b}_R \rightarrow b_L H, \; t_L W_{ long. }
\nonumber 
\end{equation}
\begin{equation}
\nonumber
\tilde{b}_R  \rightarrow  t_R \, W^-  \ \ \  \mbox{via} \  W_R \mbox{-}W_L \ \ \mbox{mixing}
\end{equation}
\begin{equation}
t^{\prime}_L  \rightarrow  t_R \, \nu_R^{\prime} \, \overline{\nu}_R^{\prime}  \ \ \  \mbox{via} \ X^{\prime}\mbox{-}X_s  \ \ \mbox{mixing}
\nonumber
\end{equation}
\begin{equation}
b^{\prime}_L  \rightarrow  t_R \, \nu_R^{\prime} \, \overline{\nu}_R^{\prime} \, W^- \ \ \  \mbox{via} \ X^{\prime}\mbox{-}X_s  \ \ \mbox{mixing}
\nonumber
\end{equation}
\begin{equation}
\nonumber
\tau^{\prime -}_R  \rightarrow  \nu^{\prime}_R \, W^-  \ \ \  \mbox{via} \  W_R \mbox{-}W_L \ \ \mbox{mixing}
\end{equation}
\begin{equation}
\nonumber
\tau^{\prime -}_L  \rightarrow  \nu^{\prime}_R \, t_R \, \overline{t}_R \, W^-  \ \ \  \mbox{via} \   X^{\prime}\mbox{-}X_s \ \ \mbox{mixing}
\end{equation}
\begin{equation}
\nonumber
\nu^{\prime }_{\tau,L}  \rightarrow  \nu^{\prime}_R \, t_R \, \overline{t}_R  \ \ \  \mbox{via} \   X^{\prime}\mbox{-}X_s \ \ \mbox{mixing}
\end{equation}
The effective couplings of $t_L^{\prime}$ to $\nu_R^{\prime}$
and   $\nu_L^{\prime}$ to $t_R$ due to $X^{\prime} - X_s$ mixing are
\begin{eqnarray}
g_{ \nu^{\prime}_R , t_L^{\prime} , X_s}&=&  \frac{g_{10}}{\sqrt{2}}\sqrt{k \pi r_c} \times {\cal F}_{ \nu^{\prime}_R , t_L^{\prime}}\times P_R \times {\cal M}_{X^{\prime} - X_s}  \\
\label{LZPtop}
g_{ \nu^{\prime}_{\tau,L} , t_R , X_s}&= & g_{ \nu^{\prime}_R , t_L^{\prime} , X_s} \times \frac{{\cal F}_{ \nu^{\prime}_{\tau,L} , t_R}}{{\cal F}_{ \nu^{\prime}_R , t_L^{\prime}}} \\
{\cal M}_{X^{\prime} - X_s} &\sim&\frac{M_{GUT}^2}{M_{KK}^2} \, k \lambda_5 \frac{v}{\Lambda}
\end{eqnarray}
The ${\cal F}\sim 1$ are the form factors reflecting the overlap between the wave functions. $P_R$ is the projector to remind that we focus on only one chirality, i.e., $(-+)$ of
the LZP (the other chirality is localized near the Planck brane and its 
coupling to $X_s$ is suppressed).  ${\cal M}_{X^{\prime} - X_s}$ is the mixing factor due to the GUT breaking vev of the bulk scalar field $\Sigma$.
$M_{GUT}/M_{KK}\lesssim 1/5$ is a measure of bulk GUT breaking. 
$k \lambda_5 \sim {\cal O}(1)$, $v$ is the Higgs vev and $\Lambda \sim$ 10 TeV is the cut off on the IR brane.

 The coupling of $\tilde{b}_R$ to the Higgs and $(t,b)_L$ is $\lambda_{ t \; 4D } 
f \left( c _R \right)$ (see section \ref{subsection:bulkbreakingPS}). 
The effective couplings of $\tau^{\prime -}_R$ to $W^-$ and $\tilde{b}_R$ to 
$W^-$ due to $W_R-W_L$ mixing resulting from EW symmetry breaking are:
\begin{eqnarray}
g_{ \nu^{\prime}_R , \tau_R^{\prime -} , W^-}&=&  \frac{g_{10}}{\sqrt{2}}\sqrt{k \pi r_c} \times {\cal F}_{ \nu^{\prime}_R , \tau_R^{\prime}} \times P_R \times {\cal M}_{W_R - W_L} \\
g_{ \tilde{b}_R , t_R , W^-}&= & g_{ \nu^{\prime}_R , \tau^{\prime-}_R , W^-} \times \frac{{\cal F}_{ \tilde{b}_R , t_R}}{{\cal F}_{ \nu^{\prime}_R , \tau^{\prime -}_R}}\\
{\cal M}_{W_R - W_L} &=& \frac{g_{10}}{g}\sqrt{2 k \pi r_c} \frac{M_W^2}{M_{KK}^2}
\end{eqnarray}
Again, the projector expresses the fact that only one chirality of the LZP has a non-suppressed coupling
 to $W_R$. Note that, in contrast, the direct interactions of our $(-+)$ KK fermions to zero-mode gauge 
 bosons,  as shown in the production process  in the  figures, are vector-like. 
This is because zero-mode gauge bosons have a flat profile 
(unlike KK modes) and couple identically to both chiralities of KK fermions.

\section{Baryogenesis}
\subsection{Relating dark matter to the baryon asymmetry}

Since our colorless dark matter particle carries baryon number, it is very tempting to investigate whether 
the origin of the apparent matter-antimatter asymmetry of the universe is that antimatter is stored in dark matter. In other words, in
a universe where baryon number is a good symmetry,  
the negative baryonic charge would be carried by DM, while the equal and opposite baryonic charge would be carried by ordinary SM quarks. This would provide a beautiful 
common explanation for these two major cosmological puzzles. 

Imagine that an asymmetry between $\nu^{\prime}_R$ and $\overline{\nu^{\prime}}_R$ 
is created due to the CP-violating out of 
equilibrium decay of KK gauge boson $X^{\prime}$: $X^{\prime}$ does not
carry baryon number but decays into
the LZP and an anti-top quark. 
The resulting asymmetry between the LZP and 
$\overline{LZP}$ (chosen to be negative) is equal to the asymmetry between the quark and antiquark. Consequently, dark matter would store the overall negative baryonic charge which is missing in the visible quark 
sector\footnote{While this paper was being finalized, reference 
\cite{Kitano:2004sv} appeared which has a comparable idea.}. 
The calculation of the relic density of dark matter is now quite different. It does not depend directly on the calculation of the annihilation cross section of the LZP but rather on the abundance of 
$X^{ \prime }$ at the time of its decay. Indeed, in the out-of equilibrium 
decay scenario, $Y_{asym}\sim \epsilon \ Y_{X^{\prime}}/g_*$
where 
$Y_{asym}$ is $(\overline{n}_{LZP}-{n}_{LZP})/s$,
$Y_{X^{\prime}}=n_{X^{\prime}}/s$ is the relic abundance $X^{\prime}$ would have today if it had not decayed, $s$ is the entropy density, $g_*$ is the number of relativistic degrees of freedom at the time of the decay and $\epsilon$ denotes CP-violation in the decay 
of $X^{\prime}$.

It is well-known that to explain the baryon asymmetry of the universe from an out of equilibrium decay, it is necessary that the decaying particle be overabundant. That is for instance what is required in leptogenesis where one needs the RH neutrino to 
be out of thermal equilibrium so that its number density can be large. The problem with the $X^{\prime}$ particle is that it has large gauge interactions and 
therefore 
large annihilation cross sections (via the s-channel gluon exchange) and, if we assume that it was at thermal equilibrium initially, 
then it
 will not be overabundant after freeze-out.
 
Remember that $X'$ would freeze out at a temperature $T_F\sim m_{X'}/x_F$ where $x_F \sim 25$ so for  $m_{X'}\gtrsim$3 TeV,  $T_F\gtrsim 120$ GeV. Then,
the only situation we can consider is that the reheat temperature of the universe is below the ``would-be'' freeze-out temperature of  $X'$ and assume that $X'$  is produced non thermally and abundantly at the end of inflation.
The thermal history is poorly known in RS1 geometries \cite{Creminelli:2001th} and we are unable at this point to make any more precise statement.
However, assuming that $X^{\prime}$ is indeed overabundant, there is a potentially interesting mechanism for relating dark matter and the baryon asymmetry as follows.

Baryon number conservation leads to
$\frac{1}{3}(n_{\mbox{\tiny LZP}}-\overline{n}_{\mbox{\tiny LZP}})=\overline{n}_b- n_b$.
Therefore,
\begin{equation}
\frac{1}{3} \ Y_{asym}\approx \frac{{n}_b}{s}  \sim 10^{-10}
\end{equation}
where in the last step we have assumed that annihilation of baryons with antibaryons 
after the production of the asymmetry is efficient enough so that the left over $\overline{n}_{b}$  are negligible compared to the excess of baryons. 
What we have to ensure is that  $X^{\prime}$ decays out of equilibrium, but 
before the baryons stop annihilating so that  the relic density 
of baryons is indeed given by the 
asymmetry.
$X^{ \prime }$ can decay because of the mixing with $X_s$ (see Fig.~\ref{fig:mixings}) and 
the size of this mixing depends on the size and
profile along the extra dimension of the vev of
the scalar $\Sigma$ which breaks $SO(10)$ in the bulk. 
Thus, it is possible that the above condition on the decay of $X^{ \prime }$
is satisfied.
 
If the LZP and anti-LZP can annihilate sufficiently after the 
$X^{\prime}$-decay, then the only dark matter left is given by the asymmetry,
i.e., $Y_{DM} = Y_{asym}$.
Clearly, we  need  a large LZP  annihilation cross section
 for this to happen. This will occur when the LZP annihilation  takes place  
near the resonances  (see Figs. \ref{fig:compare} and \ref{fig:relicdensity}). 
%
%
%
In the case that only the excess 
of anti-LZPs remain, since $\rho_{\mbox{\tiny DM}} \approx 6 \  \rho_b$ and 
$\rho_{\mbox{\tiny DM}}=m_{\mbox{\tiny LZP}} Y_{asym} s= 3 m_{\mbox{\tiny LZP}} 
{Y_b}s$,  we obtain that 
\be
\label{2GeVprediction}
m_{\mbox{\tiny LZP}}\approx 2 \ \mbox {GeV}.
\ee

This mass is not near a resonance so that, within our strict framework, 
we cannot guarantee that DM is just given by the LZP asymmetry. 
 Increasing the LZP-$Z$ coupling is a way to increase the annihilation cross section. Note that such a low mass is not constrained at all  by direct detection experiments. However, there are constraints from the $Z$ width.
  It is quite clear from Fig.~\ref{fig:relicdensity}  that even for the largest allowed LZP-Z coupling, the relic density of a 2 GeV LZP is still too large by  two orders of magnitude.
 
 So, one has to modify the model in such a way to increase the annihilation cross section at LZP masses $\approx 2$ GeV. There could be additional
annihilation channels playing a role even
for small LZP masses.
It might be possible to
open the annihilation channel
  via the Higgs exchange.
An alternative would be to assume that the reheat temperature is quite low $\lesssim {\cal O}$(1 GeV) so that the LZP never reaches thermal equilibrium and its initial abundance is suppressed. This solves the problem of overclosure of the universe by LZPs. We still have to guarantee a large enough annihilation cross section of the LZP for the validity of Eq.~\ref{2GeVprediction}.

Away from the regions where the annihilation cross section is large ($>$ 1 pb), 
we can neglect the effect of the asymmetry on the relic 
density of dark matter and the relation between $\Omega_{DM}$ and $\Omega_b$ is less straightforward.

Note that we did not consider the decay of $X_s$ because it has too 
large couplings to $t_R$ and the LZP so that it would not decay out of equilibrium.

Finally, because of  $Z_3$ conservation, 
the LZP cannot annihilate with itself into a SM final state 
(it can annihilate only with the anti-LZP).
Thus, there is no annihilation diagram
of the simple form LZP LZP $\rightarrow A+B$ where $A$  
and $B$ are SM particles 
 leading to the 
transfer 
of baryon number from
 the dark sector to the visible sector 
(hence washout of the baryon asymmetry)\footnote{We thank
R.~Kitano and I.~Low for discussions on this issue.}. Processes such 
as LZP LZP $\rightarrow$ anti-LZP $+$ SM are allowed by the $Z_3$ symmetry:
a detailed study would be required to see if they can lead to significant
washout of the baryon asymmetry or not.

\subsection{GUT baryogenesis at the TeV scale}
\label{GUTbaryo}
We now turn to a very different idea. In this paper, we mainly talked about the model with gauged baryon number. 
However, we stressed in subsection \ref{leptonnumbersymmetry} the fact that proton stability can be guaranteed by 
assuming lepton number symmetry instead of baryon number. This is actually an economical solution since it also 
forbids dangerous lepton number violation due to Majorana masses on the TeV brane. Unfortunately, there is no DM 
particle in this case. An intriguing feature of this model though is the possibility to observe baryon number 
violation at colliders, for instance via the production of 
(perhaps off-shell) KK $X$ gauge boson  which then decays into $u$ and
$d$ quarks, violating $B$.

We now want to expose some potentially interesting baryogenesis idea, even if at first sight it seems difficult 
to achieve in our particular framework.
In traditional GUT baryogenesis, one uses the out of equilibrium $B$-violating decay of GUT scale mass $X/Y$
gauge bosons.
However, if the $X/Y$ gauge bosons have a mass at the TeV scale, for their
decay to be out-of equilibrium, the gauge coupling has to be
smaller than about $10^{-8}$. This comes from equating the decay rate to the expansion rate: $g^2 M_X/(8\pi) 
\sim T^2/ M_{ Pl }$. In our model, the decays which violate $B$ 
(without violating $L$) are $X^{\prime} \rightarrow d d$, 
$Y^{\prime }\rightarrow u d $ and $X \rightarrow u u $. 

As mentioned in section \ref{subsection:gaugedlepton}, quarks of the third 
generation originate in
different multiplets to account for their different $c$'s. Whereas,
quarks of the 1st and 2nd generations can come from the same multiplet. So, $X$-type gauge
bosons can decay into two quarks from the 1st or 2nd generation only.
However, even though the 1st and 2nd generation quarks are localized near the Planck brane
(while the $X$-type gauge bosons are near the TeV brane),
their coupling to $X$-type gauge boson is typically $g\sim 10^{-5}$ for 1st generation
(and larger for 2nd generation)
to be compatible 
with their Yukawa couplings. Thus, it is likely that the decays will not be out of 
equilibrium. 

In any case, assuming some mechanism to further suppress the couplings of $X$-type gauge bosons, 
one would also have to check that 
the new CP violating sources required for baryogenesis in the X/Y sector are not in conflict with current 
experimental constraints.
And again, we would need a reheat temperature below, say, 
$120$ GeV to insure that $X,Y$
can be overabundant when they decay.

\section{Conclusion}

In this paper, we have studied the issue of baryon-number violation in a GUT
in a non supersymmetric warped extra dimension.
One way to suppress proton decay is 
to impose gauged baryon-number symmetry.
We showed how this solution 
to the proton stability problem leads to a stable KK particle, the stability being
guaranteed
by a combination of baryon-number and color, named $Z_3$. 
$Z_3$ is already present in the Standard Model at the renormalizable level, 
though SM particles are not charged under it.
This is similar to $R$-parity
leading to a stable particle in SUSY models.

Our firm prediction is that the lightest $Z_3$-charged particle (LZP) is
a GUT partner of the top quark and that 
its lightness is related to the top quark's heaviness. 
At this stage, 
we are not able to 
predict which particle in the top multiplet is the LZP,
but
as far as dark matter is concerned, the LZP must have gauge quantum numbers 
of a RH neutrino.  We can ensure that this is the case due to the breaking of
the GUT in the bulk. 
We showed in detail how this exotic RH neutrino acts as a WIMP and why 
its relic density is of the right value for masses in the 10 GeV -- a few TeV range.
We also explained why 
 the entire parameter space of this DM candidate
will be tested at near future direct detection experiments.

The breaking of the GUT has direct observable effects at the TeV scale.
The other KK modes in the top quark multiplet are also light. Because of their 
strong (QCD) 
coupling, the quark-like
states can easily be produced  at colliders and be detected via their distinctive decay into the LZP.
The production of these
other exotic, light partners of the top quark at high-energy colliders is an interesting manifestation of unification in AdS. 
 
We studied both models with the Higgs localized {\em on} the TeV brane and
with a profile for the Higgs in the bulk (but still
localized {\em near} the TeV brane).
Our qualitative results apply to any warped 
extra-dimensional GUT with electroweak symmetry breaking localized 
near the TeV brane. For example, our models can be seen as a GUT embedding of the
recently studied Higgsless models in warped space 
\cite{custodial2}.  
The choice of Pati-Salam or $SO(10)$ gauge group over $SU(5)$ 
is dictated by the need to incorporate custodial isospin symmetry, $SU(2)_R$, 
to satisfy electroweak precision constraints. 
Actually, up to factors of $O(1)$, we expect that 
our quantitative results also apply to Higgsless models. Of course, in these models,
the KK scale is no longer a free parameter since it is related to the $W$ and $Z$ masses.
%
%
 
There are many issues one could investigate further.  It would
be interesting to consider variations of the model we presented. For instance,
as far as bulk GUT breaking is concerned, it would be instructive
to see what happens when the bulk scalar field $\Sigma$ has a profile.  
One could also consider the case with no bulk breaking of GUT (with some 
alternative source for threshold corrections
required
for unification). In this case, 
the mass splitting between different partners of the top quark
will arise from radiative corrections. One-loop corrections to the lightest 
KK masses would be a useful calculation
to determine the identity of the LZP and NLZPs.
Can we obtain a weak-scale mass for the RH neutrino
and simultaneously a realistic phenomenology for other light partners?
Inclusion of brane kinetic terms \cite{braneterm} could modify
our results for dark matter relic density and detection.
Lastly, it would be interesting to study 
signatures for indirect detection, due to 
annihilation of the LZP in the Sun or in the galactic center. 

  Other models are also of interest.
For instance, 
as briefly mentioned, if $SO(10)$ is broken on the TeV brane rather than on the
 Planck brane, there is also
a light stable KK fermion, in this case coming from
one of the light fermion multiplets. A detailed study of this possibility  could be interesting.
We also reiterate that there is the alternative option of imposing lepton 
number instead of baryon number. Although there is no stable particle in this case, 
there could still be some other interesting phenomenology to study, like the possibility of observing baryon number violating decays at high energy colliders. 
Finally, the issue of baryogenesis certainly deserves more attention.

\acknowledgments

This work benefitted from diverse discussions with 
C. Balazs, M. Battaglia, A.~Birkedal,
C. Cs\'aki, A. Delgado, H. Frisch, C. Grojean,
I. Hinchliffe, D.E. Kaplan, R.~ Kitano, I.~ Low, J.~Lykken, D. Morrissey, F.~Petriello,
R. Sundrum, T. Tait and N. Weiner.
We thank the hospitality of the Aspen Center for 
Physics where this work was initiated. 
G.S thanks ITP, Santa 
Barbara and University of Michigan where part of this work was carried out.
K.~A.~is supported by the Leon Madansky fellowship
and NSF Grant P420D3620414350.
G.~S.~was supported in part by the US 
Department of Energy, High Energy Physics Division, under contract 
W-31-109-Eng-38 and also by the David and Lucile Packard Foundation.  


\section*{Appendix}
\begin{appendix}

\section{Couplings of KK modes}
In this part, the reader is referred to \cite{neubert,gp,choi} for more details.
 We assume $SO(10)$ as a gauge group so that there is only one $5D$ gauge coupling.
It is easy to extend these formulae for 
the three different bulk gauge couplings in the case of Pati-Salam.

\subsection{Coupling of two zero-mode fermions to a gauge KK mode}
\label{fermionzero}

The couplings of KK/zero modes are given by overlap of their wavefunctions.
Decomposing the $5D$ fermion as $\Psi ( x , z ) = \sum_n \psi^{ (n) } (x)
\chi_n ( c , z )$, the wavefunction of the zero-mode fermion is
\begin{equation}
\chi_0 (c, z) = \sqrt{ 
\frac{1-2c}{ z_h \left( e^{ k \pi r_c ( 1-2c) }
-1 \right) } } \left( \frac{z}{z_h} \right)^{2-c}
\label{zero}
\end{equation}
Similarly, decomposing the $5D$ gauge fields as
$A_{ \mu } 
( x, z ) = \sum_n A^{ (n) } (x) f_n (z)$, 
the wavefunction of gauge KK mode is given by
\begin{equation}
f_n ( z ) = 
\sqrt{ \frac{1}{z_h} } \frac{z}{ N_n } 
\left[ J_1 \left( m_n z \right) + b_n
Y_1 \left( m_n z \right) \right] 
\label{gaugekk}
\end{equation}
For 
KK modes
of $Z^{ \prime }$ (i.e.,
$(-+)$ boundary condition for gauge 
field), the normalization
factor is given by
\begin{eqnarray}
N_n^2 & = & \frac{1}{2} \left[ z_v^2 \big[ J_1 \left( m_n z_v \right) + b_n
Y_1 \left(  m_n z_v \right) \big]^2 - 
z_h^2 \big[ J_0 \left(  m_n z_h \right) + b_n
Y_0 \left(  m_n z_h \right) \big]^2 \right]  
\nonumber \\
\end{eqnarray}
and the masses of gauge KK modes and $b_n$ are given by
\begin{eqnarray}
\frac{ J_1 \left( m_n z_h \right) }{ Y_1 \left( m_n z_h \right) } & = & 
\frac{  J_0 \left( m_n z_v \right) }{  Y_0 \left( 
m_n z_v
\right) } \equiv - b_n
\end{eqnarray}
so that, for $m_n z_h \ll 1$, we get
$m_n z_v \approx $ zeroes of $J_0$.
In particular, as mentioned in
the main text, we define the KK scale of the model, $M_{ KK }$ to be
the mass of the lightest gauge KK mode:
\begin{equation}
M_{ KK } \equiv m_1 \approx 2.4 \ z_v^{-1}.
\label{lightestgaugeKK}
\end{equation}
For $m_n z_v \; ( \approx \; \hbox{zeroes of} \; J_0 ) \gg 1$, i.e., $m_n z_v
\approx \pi \left( n - 1/4 \right)$, we can show
that
\begin{eqnarray}
N_n^2 & \approx  & \frac{ z_v }{ \pi m_n } 
\end{eqnarray}

As discussed in subsection \ref{g10}, we define $g_{ 10 }$ as:
\begin{equation}
g_{ 10 } \equiv \frac{ g_{ 5 \; ren. } }{ \sqrt{ \pi r_c } },
\label{0mode}
\end{equation}
where
$g_{ 5 \; ren. }$ is the renormalized $SO(10)$ $5D$ gauge coupling.
The coupling of zero-mode of fermion $f$ to $n^{th}$ gauge KK mode 
of $Z^{ \prime }$  
is then given by (group theory factors can easily be generalized to the 
case of  other $(-+)$ or $(++)$ KK gauge fields)
\begin{eqnarray}
\frac{ g_{ Z^{ \prime \; (n) } }^{ f^{ (0) } } (c) }{ 
Q^f_{ Z^{ \prime } } \sqrt{5/2} \; g_{ 10 }
} & = &
\sqrt{ \pi r_c }
\int dz \sqrt{-G} \; \frac{z}{z_h} \; \chi_0^2 ( c, z ) f_n (z), 
\label{zero1kk}
\end{eqnarray}
where
$z / z_h$ is the funfbein factor, $-G = 
\left( z / z_h \right)^{-5}$ is the determinant of the metric.

We have used the fact that, in $SO(10)$, the coupling of the ``would-be'' zero-mode
of $Z^{ \prime }$ is given by
\begin{equation}
g_{ Z^{ \prime } } = \sqrt{5/2} \; g_{ 10 }
\end{equation}
Also,  
the mixing angle (analogous to
$\sin^2 \theta _W$), 
$\sin^2 \theta^{ \prime } 
\equiv g_{ 5 \; B - L }^2 / g_{ 5 \; Z^{ \prime } }^2 = 3/5$ in $SO(10)$, where 
$g_{ 5 \; B - L }$ is the coupling of the gauge boson which couples
to the charge $1/2 (B - L)$.
Thus,
the charge of the fermion under $Z^{ \prime }$, 
$Q_{ Z^{ \prime } } = \tau_R^3 - \sin^2 \theta ^{ \prime } \;
Y = \tau_R^3 - 3/5 \; Y$, where the last
relation is for $SO(10)$ (this is similar to 
the charge under $Z$ is $Q_{Z} = \tau_L^3 - Q \sin^2 \theta _W$).

For completeness,
the wavefunction of a KK mode of a 
$(++)$ gauge boson is as in Eq. \ref{gaugekk}, except
the normalization factor which is given by
\begin{eqnarray}
N_n^2 & = & \frac{1}{2} \left[ z_v^2 \big[ J_1 \left( m_n z_v \right) + b_n
Y_1 \left(  m_n z_v \right) \big]^2 - 
z_h^2 \big[ J_1 \left(  m_n z_h \right) + b_n
Y_1 \left(  m_n z_h \right) \big]^2 \right]   
\nonumber \\
\end{eqnarray}
and the masses of gauge KK modes and $b_n$ are given by
\begin{eqnarray}
\frac{ J_0 \left( m_n z_h \right) }{ Y_0 \left( m_n z_h \right) } & = & 
\frac{  J_0 \left( m_n z_v \right) }{  Y_0 \left( 
m_n z_v
\right) } \equiv - b_n
\end{eqnarray}
so that, for $m_n z_h \ll 1$, we get
$m_n z_v \approx $ zeroes of $J_0 + O \left( 1 / 
\big[ 
\log m_n z_h 
\big] \right)$.
For $m_n z_v \; ( \approx \; \hbox{zeroes of} \; J_0 ) \gg 1$, i.e., $m_n z_v
\approx \pi \left( n - 1/4 \right)$,
we can show that
$N_n^2 \approx z_v / \left( \pi m_n \right)$ as before.

The coupling of gauge KK modes to the Higgs
is obtained by evaluating the wavefunction on the TeV brane. We can show that
gauge KK modes with both $(++)$ and $(-+)$ BC approximately have the
same wavefunction on the TeV brane. For example, coupling of the Higgs to 
the $n^{ th }$ $Z^{ \prime }$ 
KK mode is given by:
\begin{eqnarray}
\frac{ g_{ Z^{ \prime \; (n) } }^H }{ Q^H_{ Z^{ \prime } } 
\; \sqrt{ 5/ 2 } \; g_ { 10 } } & \approx &
(-1)^{ ( n - 1 ) } \sqrt{ 2 k \pi r_c } 
\label{gaugeKKHiggs}
\end{eqnarray}
Here, $Q^H_{ Z^{ \prime } } = \pm 1/2 \left( 1 - \sin^2 \theta^{ \prime} \right) =
\pm 1/5$ (where the last relation is for $SO(10)$).
 
\subsection{Coupling of two KK fermions to a gauge KK mode}
\label{fermionkk}

The $(-+)$ helicity of the fermion KK mode (of mass $m_n$) has wave function
\begin{eqnarray}
\chi_n (c, z) = \frac{ \left( \frac{z}{ z_h } \right)^{5/2} }{ N_n \sqrt{ \pi r_c } }
\big[ J_{ \alpha } \left( m_n 
z \right) + 
b_{ \alpha } (m_n) 
Y_{ \alpha } \left( m_n z
\right)
\big],
\label{fermion-+}
\end{eqnarray}
where 
$\alpha = | c  + 1/2 |$,
$m_n$ and $b_{ \alpha}$ are given by 
\begin{eqnarray}
\frac{ J_{ \alpha } \left( m_n z_h \right) }
{ Y_{ \alpha } \left( m_n z_h \right) } & = & 
\frac{ J_{ \alpha \mp 1} \left( m_n z_v \right)  }
{ Y_{ \alpha \mp 1} \left( m_n z_v \right) } \equiv - b_{ \alpha } (m_n), 
\label{keyequation}
\end{eqnarray}
with upper (lower) signs for $c > -1/2$ ($c < -1/2$)
and
\begin{eqnarray}
N_n^2 & = & \frac{1}{ 2 \pi r_c z_h } 
\left[ z_v^2 \big[ J_{ \alpha } \left( m_n z_v \right) 
+ b_{ \alpha } (m_n) Y_{ \alpha } \left( m_n z_v \right) \big]^2 - 
z_h^2 \big[ J_{ \alpha \mp 1 } \left( m_n z_h \right) 
+ b_{ \alpha } (m_n) Y_{ \alpha \mp 1 } \left( m_n z_h \right) \big]^2 \right].
\nonumber \\
\end{eqnarray}
We will always assume that
$m_n z_h \ll 1$. Then, 
for $c > - 1/2 + \epsilon$
(where $\epsilon 
\sim 0.1$), we get $m_n z_v \approx$ zeroes of $J_{ c - 1/2 }$
(since $J_{ \alpha } (0) \rightarrow 0$ and $Y_{ \alpha } (0)\rightarrow \infty$ so
that LHS of \ref{keyequation} $ \sim 0$).

In particular, 
for $c > -1/4$, we can show that
zeroes of $J_{ c - 1/2 } > 1$ so that we use a large argument
approximation for $J_{ c - 1/2 }$ which is
$J_{ \nu } (x) \propto \cos \left( x - \pi / 2 \nu - \pi / 4 
\right)$. This gives $m_n z_v \approx \pi ( n - 1/2 + c / 2 )$
so that the lightest KK mode has mass $\approx z_v^{ -1 } \pi ( 1 + c ) / 2$.
Whereas, for $- 1/2 + \epsilon < c < -1/4$, the  smallest
zero of $J_{ c - 1/2 } < 1$ so that we use a small argument
approximation for $J_{ c - 1/2 }$ which is $J_{ \nu } (x) \approx
x^{ \nu } \big[ 1 / \left( 2^{ \nu } \Gamma ( \nu + 1 ) \right) - x^2 /
\left( 2^{ 2 + \nu } \Gamma ( \nu + 2 )\right) \Big]$ leading to $m_1 z_v \approx 2 \sqrt{ c + 1/2 }$.

For $c = - 1/2$,
the above equation gives $J_0 \left( m_n z_h \right) /
Y_{0} \left( m_n z_h \right) =  
J_{1} \left( m_n z_v \right) / Y_{1} \left( m_n z_v \right)$. 
We can show that there is a mode much lighter
than $1/z_v$ when arguments of both LHS and RHS 
of Eq.~\ref{keyequation}  are small with mass
$m \approx  z_v^{ -1 } \times \sqrt{ 2 /  \left( k \pi r_c \right) }$.
The masses of other modes are given by 
$m_n z_v \approx$ zeroes of 
$J_1 \approx \pi \left( n + 1/4 \right)$.

Finally, for $c < -1/2 - \epsilon$, we get 
$m_n z_v \approx$ zeroes of $J_{ - c + 1/2 } \approx \pi \left(
n - c / 2 \right) $ .
In addition, there is a mode much lighter than 
$1/z_v$ when arguments of both LHS and RHS of  \ref{keyequation} are small, given
by $m z_v \approx 2 \sqrt{ \alpha ( \alpha + 1 ) }  
\left( z_h/z_v \right)^{ \alpha }$. The $(-+)$ helicity
 of this light mode is localized near the TeV  brane.

For $m_n z_v \gg 1$ (and for all $c$),  
\begin{eqnarray}
N_n^2 & \approx & 
\frac{ z_v }{ z_h } \frac{1}{ \pi^2 m_n r_c }
\end{eqnarray}
Whereas, for light mode with $c < -1/2 - \epsilon$, we get
\begin{eqnarray}
N_n^2 & \approx & \frac{ z_v }{ z_h } \frac{1}{ \pi^2 m_n r_c } 
\frac{2}{ 1 - 2 c }
\end{eqnarray}

Using these wavefunctions, we can show
that (for other than light mode) 
%
\begin{eqnarray}
\hbox{wavefunction of KK fermion} |_{ \hbox{TeV brane} } & \approx &
\frac{ \sqrt{2} z_v^{ \frac{3}{2} } }{ z_h^2 }
\label{fermion-+TeV1}
\end{eqnarray}
and for light mode for $c \lesssim -1/2 - \epsilon$,
\begin{eqnarray}
\hbox{wavefunction of KK fermion} |_{ \hbox{TeV brane} } & \approx &
\frac{1}{ f (c ) } \frac{ \sqrt{2} z_v^{ \frac{3}{2} } }{ z_h^2 },  
\label{fermion-+TeV2}
\end{eqnarray}
where $f (c) \approx \sqrt{ \Big[ 2 / \left( 1 - 2 c \right) \Big] }$.

The wavefunction of the other $(+-)$ helicity  of KK fermion is as above, except
that 
its ``effective'' $c$\footnote{This is the $c$
entering the equation of motion of fermion.} 
is {\em opposite} to that of the $(-+)$ helicity. So, we get
$\alpha = | - c + 1/2 |$ and since its boundary condition is $(+-)$, 
$m_n$ and $b_{ \alpha }$ are given by
\begin{eqnarray}
\frac{ J_{ \alpha \pm 1} \left( m_n z_h \right)  }
{ Y_{ \alpha \pm 1} \left( m_n z_h \right) } & = &
\frac{ J_{ \alpha } \left( m_n z_v \right) }
{ Y_{ \alpha } \left( m_n z_v \right) } 
\equiv - b_{ \alpha } (m_n),
\end{eqnarray}
with upper (lower) signs for $c > ( < ) 1/2$
and
\begin{eqnarray}
N_n^2 & = & \frac{1}{ 2 \pi r_c z_h } 
\left[ z_v^2 \big[ J_{ \alpha \pm 1 } \left( m_n z_v \right) 
+ b_{ \alpha } (m_n) Y_{ \alpha \pm 1 } \left( m_n z_v \right) \big]^2  
- 
z_h^2 \big[ J_{ \alpha } \left( m_n z_h \right) 
+ b_{ \alpha } (m_n) Y_{ \alpha } \left( m_n z_h \right) \big]^2 \right].
\nonumber \\
\end{eqnarray}
Of course, $m_n$ obtained from above is the same as for the $(-+)$ helicity.
For $c \lesssim -1/2$, this helicity 
of the light mode is localized near the Planck brane.
Of course, for all $c$'s, the wavefunction of 
this helicity vanishes at the TeV brane.  

The coupling of a $(-+)$ or $(+-)$ $n^{ th }$ KK mode of 
fermion $f$ to $m^{ th }$ KK mode
of $Z^{ \prime }$  is given by (this formula, including 
group theory factors, can be generalized to the coupling 
of $2$ different
KK fermions to KK mode of 
other gauge fields
with $(-+)$ or $(++)$ boundary condition):
\begin{eqnarray}
\frac{ g_{ Z^{ \prime \; (m) } }^{ f^{ (n) } } (c) }
{ Q^f_{ Z^{ \prime } } \sqrt{5/2} g_{ 10 }
} & = &
\sqrt{ \pi r_c }
\int dz 
\sqrt{-G} \; \frac{z}{z_h} \;
\chi^2_n ( c , z ) f_m (z), 
\label{3kk}
\end{eqnarray}
where $Q_{ Z^{ \prime } } = 1/2$ for $\nu^{ \prime} _R$.

\subsection{Coupling of a zero-mode fermion and KK fermion to a gauge KK mode}

Similarly, the coupling of $m^{th}$ KK mode of $X_s$ and zero-mode of $t_R$ 
to $n^{ th }$ mode of LZP is given by (again, 
group theory factors can be generalized to the coupling 
of the zero-mode fermion and KK  fermion to the KK mode
of  other $(-+)$ or $(++)$ gauge fields):
\begin{eqnarray}
\frac{ g_{ X_s^{ (m) } }^{ \nu^{ \prime \; (n) }_R } (c) }{ 
\sqrt{ \frac{1}{2} } 
} & = &
\sqrt{ \pi r_c }
\int dz 
\sqrt{-G} \; \frac{z}{z_h} \; \chi_0 ( c, z ) \chi_n ( c , z ) f_m (z). 
\label{zero2kk}
\end{eqnarray}

\section{Profile for the Higgs}
\label{sec:profileHiggs}

In the model with the Higgs on the TeV brane, 
the Higgs mass gets a {\em divergent} contribution from loops of gauge and top quark KK modes
in addition to loops of zero-modes. The KK 
contribution dominates due to the large
multiplicity of KK modes and also due to 
the couplings of 
KK modes to the Higgs being enhanced compared to those of zero-modes.
This
results in a fine-tuning at the $1 \%$ level \cite{custodial1}
(as expected, the effect of the top quark modes is larger than that of gauge modes).
In the CFT picture, the Higgs 
is a ``regular'' composite state. The natural size for its mass 
is the same as other composites (i.e., few TeV) so that a light Higgs
(as required to fit electroweak data) is fine-tuned.

We can introduce a symmetry protection for the Higgs
mass 
from loops of KK modes (as opposed to zero-modes which
are inescapable) as follows.
In the CFT picture, the Higgs can be a 
pseudo-Goldstone boson of a spontaneously broken
global symmetry. It  is naturally lighter than other bound states
just like the pion in QCD. The $5D$ dual of this CFT picture is an extended
bulk
gauge symmetry with $(--)$ BC for $A_{ \mu }$ of a non-SM gauge field and 
$(++)$ for the corresponding $A_5$, i.e., there is a massless scalar 
(at tree-level) in
the spectrum \cite{cnp}. The pseudo-Goldstone boson acquires a {\em finite}
mass at the loop
level (i.e., the quadratic divergence in Higgs mass is cut-off at the KK scale
instead of the $5D$ cut-off scale)  and 
thus is naturally lighter than other KK states, improving the fine-tuning to
$\sim 10 \%$ \cite{acp}.

For our purpose, the only resulting modifcation is in the Higgs couplings
 as follows.
The Higgs (which is the zero-mode of $A_5$) has the following profile \cite{cnp}:
$A_5 ( x, z ) \ni H f_H ( z )$
where
\begin{eqnarray}
f_H ( z ) & = & \frac{ 
2 \left( \frac{z}{ z_h } \right)^2 
}
{ 
\left( 1 - \frac{z_h^2 }{ z_v^2 } \right) 
} 
\label{HiggsasA5}
\end{eqnarray}
In this case, the $4D$ Yukawa coupling is modified to
\begin{eqnarray}
\lambda_{ 5D } 
\int dz \sqrt{ -G } \frac{z}{ z_h } f_H (z) \chi (z) \chi^{ \prime } (z)
\end{eqnarray}
where $\lambda_{ 5D }$ has now dimensions of 
(mass)$^{-1/2}$ just like the $5D$ gauge coupling\footnote{This is expected
since the Higgs is the component of a gauge field, although we can show that,
in general,
the effective $\lambda_{ 5D } \neq g_{ 5D }$ due to mixing
between bulk fermion multiplets on the TeV brane \cite{acp}.},
$\chi (z)$'s are 
wavefunctions of zero and KK mode fermions and $\frac{z}{ z_h }$ is funfbein. 
The funfbein factor
appears since the Higgs is a component of a gauge field so that the coupling of
fermions to the Higgs is similar to the
coupling of fermions to gauge modes.

The Higgs coupling to gauge KK modes, for example, to $Z^{ \prime }$
is also modified to
\begin{eqnarray}
\frac{ g_{ Z^{ \prime \; (n) } }^H }{ Q^H_{ Z^{ \prime } } 
\; \sqrt{ 5/ 2 } g_ { 10 } } & = & \sqrt{ \pi r_c } \int dz \sqrt{ - G } 
\left( \frac{z}{ z_h } \right)^4 f_n (z) f^2_H (z)
\end{eqnarray}
where factors of $z/z_h$
come from  the inverse  metric.

\section{Contributions to the $S$ and $T$ parameters from light KK states}

The presence of light KK states in the $t_R$ multiplet
raises the question of enhanced contributions to $S$ and $T$ parameters.
Consider first the contribution to the
$S$ parameter which is the kinetic mixing between $Y$ and $W_3^L$
($S = 16 \pi \Pi^{ \prime }_{ 3 Y }$)
and  requires EWSB. We have to 
use the mass term $\sim \left( 2 \lambda_{ 5D } k \right) v$ to flip from light KK states
from $t_R$ multiplet to heavier
KK states from $\left( t, b \right)_L$ multiplet. For example, the contribution
of $L_R^{ \prime }$ KK states from $t_R$ multiplet and $L^{ \prime }_L$ KK states from $t_L$
multiplet is estimated to be:
\begin{eqnarray}
\Pi^{ \prime }_{ 3 Y } & \sim & 
\frac{ \left( 2 \lambda_{ 5D } k v \right)^2 }
{ 16 \pi^2 m_{ L^{ \prime }_L }^2 }
\log \left( \frac{ m_{ L^{ \prime }_R } }
{ m_{ L^{ \prime }_L } } \right) 
\log \left( 
\frac{ \Lambda }{ m_{ KK } }
\right)
\end{eqnarray}
where
$\log \left( m_{ e^{ \prime }_R } / m_{ L^{ \prime }_L } \right)$ comes from the IR divergence in the loop integral
(which is UV finite for each pair of KK modes) 
and $\log \left( \Lambda / m_{ KK } \right)$ comes from the sum over $2$ KK towers. 
Crucially, the loop diagram is not
significantly enhanced due to the presence of light KK states
(the logarithms are $O(1)$). This loop contribution is smaller by a loop factor
$\sim \left( 2 \lambda_{ 5D } k \right)^2 / \left( 16 \pi^2 \right)$
than the tree-level contribution
to the $S$ parameter from gauge KK modes $\sim 16 \pi v^2 / m_{ KK } ^2$ \cite{custodial1}.

Next, consider, the contribution to the 
$T$ parameter.
With no bulk breaking of $SO(10)$ (hence of custodial isospin), 
$c$ for $e^{ \prime }_R$ and $\nu^{ \prime }_R$ 
are the same. They both have $(-+)$ BC and the same spectrum. Thus,
there is no loop contribution
to $T$ from these states. In contrast,  $t_R$ and $\tilde{b}_R$ 
(which have the same $c$) have different BC and 
hence
different spectra so that they do give a loop contribution
to $T$ \cite{custodial1}. 

With bulk breaking
of custodial isospin, $c$'s for $e^{ \prime }_R$ and $\nu^{ \prime }_R$ are different
so that there is a loop contribution to $T$ from these KK states 
also.
However, 
just like for $S$, 
contribution to $T$  requires EWSB. We have to use $\left( 2 \lambda_{ 5D } k \right) v$ to flip to
heavier $L^{ \prime }_L$ states so that there is no  enhancement due to light 
KK states in the loop.
This contribution to $T$ is smaller than the
 tree level one from splitting in $W^{ \pm }_R$-$W^3_R$ spectrum
due to breaking of custodial isospin in the bulk
(in addition to the mass splitting due to 
different BC's for $W^{ \pm }_R$ and $W^3_R$) \cite{custodial1}.

\section{Annihilation cross sections}
\label{Appendix:anni}

We consider annihilation due to the exchange of the lightest gauge KK modes only
since the effects decouple very fast with increasing KK masses. As for fermions, we restrict ourselves to
zero-modes (SM fermions) and the lightest
KK modes of other fermions. Hence, in what follows, we omit the
superscript $(n)$ on all the modes. Let us
denote the annihilation cross section for LZP and anti-LZP
into $t_R$ through the $t$-channel 
exchange of $X_s$ by $\sigma_1$, the annihilation into any fermion  through the $s$-channel 
exchange of $Z$ by $\sigma_2$ and the annihilation into top and bottom via 
the s-channel exchange of 
$Z^{ \prime }$ by $\sigma_3$. Also, $\sigma_{12}$ ($\sigma_{13}$) denotes
the interference between the $s$-channel
$Z$ ($Z^{ \prime }$) exchange
and $t$-channel exchange of $X_s$ for the annihilation into RH top quark. 
$\sigma_{23}$ is the interference between the $Z$ and $Z^{\prime}$ exchanges for 
the annihilation into top and bottom. Their exact expressions are given below. 
Here,
$m$, $m_t$ and $M_s$ denote the LZP, top and $X_s$ masses respectively. $N_c$ is the number of QCD colors. 
$g^{ \nu^{ \prime }_R }_{ X_s}$ 
is the effective $X_s \nu'_R t_R$ coupling 
given in Eq.~\ref{zero2kk}. 
Basically, it is the effective 4D $SO(10)$ coupling times a 
factor reflecting the overlap of wave functions of $X_s$, $t_R$, and $\nu^{\prime}_R$. We get 
\begin{equation}
\sigma_1= \frac{ \left( g^{ \nu^{ \prime }_R }_{ X_s} \right)^4  
\ N_c  \ ( \ \beta \   \beta_t  \ s \   {\cal E}  \ - \  {\cal F} \  L \  )}{ 256 \  \pi \  M_s^4  \  \beta^2 \   s^2 \  {\cal G}}
\end{equation}

\begin{equation}
\mbox {where} \ \ \beta=\sqrt{1-\frac{4 m^2}{s}} \ \ \ , \ \ \  \beta_t=\sqrt{1-\frac{4 m_t^2}{s}}
\end{equation}

\begin{equation}
\mbox {and} 
\ \ L=\ln\left[\frac{1+\gamma}{1-\gamma}\right]  \ \ \ , \ \ \  \gamma=\frac{s \  \beta \ \beta_t}{2 \ (m^2+m_t^2-M_s^2)-s}.
\end{equation}
Also,
\begin{eqnarray}
{\cal E}&=&
\nonumber
2\,m^8 + 8\,{{M_s}}^8 + 2\,{{m_t}}^8 - 4\,m^6\,\left( {{M_s}}^2 + {{m_t}}^2 \right)  - 
  4\,{{M_s}}^6\,\left( 4\,{{m_t}}^2 - 3\,s \right)  
+  {{M_s}}^2\,\left( -4\,{{m_t}}^6 + 5\,{{m_t}}^4\,s \right)  \\
\nonumber
 && + 2\,{{M_s}}^4\,\left( 5\,{{m_t}}^4 - 4\,{{m_t}}^2\,s + 2\,s^2 \right)  + 
  m^4\,\left( 10\,{{M_s}}^4 + 4\,{{m_t}}^4 + {{M_s}}^2\,\left( -4\,{{m_t}}^2 + 5\,s \right)  \right)
  \\ && - 
  2\,m^2\,\left( 8\,{{M_s}}^6 + 2\,{{m_t}}^6 + 2\,{{M_s}}^4\,\left( {{m_t}}^2 + 2\,s \right)  + 
     {{M_s}}^2\,\left( 2\,{{m_t}}^4 + 3\,{{m_t}}^2\,s \right)  \right),
\end{eqnarray}
\begin{eqnarray}
\frac{{\cal F}}{2}&=&
\nonumber
 m^{10} - 4\,{{M_s}}^{10} + {{m_t}}^{10} - 3\,m^8\,\left( {{M_s}}^2 + {{m_t}}^2 \right)  + 
    4\,{{M_s}}^8\,\left( 3\,{{m_t}}^2 - 2\,s \right)  + 
    {{M_s}}^4\,\left( 7\,{{m_t}}^6 - 5\,{{m_t}}^4\,s \right)  \\
    \nonumber
   &+&  {{M_s}}^2\,\left( -3\,{{m_t}}^8 + {{m_t}}^6\,s \right)  + 
    {{M_s}}^6\,\left( -13\,{{m_t}}^4 + 12\,{{m_t}}^2\,s - 4\,s^2 \right)  \\
&+& m^6\,\left( 7\,{{M_s}}^4 + 2\,{{m_t}}^4 + {{M_s}}^2\,\left( 4\,{{m_t}}^2 + s \right)  \right)\\
\nonumber
      &-&  m^4\,\left( 13\,{{M_s}}^6 - 2\,{{m_t}}^6 + 
       {{M_s}}^2\,{{m_t}}^2\,\left( 2\,{{m_t}}^2 + s \right)  + 
       {{M_s}}^4\,\left( 7\,{{m_t}}^2 + 5\,s \right)  \right)\\
       \nonumber
  &+&   m^2\,\left( 12\,{{M_s}}^8 - 3\,{{m_t}}^8 - 6\,{{M_s}}^6\,\left( {{m_t}}^2 - 2\,s \right)  + 
       {{M_s}}^4\,\left( -7\,{{m_t}}^4 + 10\,{{m_t}}^2\,s \right)  + 
       {{M_s}}^2\,\left( 4\,{{m_t}}^6 - {{m_t}}^4\,s \right)  \right)  
\end{eqnarray}
and finally
\begin{eqnarray}
{\cal G}&=&m^4+M_s^4+m_t^4-2m^2(M_s^2+m_t^2)+M_s^2(-2m_t^2+s)
\end{eqnarray}
The s-channel exchange of $Z$ is given by
\begin{equation}
\sigma_2( g^{ \nu^{ \prime }_R }_Z  ,g_Z^t, M_Z,\Gamma_Z)=
\frac{ { g^{ \nu^{ \prime }_R }_Z }^2 { g_Z^t }^2 
{N_c} \beta_t
    \left( {{M_Z}}^4 s \left( -{{m_t}}^2 + s \right)  + 
      m^2 \left( - {{M_Z}}^4 s  + 
         {{m_t}}^2 \left( 4 {{M_Z}}^4 - 6 {{M_Z}}^2 s + 3 s^2 \right)  \right)  
\right) }
{48\, {{M_Z}}^4\,\pi  \beta s
    \left( ({{M_Z}}^2-s)^2 + {{M_Z}}^2 {{\Gamma_Z}}^2  \right) }
\end{equation}
and the s-channel exchange of $Z^{ \prime }$ is obtained by the substitution $Z \rightarrow
Z^{ \prime }$ in the formula for $\sigma_{ 2 }$.
%
%
Here, $g^{ \nu^{ \prime }_R }_Z$ is the LZP-$Z$ coupling defined in 
Eq. \ref{LZPZ},
$g^{ \nu^{ \prime }_R }_{ Z^{ \prime } }$ is the LZP-$Z^{\prime}$ coupling
(obtained from Eq.\ref{3kk}),
 $g_Z^t$ is the usual top-$Z$ coupling and 
$g_{ Z^{ \prime  } }^t$ is the top-$Z^{\prime}$ coupling obtained from
Eq. \ref{zero1kk}.   

The interference terms between the $t$ and $s$-channel exchanges read  
\begin{equation}
\sigma_{12}( g^{ \nu^{ \prime }_R }_Z, g_Z^t, M_Z, \Gamma_Z)=
 \frac{ \left( g^{ \nu^{ \prime }_R }_{ X_s} \right)^2 \, \, g_Z^{ \nu^{ \prime }_R } 
\, \, g_Z^t \, \, N_c \, \, \beta_t \, (M_Z^2-s) 
(\frac{2 \, L \, }{\beta \beta_t s} \,  {\cal I}  +   {\cal J}  )  }
{64 \, \pi \, M_s^2 \, M_Z^2 \, \beta  \, s \, ( \, \Gamma_Z^2  \, M_Z^2 \, + \, ( \, M_Z^2 - s \, )^2)}
\end{equation}    
and $\sigma_{ 13 }$ is obtained by the substitution
$Z \rightarrow Z^{ \prime }$ in the above formula.
%
%
Here,
\begin{eqnarray}
\nonumber
{\cal I}(M_Z)&=&-\left( m^4\,{{M_{Z}}}^2\,\left( 2\,{{M_s}}^2 + s \right)  \right)  - 
  {{M_{Z}}}^2\,\left( 2\,{{M_s}}^6 - 4\,{{M_s}}^4\,\left( {{m_t}}^2 - s \right)  + 
     2\,{{M_s}}^2\,{\left( {{m_t}}^2 - s \right) }^2 + {{m_t}}^4\,s \right)  \\
     &+& 
  m^2\,\left( 4\,{{M_s}}^4\,{{M_{Z}}}^2 + 2\,{{m_t}}^2\,{{M_{Z}}}^2\,s + 
     {{M_s}}^2\,\left( 4\,{{M_{Z}}}^2\,s + {{m_t}}^2\,\left( -2\,{{M_{Z}}}^2 + s \right)  \right)  \right)
\end{eqnarray}
and
\begin{eqnarray}
{\cal J}(M_Z)=m^2\,\left( 4\,{{M_s}}^2\,{{M_{Z}}}^2 + 2\,{{m_t}}^2\,\left( 2\,{{M_{Z}}}^2 - s \right)  \right)  - 
  2\,{{M_s}}^2\,{{M_{Z}}}^2\,\left( 2\,{{M_s}}^2 - 2\,{{m_t}}^2 + 3\,s \right)
\end{eqnarray}
We end with the interference between the $Z$ and $Z^{ \prime }$ exchange
\begin{eqnarray}
\sigma_{23} \left( g_{ Z^{ \prime } }^t, g_Z^t \right) 
=\frac{ { \beta_t } \, { g^{ \nu^{ \prime }_R }_Z } \,
{ g_{ Z^{ \prime } }^{ \nu^{ \prime }_R } }
\,{ g_{ Z^{ \prime } }^t }\,{ g_Z^t }\,{Nc}\,
    \left( {\Gamma_{Z'}}\,{\Gamma_{Z}}\,{M_{Z}}\,{M_{Z^{ \prime }}} + 
      \left( {{M_{Z}}}^2 - s \right) \,\left( {{M_{Z'}}}^2 - s \right)  \right) \,
  {\cal P} }{24\,\beta\,{{M_{Z}}}^2\,{{M_{Z^{ \prime }}}}^2\,
    \pi \,s\,\left( {{\Gamma_{Z}}}^2\,{{M_{Z}}}^2 + {\left( -{{M_{Z}}}^2 + s \right) }^2 \right) \,
    \left( {{\Gamma_{Z^{ \prime }}}}^2\,{{M_{ Z^{ \prime } }}}^2 + {\left( -{{M_{Z^{ \prime }}}}^2 + s \right) }^2 \right) } \nonumber \\
\end{eqnarray}
\begin{eqnarray}
\nonumber
  {\cal P}={{M_{Z}}}^2\,{{M_{Z^{ \prime }}}}^2\,s\,\left( -{{m_t}}^2 + s \right)  + 
      m^2\,\left( -\left( {{M_{Z}}}^2\,{{M_{Z'}}}^2\,s \right)  + 
         {{m_t}}^2\,\left( {{M_{Z}}}^2\,\left( 4\,{{M_{Z^{ \prime }}}}^2 - 3\,s \right)  + 
            3\,s\,\left( -{{M_{Z^{ \prime }}}}^2 + s \right)  \right)  \right)  \\
\end{eqnarray}            
We can now list the cross-sections
for the annihilation processes into different 
final states, in terms of  the cross sections defined above:
\subsection{Annihilation into $t_R$}

\begin{eqnarray}
\sigma_{ \nu_R^{ \prime } \overline{\nu}_R^{ \prime } 
\rightarrow t_R \overline{t}_R} & = &
\sigma_1+ \sigma_2( g_Z^t =g_Z^{ t_{ \mbox{\tiny R} } } ) +
\sigma_3( g_{ Z^{ \prime } }^t = g_{ Z^{ \prime } }^{ t_{ \mbox{\tiny R} } } )+
\sigma_{12} \left( g_Z^t =g_Z^{ t_{ \mbox{\tiny R} } } 
\right) + \sigma_{13} \left( g_{ Z^{ \prime } }^t =
g_{ Z^{ \prime } }^{ t_{ \mbox{\tiny R} } } \right) + \nonumber \\
 & & \sigma_{23} \left( 
g_Z^t =g_Z^{ t_{ \mbox{\tiny R} } }, 
g_{ Z^{ \prime } }^t =
g_{ Z^{ \prime } }^{ t_{ \mbox{\tiny R} } } \right)
\end{eqnarray}

\subsection{Annihilation into $t_L$}

\begin{eqnarray}
\sigma_{ \nu_R^{ \prime } \overline{\nu}_R^{ \prime } 
\rightarrow t_L \overline{t}_L }=  \sigma_2(
g_Z^t = g_Z^{ t_{ \mbox{\tiny L} } } ) +
\sigma_3( g_{ Z^{ \prime } }^t = g_{ Z^{ \prime } }^{ t_{ \mbox{\tiny L} } } ) 
+\sigma_{23} \left( g_Z^t = g_Z^{ t_{ \mbox{\tiny L} } }, 
g_{ Z^{ \prime } }^t = g_{ Z^{ \prime } }^{ t_{ \mbox{\tiny L} } }
\right)
\end{eqnarray}

\subsection{Annihilation into light fermions}

 As mentioned in the main text, we neglect the coupling of $Z^{ \prime }$ to
all the SM fermions (denoted by $f$) other than the 
top or {\em left}-handed bottom so that:
\begin{eqnarray}
\sigma_{ \nu_R^{ \prime } \overline{\nu}_R^{ \prime } \rightarrow 
f \overline{f} } = \sigma_2( g_Z^t = 
g_Z^{ f_{ \mbox{\tiny L} } } ) 
+ \sigma_2 ( g_Z^t = g_Z^{ f_{ \mbox{\tiny R} } } ) 
\end{eqnarray}

\subsection{Annihilation into bottom}
\begin{eqnarray}
\nonumber
\sigma_{\nu_R^{ \prime } 
\overline{\nu}_R^{ \prime } \rightarrow b \overline{b}}= 
\sigma_2( g_Z^t = g_Z^{ b_{ \mbox{\tiny L} } } ) 
+ \sigma_2( g_Z^t = g_Z^{ b_{ \mbox{\tiny R} } } ) + 
\sigma_3( g_{ Z^{ \prime } }^t = g_{ Z^{ \prime } }^{ b_{ \mbox{\tiny L} } } ) 
+\sigma_{23} \left( g_Z^t = g_Z^{ b_{ \mbox{\tiny L} } }, 
g_{ Z^{ \prime } }^t = g_{ Z^{ \prime } }^{ b_{ \mbox{\tiny L} } }
\right) \\
\end{eqnarray}

\subsection{Annihilation into $W^+$ $W^-$ and $Z$ $H$}

\begin{eqnarray}
\sigma_{\nu_R^{ \prime } \overline{\nu}_R^{ \prime } \rightarrow WW}
= \frac{ \left( g_{ Z^{ \prime } }^H \right)^2 \,
\left( g^{ \nu^{ \prime }_R }_{ Z^{ \prime } } \right)^2 \, 
\beta_W \, \left( -m^2 + s \right) \,
    \left( -4\,{{m_{W}}}^2 + s \right) }{96\,\pi \, \beta \,{\left( {{M_{Z^{ \prime }}}}^2 - s \right) }^2\,s}
\end{eqnarray}
and
\begin{eqnarray}
\sigma_{\nu_R^{ \prime } \overline{\nu}_R^{ \prime } 
\rightarrow Z H } = \frac{ \left( g_{ Z^{ \prime } }^H \right)^2 \,
\left( g^{ \nu^{ \prime }_R }_{ Z^{ \prime } } \right)^2 \,
{\sqrt{\frac{{{m_H}}^4 + {\left( {{M_{Z}}}^2 - s \right) }^2 - 
          2\,{{m_H}}^2\,\left( {{M_{Z}}}^2 + s \right) }{s^2}}}\, {\cal K} }
    {96\,{{M_{Z^{ \prime }}}}^4\,\pi \, \beta \,{\left( {{M_{Z^{ \prime }}}}^2 - s \right) }^2\,s^2},
\end{eqnarray}
where
\begin{eqnarray}
{\cal K}&= & {{M_{Z^{ \prime }}}}^4\,s\,\left( {{m_H}}^4 + {\left( {{M_{Z}}}^2 - s \right) }^2 - 
         2\,{{m_H}}^2\,\left( {{M_{Z}}}^2 + s \right)  \right)  \\
         \nonumber
         &  + & 
      m^2\,\left( 2\,{{M_{Z}}}^2\,{{M_{Z^{ \prime }}}}^4\,s - {{M_{Z^{ \prime }}}}^4\,s^2 + 
         {{m_H}}^4\,\left( 2\,{{M_{Z^{ \prime }}}}^4 - 6\,{{M_{Z^{ \prime }}}}^2\,s + 3\,s^2 \right)  \right) \\
         \nonumber
         &+& m^2\, \left( {{M_{Z}}}^4\,\left( 2\,{{M_{Z^{ \prime }}}}^4 - 6\,{{M_{Z^{ \prime }}}}^2\,s + 3\,s^2 \right)  + 
         {{m_H}}^2\,\left( 2\,{{M_{Z^{ \prime }}}}^4\,s - 
            2\,{{M_{Z}}}^2\,\left( 2\,{{M_{Z^{ \prime }}}}^4 - 6\,{{M_{Z^{ \prime }}}}^2\,s + 3\,s^2 \right)  \right)  \right)  
\end{eqnarray}

\section{Annihilation via Higgs exchange}
\label{section:Higgsexchange}
The LZP can annihilate through Higgs exchange into (i) top pair
via the top Yukawa coupling, $H t_L t_R \lambda_t$, (ii) two transverse $W$'s via
the coupling $m^2_W / v \; H W_{ trans. } W_{ trans. }$ (from $| H |^2 W^2$),
(iii) one transverse $W$ and one longitudinal $W$ via the
coupling $g \partial_{ \mu } H W_{ trans. }^{ \mu } W_{ long. }$ 
(from $ \partial _{ \mu } H W^{ \mu } H$)  and (iv) 
two longitudinal $W/Z$'s  via the coupling 
$m_H^2 / v \; H W_{ long. } W_{ long. }$ (from Higgs quartic).
It is easy to check that (ii)
is sub-dominant to (iv) while 
(iii) is
subdominant to (i).
We can estimate the cross-sections as follows
(up to factors of $2 \pi$ from phase space)
\begin{eqnarray}
\sigma_{ H \rightarrow t \bar{t} }  \sim  
N_c g_H^2 \lambda_t^2 
\frac{ m_{ LZP }^2 } { \left( m_H^2 + m_{ LZP }^2 \right)^2 } \ \ \ , \ \ \ 
\sigma_{ H \rightarrow W^+_{ long. } W^-_{ long. } }
 \sim g_H^2 \left( \frac{ m_H^2 }{v} 
\right)^2 \frac{1}{ \left( m_H^2 + m_{ LZP }^2 \right)^2 },
\label{HiggsWlong}
\end{eqnarray}
where $g_H$ is coupling of the two chiralities of LZP to Higgs
defined in section 9.3.
The ratio of these two cross-sections is $\sim 3 m_t^2 m_{ LZP }^2 / m_H^4$ so that
the Higgs exchange into longitudinal $W$'s dominates over top pairs for $m_{LZP} 
\lesssim m_H^2 / m_t$.
For $m_{ LZP } \lesssim m_Z$, 
the Higgs exchange is very small since the top
or $W/Z$ channel is not open. For
$m_Z \lesssim m_{ LZP } \lesssim m_t$, the
Higgs exchange is dominantly into longitudinal
$W/Z$'s. 
The annihilation via $Z^{ \prime }$ exchange is small since 
$Z^{ \prime } \rightarrow t \bar{t}$ is not open.
So, we compare Higgs exchange to $Z$ exchange into light fermions which is
enhanced by multiplicity factor $N \sim 20$
(counting color and generation factors):
\begin{eqnarray}
\sigma_{ Z \rightarrow f \bar{f} }  \sim  
N \left( g^{ \nu^{ \prime }_R }_Z \right)^2 g_Z^2 \frac{1}{ m_{ LZP }^2 } 
\sim 
N g_{ Z^{ \prime } }^4 \left( k \pi r_c 
\right)^2 \frac{ m_Z^4 }{ m_{ LZP}^2 m_{ Z^{ \prime } }^4 } 
\label{Zff}
\end{eqnarray}
where for the LZP coupling to $Z$, i.e., $g^{ \nu^{ \prime }_R }_Z$
we 
have used the coupling induced by $Z - Z^{ \prime }$ mixing
$\sim \frac{g^2_{ Z^{ \prime } } }{ g_Z} \; k \pi r_c \; 
\frac{m_Z^2 }{ m_{ Z^{ \prime } }^2}$. Here, $g_Z$ and $g_{ Z^{ \prime } }$
are the zero-mode (``would-be'' in case of $Z^{ \prime }$) 
gauge couplings.
Assuming $m_{ \nu^{ \prime }_L } \sim m_{ Z^{ \prime } }$ and $k \pi r_c \sim 30$, we get
from Eqs. \ref{HiggsWlong} and \ref{Zff}
\begin{eqnarray}
\frac{ 
\sigma_{ H \rightarrow W^+_{ long. } W^-_{ long. } } 
}
{ 
\sigma_{ Z \rightarrow f \bar{f} } 
} 
& \sim & 
\left( \frac{
2 \lambda_{ 5D } k \; m_{ \nu^{ \prime }_L \nu^{ \prime }_R } / 30 
}
{ m_Z } \right)^2
\left( 
\frac{
m_{ LZP }
}
{
m_Z 
} \right)^4
\frac{ g_Z^2 }
{ g_{ Z^{ \prime } }^4 N }
\end{eqnarray}
%
so that Higgs exchange is $\sim 1/4$ for $m_{ LZP } \sim m_t$ and much
smaller for $m_{ LZP } < m_t$.

For $m_{ LZP } \gtrsim m_t$, $Z^{ \prime }$ exchange
into top pair
is open and dominates over $Z$ exchange 
since 
\begin{eqnarray}
\sigma_{ Z^{ \prime } \rightarrow t \bar{t} }  \sim  3 {g_Z^{ \prime }}^4 
\left( k \pi r_c \right)^2 \frac{ m_{ LZP }^2 }{ m_{ Z^{ \prime } }^4 }\ \ \
\rightarrow \ \ \
\frac{ 
\sigma_{ Z \rightarrow f \bar{f} } 
}
{ \sigma_{ Z^{ \prime } \rightarrow t \bar{t} } }
 \sim  \frac{N}{3} \left( \frac{ m_Z }{ m_{ LZP } } \right)^4
\end{eqnarray}
and
%
\begin{eqnarray}
\frac{ 
\sigma_{ H \rightarrow W^+_{ long. }W^-_{ long. } }
}
{ 
\sigma_{ Z^{ \prime } \rightarrow \bar{t} t } 
}
& \sim & 
\frac{ 
g^2_Z 
}
{ 3 \; 
g^4_{ Z^{ \prime } }
} 
\left( 
\frac{
2 \lambda_{ 5D } k \; m_{ \nu^{ \prime }_L \nu^{ \prime }_R } / 30
}
{ 
m_Z 
} 
\right)^2
(\hbox{for} \; m_{ LZP } \lesssim m_H) \nonumber \\
 & \sim & 
\frac{ g^2_Z 
}
{ 3 \; 
g^4_{ Z^{ \prime } }
} 
\left( 
\frac{
2 \lambda_{ 5D } k \; m_{ \nu^{ \prime }_L \nu^{ \prime }_R } / 30
}
{ 
m_Z 
} 
\right)^2
\left( \frac{ m_{H} }{ m_{ LZP } } \right)^4 \; 
(\hbox{for} \; m_{ LZP } >  m_H) \nonumber \\
\end{eqnarray}
%
So, cross section for
Higgs exchange is $\sim 1/10$ of that for $Z^{ \prime }$ exchange for $m_t
\lesssim m_{ LZP }\lesssim
m_H$ and much smaller for $m_{ LZP } > m_H$.

Finally, for $m_{LZP}\gtrsim m^2_H / m_t$, 
we should compare Higgs exchange into top pairs (since it dominates
exchange into longitudinal $W/Z$'s) with 
$Z^{ \prime }$ exchange:
\begin{eqnarray}
\frac{ 
\sigma_{ H \rightarrow t \bar{t} }
}
{ 
\sigma_{ Z^{ \prime } \rightarrow \bar{t} t } 
}
& \sim & \frac{1}{ g^4_{ Z^{ \prime } } \left( k \pi r_c \right)^2 } 
\left(
\frac{ 2 \lambda_{ 5D } k \; m_{ \nu^{ \prime }_L \nu^{ \prime }_R } }{ m_{ LZP } } 
\right)^2
\end{eqnarray}
so that Higgs exchange into top pairs is smaller
by $\sim 1/30$ 
(assuming 
$m_{ LZP } 
\gtrsim 300$ GeV).

To summarize, the cross-section for annihilation via 
Higgs exchange strongly depends on the Higgs mass. 
It is significant  (but less than $\sim 1/5$ of
$Z$/$Z^{ \prime }$
exchange) only
for $m_t \lesssim m_{ LZP }  \lesssim m_H$.
An exception is when $m_{ LZP } \approx m_H /2$, 
where there is an enhancement from Higgs resonance resulting in
suppressed relic density.
As a result,
for a first study of this DM candidate, we neglect the Higgs
exchange.

\section{CFT interpretation}
\label{cft}

As per AdS/CFT correspondence \cite{adscft}, 
the RS1 model is dual \cite{rscft}
to a strongly coupled CFT of which the minimal Higgs is a
composite arising after conformal invariance is broken at $\sim$ TeV. Since
gauge and fermion fields are in the bulk, in the dual $4D$ picture,
the SM gauge and fermions fields originate as fundamental fields, {\em external} 
to CFT, but coupled to the CFT/Higgs sector.
Due to this coupling,
these external fields mix with CFT composites, 
the resultant massless states corresponds to the 
SM gauge and fermion fields (which are dual to {\em zero}-modes
on the RS1 side).
The degree of this mixing depends on the anomalous/scaling dimension
of the CFT operator to which the fundamental fields
couple. The coupling of  SM gauge bosons and fermions 
to the Higgs goes via their composite
component 
since the
Higgs is a 
composite of the CFT.
Thus, the above coupling of fundamental gauge and fermion fields to 
CFT operators is essential for gauge boson and fermion masses to arise at 
the weak scale.

In the following sections, we will give details of the CFT interpretation of the
grand unified model.  
Some of the discussion appears in the literature (see, for example,
references \cite{continorunning, gns, rsgut, dual, custodial1, Contino:2004vy} in addition to
\cite{rscft}), but we review it
for completeness.

\subsection{Duality at qualitative level}
\label{cftqual}

The dual interpretation of  {\em gauge} fields in the bulk
is that the $4D$ CFT has a conserved {\em global} 
symmetry current (which is a marginal operator, i.e., with
zero anomalous dimension). In our case, 
there is a $SO(10)$ or Pati-Salam gauge symmetry in the bulk so that the dual 
CFT has global $SO(10)$ or Pati-Salam symmetry.

Since only SM gauge fields are $(+)$ (i.e. do  not vanish) 
on the Planck brane ($X_s$, $W^{ \pm }_R$, $X$, $Y$, 
$X^{ \prime }$ and $Y^{ \prime }$ 
vanish on the Planck brane), only the SM subgroup of the 
$SO(10)$ global symmetry of the CFT is   
gauged, i.e.,
only $J^{ \mu }_{ SM }$ is coupled to 
$4D$ SM gauge fields: $A^{SM}_{ \mu } J^{ \mu }_{
SM 
}$. This gauging is 
similar to the gauging of $U(1)_{em}$ {\em global} symmetry of real QCD
by coupling $J_{ em } ^{ \mu }$ to $\gamma$.
The operator $J^{ \mu }
$ 
interpolates/creates out of the vacuum
spin-$1$ hadrons/composites of CFT, including states with quantum numbers
of $X_s$ etc. These are 
similar to $\rho$-mesons in real QCD and are dual to  
gauge KK modes, including those of $X_s$ etc., on the RS1 side.

The dual interpretation of 
a bulk fermion, for example
$F^q_L$  (using the Pati-Salam notation),  is that
the CFT has a fermionic operator (in conjugate representation), 
denoted by
${\cal O}_{
F^q_L}$, and similarly for other bulk fermions. 
Since $Q_L$ is $(+)$ on the Planck brane, whereas
$L^{ \prime }_L$ is $(-)$ (i.e., vanishes on the Planck brane), in the dual CFT picture,
we add a fundamental fermion, also denoted by $Q_L$ and
couple it to the color triplet part of ${\cal O}_{
F^q_L}$, whereas there is {\em no}
fundamental $L_L^{ \prime }$ coupled to this operator. 
A fundamental $L_L$ couples
to the color singlet part of a {\em different} operator, ${\cal O}_{
F^l_L}$.
We see that fundamental fermions do not
have to be in complete $SO(10)$ multiplets since the full 
$SO(10)$ global symmetry of the CFT is not gauged,
but they do have to be parts of $SO(10)$ multiplets 
(providing understanding of their quantum numbers)
since they couple to
CFT operators which  are in complete $SO(10)$ multiplets.
The operator ${\cal O}
$ creates out of the vacuum spin-$1/2$ composites
(just like $J^{ \mu }
$ creates spin-$1$ composites). These hadrons are dual to
fermion KK modes on RS1 side (again, in complete $SO(10)$ multiplets).
Thus, fundamental gauge and fermion fields are exactly (and no more) 
as in the SM (with the addition of the right-handed neutrinos).  Up to mixing with CFT
composites, these {\em are} the SM fields and are dual to zero-modes of
fermions and gauge fields on the RS1 side.

The
scaling dimension of ${\cal O}
$
determines the mixing between fundamental fermions ($\psi$) 
and CFT composites and is dual to 
the bulk fermion mass parameter $c$ as follows. 
The choice $c > 1/2$ for light fermions is dual to the
{\em ir}relevant coupling between fundamental fermions and CFT operators
so that the
mixing between $\psi$ and CFT composites is small. Thus,
SM fermion is mostly fundamental and its coupling to composite 
$\rho$-mesons
(which goes through this mixing) is small.
Whereas, $c \lesssim  0$ for $t_R$ is dual to a {\em relevant}
coupling of fundamental $t_R$ to CFT operator corresponding to a large
mixing between fundamental $t_R$ and CFT composites. This implies 
that
SM $t_R$ contains a sizable admixture 
of composites and that its coupling to $\rho$-mesons is large.
We see that this CFT picture agrees qualitatively with the couplings to gauge
KK modes obtained on the $5D$ side as presented in subsection \ref{subsec:estimates}:
the coupling of $t_R$ to $Z^{\prime}$ is enhanced by $\sqrt{k \pi r_c} $ 
because $t_R$ and $Z^{ \prime }$
are localized near the TeV brane while the coupling of 
light fermions to $Z^{\prime}$ is suppressed due to the small overlap of their wave functions.

\subsection{Lightness of the LZP}
\label{cftlight}

Next, we consider the dual interpretation of the
lightness of the LZP for $c \sim -1/2$ or smaller. For this purpose, it is 
more convenient to consider 
a different CFT description \cite{Contino:2004vy} which is equivalent to
a
dual interpretation of the {\em other} chirality of the LZP
(not shown in Eq. \ref{table}),
denoted $\hat{ \nu^{ \prime } }_R$.

As before, since $\hat{ \nu^{ \prime } }_R$ is $(+-)$, whereas 
$\hat{ t }_R$ is $(--)$,
we add a fundamental $\hat{ \nu^{ \prime } }_R$ 
(but not $\hat{ t }_R$) in the dual CFT and couple it
to the color singlet part of
$\hat{ {\cal O} }_{ F^q_{ R \; 1 } }$. Also, on the $5D$ side, both
$\hat{ \nu^{ \prime } }_R$ and $\hat{ t }_R$ have $(-)$ boundary condition 
on the TeV brane -- this results in a zero-mode for $t_R$, but
not for $\nu^{ \prime }_R$. The dual interpretation is that 
the CFT operator $\hat{ {\cal O} }_{ F^q_{ R \; 1 } }$ 
interpolates massless composites with quantum numbers of
$\nu^{ \prime} _R$ and $t_R$\footnote{Since $SO(10)$ or Pati-Salam is
{\em not} spontaneously broken by CFT (this is dual to $SO(10)$ being
unbroken on {\em TeV} brane), the composites have to be in complete
$SO(10)$ multiplets.}. The former gets a Dirac mass with the
fundamental
$\hat{ \nu^{ \prime } }_R$, 
whereas the latter (with no fundamental fermion to marry)
is the SM $t_R$. 

Recall that, on the
$5D$ side, the
effective $c$ for $(+-)$ helicity is opposite to that of $(-+)$ helicity,
i.e. $c$ for $\hat{ \nu^{ \prime } }_R$ is $ \sim +1/2$ or larger 
meaning  that the coupling of the fundamental $\hat{ \nu^{ \prime } }_R$ to the CFT 
operator is close to marginal and the mixing of the
fundamental fermion
with composites is mild. 
The Dirac mass for the fundamental
$\hat{ \nu^{ \prime } }_R$ with the CFT composite must go through this 
mild mixing. 
Thus, this mass 
is smaller 
than the mass of other composites (like $\rho$-meson, i.e., gauge KK mass). 
In the CFT picture, the
$(+-)$ helicity of LZP is mostly the fundamental $\hat{ \nu^{ \prime} }_R$
and the $(-+)$ helicity is mostly
the massless composite 
interpolated by the CFT operator.  This provides a dual interpretation
for the fact that the
$(-+)$ helicity couples strongly to gauge KK modes (i.e., $\rho$-mesons
in the CFT picture), 
whereas the $(+-)$ couples weakly.

We see that particles localized near the TeV brane such as $t_R$ zero-mode, 
Higgs, KK modes (most of them, except, for example,
$(+-)$ helicity of LZP) are mostly composites
in the CFT picture. This is expected since the
TeV brane corresponds to the IR of the CFT so that
particles 
localized there correspond to IR degrees of freedom (i.e. composites) of the CFT.
Similarly, particles localized near the Planck brane (light fermion zero-modes,
$(+-)$ helicity of LZP) are mostly fundamental/external in  
the CFT picture.  Again, this is expected since the Planck brane
corresponds to the UV in the $4D$ picture. Particles localized there
correspond to UV 
degrees of freedom in the CFT picture, in contrast with composite states.

\subsection{Baryon number}

The dual interpretation of baryon-number symmetry is as follows. First, note that
the composite $X_s$ cannot couple SM $Q_L$ to SM $L_L$ since
these fermions have their origin in fundamental fields (and in CFT operators
since SM fermions have an admixture of CFT composites)
which are not related by the unified symmetry. So, proton decay
from exchange of $X_s$ states is absent.

Recall that to suppress $B$ violation from higher-dimensional operators,
$U(1)_B$ is  gauged in the bulk by adding spectators on the Planck brane.
The dual interpretation is that the CFT and the 
fundamental fermions
coupled to it\footnote{Spectator fermions are also fundamental,
but not directly coupled to CFT.} have 
exact  global  $U(1)_B$ symmetry (i.e.  $SO(10) \times U(1)_B$ symmetry)
which is gauged
by a $4D$ vector field.
The operator ${\cal O}_{ F^q_L }$ has $B = -1/3$ since its color triplet part is
coupled to fundamental
$Q_L$ to which we assign $B = 1/3$. We cannot couple 
fundamental $L_L$ (assigned $B = 0$) 
to color singlet part of ${\cal O}_{ F^q_L }$ since it
also has $B = -1/3$. Thus, composite fermions
interpolated by color singlet part of ${\cal O}_{ F^q_L }$ (which are 
dual to $L_L^{ \prime }$ KK modes) have $Z_3$-charge. As mentioned before, a
fundamental $L_L$ couples instead
to a different operator ${\cal O}_{ F^l_L }$ which has $B = 0$. 
Similarly, $J^{ \mu }_{ 
X_s }$ has color, but $B = 0$
and hence composite $X_s$ have $Z_3$ charge.

On the $5D$ side, the $U(1)_B$  gauge symmetry
is broken 
by the Planckian vev of a SM singlet scalar
living on the Planck brane.
In the $4D$ picture, the $U(1)_B$ gauge theory is also Higgsed near the Planck 
scale.  The gauge boson coupled to $U(1)_B$ current and spectators get
a Planckian mass.
At this scale, operators involving fundamental fields
and/or CFT operators 
violating $U(1)_B$ are 
allowed. 
For example, the coupling of fundamental
$L_L$ to the color  singlet part of ${\cal O}_{ F^q_L }$
would now be allowed. This will result in mixing of $L_L$ with composite  
fermions
interpolated by
${\cal O}_{ F^q_L }$. Recall that the
SM $Q_L$ has an admixture of composites interpolated by the
color triplet part of the {\em same} operator. Thus,
there will be a coupling of
composite $X_s$ to SM $L_L$ and SM $Q_L$ and other similar couplings.
These couplings, in turn,
will lead to too fast proton decay. 

So, just as on the $5D$ side, 
we impose the $Z_3$ symmetry in the CFT picture
so that $\Delta B \neq 1/3, 2/3$ in order to forbid the above coupling
of $L_L$ to ${\cal O}_{ F^q_L }$.
However, operators such as $Q_L^3 L_L$ are still allowed.
The central point is that these operators are
suppressed by the Planck scale, i.e., such 
violations of $U(1)_B$
are strongly irrelevant in the IR of the CFT coupled to fundamental
fermions and light gauge fields\footnote{Higher-dimensional
operators generated by the breaking of conformal
invariance and suppressed by the TeV scale (which are dual
to TeV-brane localized operators on the $5D$ side) do not break $U(1)_B$.}.
In other words, at sub-Planckian energies, $U(1)_B$ is an accidental  
and anomalous
global symmetry very much like 
in the SM. This is the dual of the fact that
$U(1)_B$ is unbroken on the RS1 side throughout the bulk
and on the TeV brane so that
baryon-number violating operators are allowed only on the Planck brane
(hence
the ones which have $\Delta B \neq 1/3, 2/3$
are suppressed by $M_{Pl}$).

Finally, the
bulk breaking of $SO(10)$ and the resulting
splitting in $c$'s within a $SO(10)$ multiplet
means that the
CFT has only approximate global $SO(10)$  
symmetry
so that different parts of fermionic operator (for example,
color singlet and triplet parts of
${\cal O}_{ F^q_L }$) can have slightly different scaling
dimensions.

\subsection{Duality at semi-quantitative level}
\label{cftquant}

So far, our CFT description was qualitative. If we assume that the CFT is like 
a large-$N$ ``QCD'' theory, i.e., $SU(N)$ gauge theory (with some ``quarks''), we can
perform a semi-quantitative check of the duality and 
even obtain
estimates for couplings of KK modes using the CFT picture.
We begin 
with the coupling of the Higgs to gauge KK mode
(see, for example, \cite{dual}). On the $5D$ side, 
this coupling is
$\approx g \sqrt{ 2 k \pi r_c }
\approx \sqrt{ 2 g^2_{ 5D } k }$
(see Eq. \ref{gaugeKKHiggs}).
All three
particles in this coupling are localized near the TeV brane.
In the CFT picture, this is a coupling of $3$ composites.
We use the naive dimensional analysis
(NDA) of large-$N$ QCD  to estimate the size of this coupling (see, for example, ref. \cite{witten}):
\begin{equation}
\hbox{coupling of} \; 3 \; \hbox{composites} \sim \frac{ 4 \pi }{ \sqrt{N} }
\end{equation}
With a coupling of this size, 
loops are suppressed
by $\sim 1/N$ compared to tree-level. 
Assuming the duality, we equate the above two couplings to obtain 
the following relation between $N$, number of colors of the CFT,
and the parameters of the $5D$ theory
\begin{equation}
\sqrt{ g_{ 5D }^2 k } \sim \frac{ 4 \pi }{ \sqrt{N} }
\label{g5DN}
\end{equation}
A consistency check of this relation can be obtained  by comparing
the low-energy gauge coupling on the two sides 
(see 6th reference of \cite{rscft}).
On the CFT side, we get
\begin{equation}
1/ g_4^2 \sim \frac{ N } { 16 \pi^2 } 
\log \left( \frac{ k }{ \hbox{TeV} } \right)
\end{equation}
This is due to contributions of CFT quarks to the running of external gauge couplings
from the Planck scale down to the TeV scale (just like the contribution of
SM quarks to the running of $\alpha_{QED}$).
Whereas, using $\log \left( k / \hbox{TeV} \right)
\sim k \pi r_c$,
we can rewrite the zero-mode low energy
gauge coupling on the $5D$ side (see Eq. \ref{0mode}) as
\begin{equation}
1/g_4^2 = \log \left( k / \hbox{TeV} \right) / \left( g_{ 5D } ^2 k \right)
\end{equation}
These two gauge couplings agree using the relation in Eq. \ref{g5DN}.
\footnote{Here, we neglected a localized kinetic term on the Planck brane. 
Small Planck brane kinetic terms 
on the RS1 side means, in the CFT picture, 
that the external gauge coupling in the CFT picture has 
a Landau pole at the Planck scale.}
In particular, we see that $N \sim 5-10$ is required to get $O(1)$ low-energy
gauge coupling.
Next, consider the 
coupling of a gauge KK mode to two KK fermions,
for example, the coupling of two LZP's to $Z^{ \prime }$. 
Again, all three particles are localized 
near the TeV brane. 
Using the CFT picture, this coupling  
is 
$\sim 4 \pi / \sqrt{N}$ 
since it is a coupling of three composites. As mentioned above (relating
$N$ to $g_{ 5D }$), this is $\sim g_4 \sqrt{ k \pi r_c }$
\footnote{On the $5D$ side, a numerical
evaluation of Eq. \ref{3kk} confirms 
that this coupling is indeed $\sim g_4 \sqrt{ k \pi r_c }$.}.
As expected, this 
is similar to 
gauge KK coupling to the Higgs.

A similar argument and estimate 
hold for the coupling of $t_R$ zero-mode to a gauge KK mode
and 
a KK fermion (for example, coupling to LZP and $X_s$ KK mode,
see Eq. \ref{zero2kk}) or coupling of two $t_R$ zero-modes to gauge KK
mode (for example, $Z^{ \prime }$, see Eq. \ref{zero1kk}). The reason is that  
$t_R$ zero-mode is localized near the TeV brane,
i.e., in the CFT picture, 
SM $t_R$ is mostly composite.

\end{appendix}


\end{document}